\documentclass[prb,aps,twocolumn]{revtex4}
\usepackage{amsmath}
\usepackage{amssymb}
\usepackage {graphicx}

\newcommand{\beq}{\begin{equation}}
\newcommand{\eeq}{\end{equation}}
\newcommand{\bea}{\begin{eqnarray}}
\newcommand{\eea}{\end{eqnarray}}
\newcommand{\e}{\varepsilon}

\newcommand{\bk}{{\bf k}}
\newcommand{\bp}{{\bf p}}
\newcommand{\bq}{{\bf q}}

\newcommand{\nn}{\nonumber}
\newcommand{\bse}{\begin{subequations}}
\newcommand{\ese}{\end{subequations}}
\newcommand{\bwt}{\begin{widetext}}
\newcommand{\ewt}{\end{widetext}}

\newcommand{\I}{\mathrm{Im}}
\newcommand{\R}{\mathrm{Re}}
\newcommand{\bsu}{\begin{subequations}}
\newcommand{\esu}{\end{subequations}}

\newcommand{\lr}{\left(}
\newcommand{\rr}{\right)}

\begin{document}

\title{Collective modes in two- and three-dimensional electron systems\\ with Rashba spin-orbit coupling}
\author{Saurabh Maiti,$^{a,b}$ Vladimir Zyuzin,$^a$ and Dmitrii L. Maslov$^a$}
\affiliation {$^a$Department of Physics, University of Florida, Gainesville, FL 32611\\
$^b$National High Magnetic Field Laboratory, Tallahassee, FL 32310}
\date{\today}

\begin{abstract}
In addition to charge plasmons, a 2D electron system with Rashba-type spin-orbit coupling (SOC) also supports three collective modes in the spin sector: the chiral-spin modes. We study the dispersions of the charge and spin modes and their coupling to each other within a generalized Random Phase Approximation for arbitrarily strong SOC, and both in 2D and 3D systems. In both 2D and 3D, we find that the charge plasmons are coupled to only one of the three chiral-spin modes. This coupling is shown to affect the dispersions of the modes at finite but not at zero wavenumbers. In 3D, the chiral-spin modes are strongly damped by particle-hole excitations and disappear for weak electron-electron interaction. Landau damping of the chiral-spin modes in 3D is directly related to the fact that, in contrast to 2D, there is no gap for particle-hole excitations between spin-split subbands. The gapless continuum is also responsible for Landau damping of the charge plasmon in 3D - a qualitatively new feature of the SOC system. We also discuss the optical conductivity of clean 2D and 3D systems and show that SOC introduces spectral weight at finite frequency in a such way that the sum rule is satisfied. The in-plane tranverse chiral-spin mode shows up as dispersing peak in the optical conductivity at finite number which can can be measured in the presence of diffraction grating. We also discuss possible experimental manifestations of chiral-spin modes in semiconductor quantum wells such InGaAs/AlGaAs and 3D giant Rashba materials of the BiTeI family.
\end{abstract}

\maketitle
\section{Introduction}
\label{sec:intro}
Spin-orbit interaction lifts the spin degeneracy by coupling electron momenta and spins. This provides a possibility to manipulate electron spins by purely electrical means, which is the ultimate goal of the growing field of spintronics.\cite{spintronics1,spintronics2,spintronics3,spintronics4} Of particular interest are the Rashba- and Dresselhaus-type spin-orbit couplings (SOCs) which occur in systems without center of inversion (either local or global).
The Rashba SOC has mostly been studied in two-dimensional (2D) electron and hole gases in semiconductor heterostructures, and also in surface states of metals. This effect (whose strength may be characterized in terms of the splitting of the otherwise degenerate spin-up and spin-down levels) is usually weak in semiconductors \cite{InGaAs} but is much stronger in the surface states of noble metals\cite {Au} and of semi-metallic bismuth,\cite{Bi} and is further enhanced in surface metal alloys. \cite{giantR1,giantR2,giantR3}

The new excitement in this field is stimulated by the discovery of a number of three-dimensional (3D) materials with giant SOC; perhaps the most investigated class of such materials are polar semiconductors BiTeX (X=Br, Cl, I), both the bulk \cite{BiteI_giantR} and surface states \cite{eremeev:2012}
of which were shown to have a giant spin splitting of the Rashba type. While it is the surface-induced asymmetry that is responsible for Rashba SOC in 2D systems, the origin of the effect in 3D BiTeX is a local electric field (along the $c$-axis), which acts on the Bi-plane sandwiched between the polar Te and X layers. \cite{BiTeI_origin}  Both \emph{ab initio} calculations \cite{BiTeI_theory2} and spin polarized angular-resolved photoemission \cite{chirality_exp} have provided support to this picture. Another interesting feature of the BiTeX family is that, in contrast to semiconductor heterostructures which usually have both Rashba and Dresselhaus spin-orbit interactions, BiTeX are \emph{purely} Rashba materials with no competing Dresselhaus effect.

Investigating the electronic properties of a Rashba metal requires a sound understanding of its excitations--both of the single-particle and collective types; the focus of this paper is on the latter. Various aspects of the collective modes in 2D systems with SOC have been studied in the past. It is important to bear in mind that a typical 2D system is a quantum well (QW) formed in a semiconductor heterostructure. Quantization of electron motion in the direction perpendicular to the QW plane splits the conduction (or valence) band into subbands. As a result, the excitation spectrum has both the intra- and inter-subband parts. In the absence of SOC, the intrasubband part consists of a particle-hole continuum and a charge plasmon mode with a $\sqrt{q}$ dispersion at small $q$ [Fig.~\ref{fig:scheme}(a), bottom].\cite{ando}  (As is the case for any system with a repulsive inter-particle interaction,\cite{LL9} the intrasubband spin collective mode lies entirely within the continuum and is thus heavily damped.)  If only the lowest transverse subband is occupied, intersubband transitions occur between this and the first few unoccupied subbands. The top part of Fig.~\ref{fig:scheme}(a) depicts the intersubband spectrum for the case of transitions between the lowest and first unoccupied subband. Intersubband transitions give rise to a separate region of the particle-hole continuum and to two kinds of collective modes: an intersubband plasmon above the continuum and three degenerate spin modes (\lq\lq spin plasmons\rq\rq) below the continuum, see Fig.~\ref{fig:scheme}(a), top.\cite{Dahl,Ryan,Giuliani} The energy scales of the intersubband transitions are on the scale of tens of meV, which makes them accessible to inelastic light scattering spectroscopy (see Ref.~\onlinecite{isb} for an extensive review of the experiment in this area).

The effect of SOC on intersubband transitions has been studied both theoretically\cite{Flatte} and experimentally.\cite{Babu1,Babu2}  The main result of these studies is that SOC lifts the degeneracy of the three spin-plasmons at finite $q$ [see Fig.~\ref{fig:scheme}(b), top]. A detailed comparison between the theory and experiment was carried out in Ref.~\onlinecite{Babu1}.

The effect of Rashba SOC on the intrasubband charge plasmons has
also been studied in some detail. \cite{wu:2003,Wang,Gumbs,Kushwaha,Gritsev,Anisotropic Plasmon}
Coupled spin-charge plasmons in a helical Fermi liquid (a system with a Dirac spectrum due to SOC only) has also been investigated within the Random Phase Approximation in Ref.~\onlinecite{Raghu}. It is
by now well established that transitions between Rashba subbands give rise to an additional--``Rashba''--
continuum which lies above the charge continuum [see Fig.~\ref{fig:scheme}(b), bottom]. Also, in addition to the usual 2D plasmon with a $\sqrt{q}$ dispersion, which corresponds to the oscillations of the total charge density, a 2D Rashba metal supports also an optical plasmon mode. (A third plasmon mode lies within the  Rashba continuum and is thus unobservable.) The intrasubband $\sqrt{q}$ plasmon gets damped by particle-hole excitations within the Rashba continuum.

The spin collective modes in a 2D system with Rashba SOC arising due to transitions between the two spin-split bands have been studied only fairly recently{\cite{shekhter,ali_maslov, Zhang} and, so far, only theoretically. The main prediction of the theory is the existence of three spin modes (\lq\lq chiral-spin waves\rq\rq\/) that arise solely due to SOC [as opposed to spin plasmons which exists even in the absence of SOC), see Fig.~\ref{fig:scheme}(b), bottom]. These modes are intrinsic collective excitations of a 2D FL with Rashba SOC\cite{ali_maslov_rashba} and, to some extent, analogs of the Silin-Leggett spin modes in a partially-polarized Fermi liquid (FL). \cite{silin:1958,leggett:1970,baym} The important difference between the chiral-spin and Silin-Leggett modes is that the former exist in the absence of the external magnetic field and arise from the effective Rashba field acting on electron spins.

The primary goal of this paper is to study the nature of collective modes in both 2D and 3D metals with Rashba SOC; when dealing with the 2D case, we will be focusing entirely on
the modes arising from transitions within the lowest spin-split subbands and ignoring transitions to the \lq\lq confinement-split\rq\rq subbands.  Separate treatment of the excitations between the spin-split and confinement-split subbands
is possible if the energy splitting due to confinement is much larger than that due to SOC--this is in fact true for the semiconductor heterostructures, where the confinement energy is significantly larger than the SOC splitting. \cite{Babu1,Babu2} In what follows, the term \lq\lq intersubband\rq\rq\/ will be reserved for the spin-split subbands of a SOC system.

\begin{figure*}[htp]
$\begin{array}{cc}
\includegraphics[width=0.5\columnwidth]{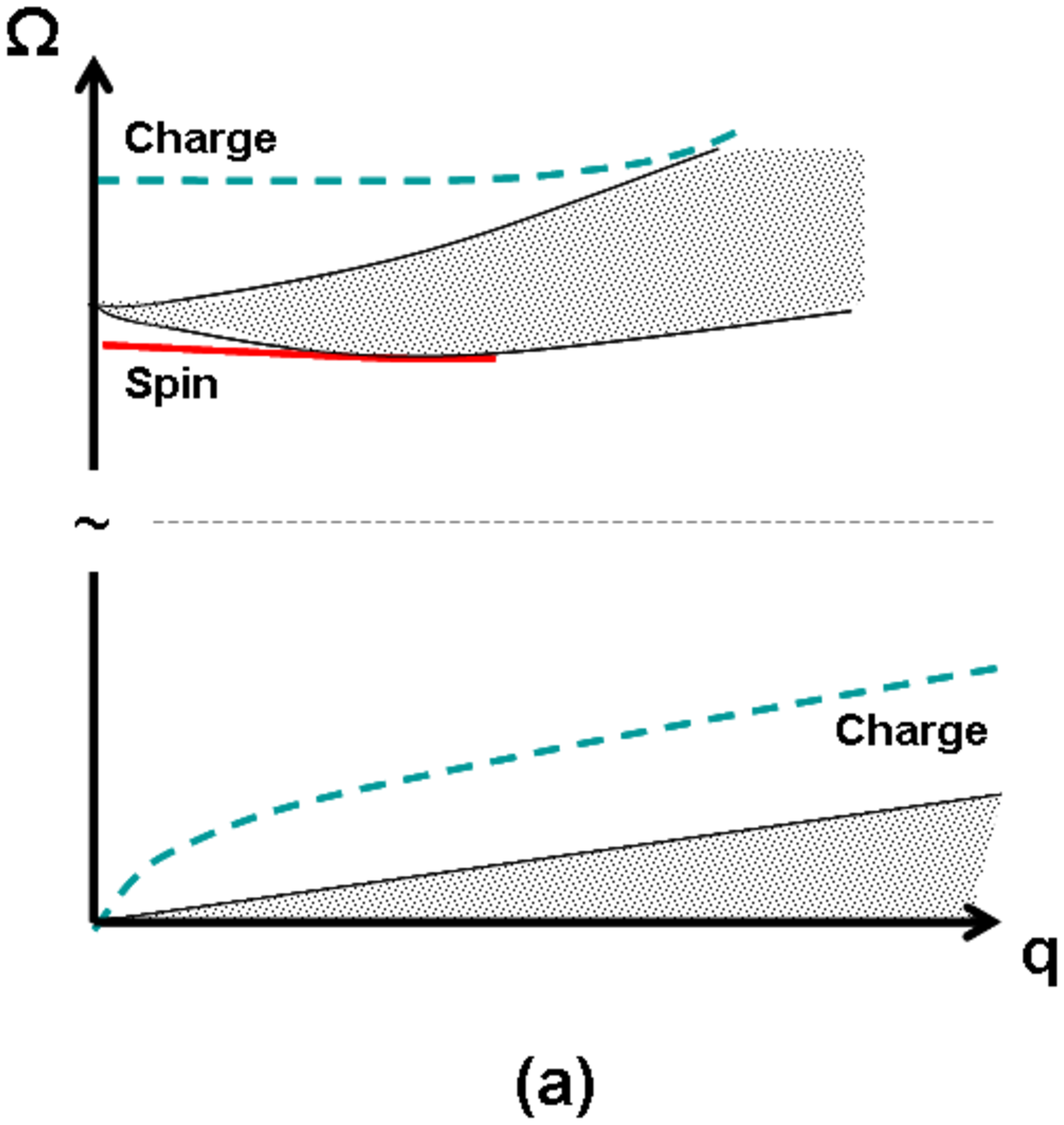}&
\includegraphics[width=0.8\columnwidth]{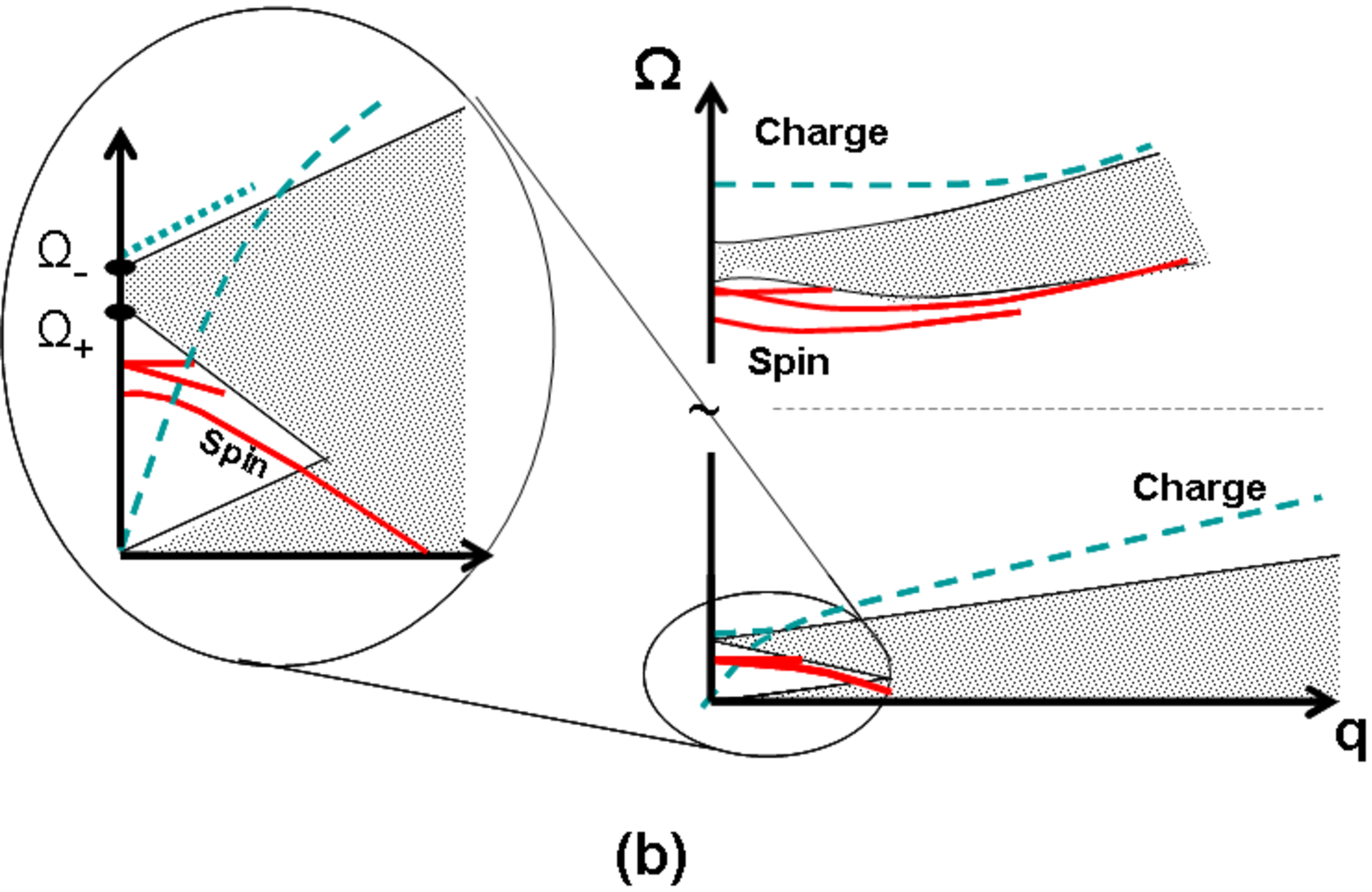}
\end{array}$
\caption{\label{fig:scheme} Schematic picture of the single-particle continua (shaded region) and collective excitations (lines) in the absence(a) and presence(b) of Rashba SOC in a 2D quantum well with the lowest subband occupied. The lower-frequency part is the intrasubband region, the higher-frequency part is the intersubband region. The solid (dashed) lines correspond to spin (charge) modes. Inset: zoom of the intersubband region. While SOC affects both inter- and intrasubband regions, it brings about qualitatively new effects, i.e., new spin and charge modes and the Rashba continuum, in the intrasubband region. }
\end{figure*}

The chiral-spin modes have been studied within a FL theory\cite{shekhter,ali_maslov} and within Random Phase Approximation (RPA) in the spin channel of a neutral system (cold atoms).\cite{Zhang} There is an infinite number of such modes  but only three of them are isotropic in the momentum space and thus couple to macroscopic electric and magnetic fields. These three isotropic modes correspond to longitudinal and transverse oscillations of magnetization in the absence of the magnetic field.

In principle, a FL theory should give a full description of the collective modes in both the charge and spin sectors. However, the spin sector of a Rashba metal with arbitrarily strong SOC cannot be described by the FL theory, at least not by its conventional version that operates
with almost free quasiparticles.\cite{ali_maslov_rashba} The reason is that to describe excitations in the spin sector one needs to take into account states located in between the Rashba subbands, and these states are strongly damped if SOC is not weak. One way to avoid this problem is to focus on the case of weak SOC, which can be then treated as a perturbation imposed on an $SU(2)$-invariant FL.  The advantage of this approach is that the electron-electron interaction can be treated non-perturbatively. This was how the chiral-spin waves at $q=0$ and finite $q$ were analyzed
in Refs.~\onlinecite{shekhter} and \onlinecite{ali_maslov}, correspondingly. If SOC is not weak, the FL approach breaks down, and one needs to retort to some kind of the perturbation theory in the electron-electron interaction while keeping SOC arbitrary. Within this approach, the zero-sound and spin modes of a neutral Rashba system with short-range interactions were studied in Ref.~\onlinecite{Zhang} using the RPA theory. The new element arising from strong SOC is that the charge and spin sectors are no longer decoupled (as they were assumed to be in Refs.~\onlinecite{shekhter} and \onlinecite{ali_maslov}).

In this work, we study the charge and chiral-spin modes, as well as coupling between them, in both 2D and 3D electron systems. We treat the electron-electron interaction within a generalized RPA, which takes into account both the long- and short-range components of the screened Coulomb interaction, while keeping SOC arbitrary.

In 2D, our results are as follows. 1)In the charge sector, we find that there are two plasmons--the first one is the usual 2D, $\sqrt{q}$ plasmon (damped by particle-hole excitations within the Rashba continuum) and the second one is an optical plasmon lying exponentially close to the upper edge of the Rashba continuum. The two-plasmon feature is generally consistent with earlier work \cite{Magarill,wu:2003,Wang,Gumbs,Kushwaha} although our result for the dispersion of the second plasmon mode disagrees with that found in Refs.~\onlinecite{Gumbs,Kushwaha}. 2) We calculate the optical conductivity and explicitly show that the spectral weight is redistributed between the Drude peak and the Rashba continuum in a such way that the sum rule is satisfied. 3) In the spin sector we find, in agreement with the previous literature,\cite{shekhter,ali_maslov,Zhang} that there are three modes split off from the lower edge of the Rashba continuum. However, we also find that SOC couples the charge  the chiral-spin modes in a very specific manner: the plasmons are coupled to only one of the chiral-spin modes while the other chiral-spin modes are coupled to each other. These couplings affect the dispersions of the respective modes but not their masses, i.e., the mode frequencies at $q=0$.
4) Within the FL approach, valid for weak SOC, the masses of the chiral-spin modes
were expressed via the FL parameters in Refs.~\onlinecite{shekhter,ali_maslov}. We show that, depending on the strength of SOC, there are, in fact, two regimes. The first one corresponds to that found within the FL theory which assumes that SOC is the weakest interaction in the system. The second one corresponds to the case when SOC is stronger than the electron-electron interaction. At the weakest electron-electron coupling, the chiral-spin modes in this case are exponentially close to the continuum boundary. 5) Thus far, all collective modes were studied for the case when both the spin-split subbands were occupied. We show that the chiral-spin modes survive even if only one the lowest subband is occupied.

In 3D, however, collective modes behave in a way that is qualitatively different from the 2D case. (By  \lq\lq 3D\rq\rq\/ here we mean a situation when the free-electron term in the Hamiltonian is extended to 3D while the Rashba term remains 2D; such a case is relevant to BiTeI.)
1) In the charge sector, there is one out-of-plane optical plasmon which is not affected at all by in-plane SOC; the other charge mode is an in-plane optical plasmon which, for
material parameters relevant to giant Rashba semiconductors of the BiTeX family, is damped by the particle-hole excitations between Rashba subbands
- a new feature of SOC systems.
2) In contrast to 2D, where the Rashba continuum starts at finite energy, the continuum in 3D is present at all energies. Therefore, the chiral-spin modes are Landau-damped by particle-hole excitations even at $q=0$. However, for a sufficiently strong electron-electron interaction, the imaginary part of spin susceptibility shows a broad dispersing peak corresponding to a damped chiral-spin mode. 3) We also calculate the optical conductivity in 3D and show explicitly that the redistribution of the spectral weight is consistent with the sum rule (just like in 2D).

The rest of the paper is organized as follows. In Sec~\ref{sec:Model}, we introduce the model and lay out the general strategy for finding the collective modes. The formalism in this section is general and holds both for the 2D and 3D cases. In Sec.~\ref{sec:2D}, we revisit the collective modes in 2D, demonstrate consistency with previous work, and point out some details missed earlier in the literature. In Sec.~\ref{sec:3D}, we consider the 3D case.
In Sec.\ref{sec:other effects}, we relate our theoretical predictions to the experiment. Sec.~\ref{sec:conclusion} summarizes our results. Appendices \ref{sec:appA}-\ref{subsec:correction} contain details of derivations not presented in the main text.
\section{Model and general strategy}\label{sec:Model}

We start with the following Rashba Hamiltonian for non-interacting electrons (we set $\hbar=1$,
unless specified otherwise):
\bse
\bea\label{eq:Hamiltonian}
\hat H_0&=&\sum_k \Psi^{\dag}_\bk \mathcal{H}_k \Psi_\bk,
\label{ham1}\\
\hat{\mathcal H}_\bk&=&\left(\frac{k_1^2+k_2^2}{2m_1}+\frac{k_3^2}{2m_3}\right)\hat\sigma_0+ \alpha(\hat{\boldsymbol{\sigma}}\times \mathbf{k})_3 \label{ham2},
\\
\Psi^{\dag}&=& (c_{\bk\uparrow}^{\dag}, c_{\bk\downarrow}^{\dag}), \eea
\ese
where the $x_1$ and $x_2$ axes of a Cartesian system are the in plane, the $x_3$ axis is along the normal to the plane,
$m_{1/3}$ is the effective in-plane/out-of-plane mass,
$\alpha$ is the Rashba parameter that encodes the strength of
the spin-orbit interaction, and $\hat{\boldsymbol{\sigma}}=(\hat\sigma_0,\hat\sigma_1,\hat\sigma_2,\hat\sigma_3)$ is a vector of Pauli matrices with $\sigma_0=\hat\sigma_0$. (Later on, in Sec.~\ref{sec:other effects}, we will also take Dresselhaus SOC into account.)
Upon diagonalizing the Hamiltonian, one obtains two branches of the energy spectrum corresponding to the opposite chiralities:
\beq\label{eq:Chiral spectrum}
\e_{\bk}^{\pm}=\frac{k_{\parallel}^2}{2m_1} +
\frac{k_{3}^2}{2m_3}\pm \alpha k_{\parallel},\eeq
where $k_{\parallel}^2=k_1^2+k_2^2$.
It is worth noting that Rashba SOC in 3D can have various forms depending on the lattice symmetries of the material.\cite{vorontsov} In our continuum model, we restrict our consideration to Rashba SOC that couples only in-plane components of the electron spin and momentum. The 3D ellipsoidal dispersion corresponds to the case of BiTeI, where the Fermi energy is smaller than the inter-plane hopping and thus the Fermi surface is closed. In this case, the 2D regime is obtained by putting $k_3=0$ (rather than taking the limit $m_3\to \infty$ in the final results for the 3D case).

The Matsubara Greens' function for the noninteracting system is
then given by
\begin{subequations}
\bea
\hat{G}(K)&=&\sum_s
\hat{\Omega}_s(\mathbf{k})g_s(K),\label{eq:Greens Function_a} \\
\hat{\Omega}_s(\mathbf{k})&=&\frac{1}{2}\left[ \hat\sigma_0+s\left( \hat\sigma_1 \sin\theta_{\bk}-\hat\sigma_2 \cos\theta_{\bk}\right)  \right] ,\label{eq:Greens Function_b}\\
g_s(K)&=&\frac{1}{i\omega_m-\e_{\bk}^s+\mu},\label{eq:Greens Function_c} \eea
\end{subequations}
where $K\equiv(i\omega_m,\mathbf{k})$, $s=\pm$ is the chirality index, $\mu$ is the chemical potential (measured from the Dirac point), and $\theta_{\bk}$ is the angle
between the projection of $\bk$ onto the $x_1x_2$ plane and the $x_1$-axis. (In 2D, the $\mathbf{k}$ vectors are always in the $x_1x_2$ plane.)

The collective modes of an interacting system show up as poles of the full susceptibilities defined as
\bea\label{eq:susceptibilities1}
\chi_{ij}(\mathbf{r},\mathbf{r}')&=&-\int_0^{1/T}~~d\tau \langle
T_{\tau}
\mathcal{O}_i(\mathbf{r},\tau)\mathcal{O}_j(\mathbf{r}',0) \rangle,
\eea
where $\mathcal{O}_i=\Psi^{\dag}\sigma_i\Psi$ with $i=0\dots 3$ are the charge and spin densities. Equations (\ref{eq:susceptibilities1}) will be evaluated within the perturbation theory in the electron-electron interaction but for an arbitrary spin-orbit strength
$\alpha$. It is useful to define bare susceptibilities as $\chi_{ij}^{0}(Q)=-\Pi_{ij}^0(Q)$, where
\bea\label{eq:susceptibilities2}
\Pi^0_{ij}(Q)&=&\int_K \text{Tr}\left[\hat\sigma_i
\hat G(K)\hat\sigma_j \hat G(K+Q)\right],\eea
with $Q=(\textbf{q},i\Omega_n)$ and $\int_K\equiv T\sum_{\omega_m}\int\frac{d^D k}{(2\pi)^D}$, $D=2,3$. To obtain
the full susceptibilities, we perform a generalized RPA sum
as illustrated in Fig.~\ref{fig:RPA_scheme}. The generalized RPA sums up
a chain-like series of polarization bubbles which, in turn, contain ladder series of vertex corrections. The interaction
vertex (due to the Coulomb interaction) in the $\Psi$ basis is given by
\bea\label{eq:int_vertex}
\Gamma_{\alpha\gamma;\beta\delta}(\textbf{q})&=&V(q)\delta_{\alpha\beta}\delta_{\hat\gamma\delta},\nonumber\\
V(q)\equiv V&=&\left\{\begin{array}{cl}\frac{2\pi e^2}{q}~~~\text{in 2D}\\
\frac{4\pi e^2}{q^2}~~~\text{in 3D.}
\end{array}\right.\eea
\begin{figure}[htp]
$\begin{array}{c}
\includegraphics[width=0.9\columnwidth]{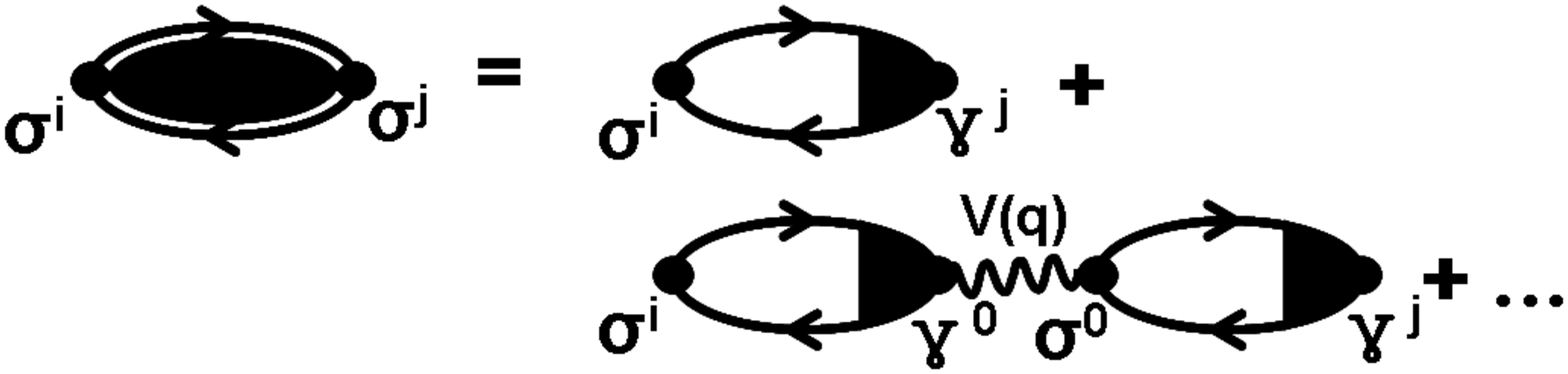}\\
\includegraphics[width=0.9\columnwidth]{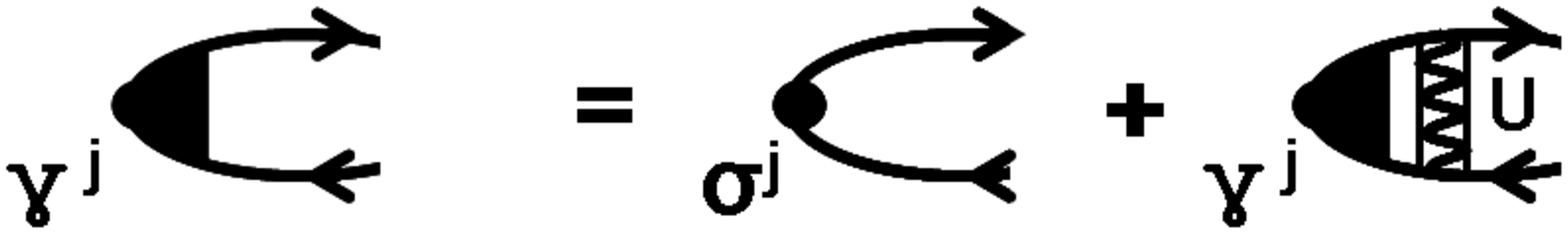}
\end{array}$
\caption{\label{fig:RPA_scheme}(Top): RPA sum for $\chi_{ij}$. The wavy line is the bare coulomb interaction $V(q)$ which carries the external momentum $\mathbf{q}$. The shaded corners denote the vertex corrections to each bubble which is obtained by summing a ladder series. (Bottom): The ladder series for the vertex corrections. Each boxed wavy line is a screened Coulomb interaction.}
\end{figure}

Making use of the tensor structure of the bare interaction vertex,
it is straightforward to carry
out the sum in Fig. \ref{fig:RPA_scheme} with the result
\bea\label{eq:RPA_sum} \chi_{ij}(Q)&=&-\left[\Pi_{ij}^U+\Pi_{i0}^U
\frac{V}{1-V\Pi_{00}^U}\Pi_{0j}^U \right], \eea
where $\Pi_{ij}^U$ is the bubble that contains the vertex corrections. This object
includes all diagrams that cannot be split into two by cutting
just one interaction line. Notice that the summation scheme in  Fig.~\ref{fig:RPA_scheme} is
so far exact. However, one needs to resort to some kind of an approximation to actually compute the vertex-corrected bubble.
The interaction inside a bubble is supposed to be a screened Coulomb potential. Here, we adopt an approximation in which this interaction is replaced by a momentum-independent constant ($U$). We will show that this approximation reproduces the known results obtained within the FL theory once $U$ is properly identified with the FL parameter. With the assumptions formulated above, it is now possible  to evaluate the ladder sum in the following way:
\bse
\bea\label{eq:Ladder sum0} \Pi_{ij}^U(Q)&=&\text{Tr}\int_K
\hat\sigma_i \hat G(K)
\hat\gamma_j \hat G(K+Q),\\
\label{eq:Ladder sum} \hat\gamma_j&=&\hat\sigma_j
-U\int_K
\hat G(K)\hat\gamma_j \hat G(K+Q), \eea
\ese
where $\hat\gamma_j$ is a  $2\times 2$ vertex which,
due to isotropy of the interaction, can only be a function of the transferred momentum $\mathbf{q}$ (which is the same as the external
momentum). We now expand $\hat\gamma_j$ over a complete set of Pauli matrices
\bea\label{eq:gamma} \hat\gamma_j&=&M_j^a\hat\sigma_a, \eea
where $a\in {0,1,2,3}$ and the coefficients $M_j^a$ form a $4\times 4$ matrix. Substituting
this back into Eq.~(\ref{eq:Ladder sum}), we find
\bea
\hat{\Pi}^U&=&\hat{\Pi}^0\hat{M}, \nonumber\\
\hat{M}&=&\left(\hat\sigma_0+\frac{U}{2}\hat{\Pi}^0 \right)^{-1}.
\label{eq:vertex chi}
\eea
In the absence of SOC, $\Pi^0_{0j}=0$
for $j=1,2,3$. The collective modes in the charge sector are
given by the roots of $1-V(q)\Pi_{00}^U=0$, while the collective modes in the
spin sector are given by the poles of $\Pi_{ij}^U$ which are
solutions of Det$(\hat M^{-1})=0$. To associate the modes with the corresponding
susceptibilities, one can first find the poles at $q=0$, when all the modes are decoupled
and then trace the dispersions at finite $q$. If the modes do not intersect, as it will be shown to be
the case here, such an identification is unique.

The problem is thus reduced to calculating $\Pi_{ij}^0(Q)$.
Using the definition of the Greens' function in Eq.~(\ref{eq:Greens
Function_a}), the sixteen
components
$\Pi_{ij}^0(Q)$ can be expressed in the
following compact form
\bea\label{eq:pi} \Pi_
{ij}^0(Q)&=&
\frac{1}{2}\int_K~~\mathcal{T}_{ij},\nonumber\\
\mathcal{T}_{ij}&=&\sum_{r,s\in{\pm}}~g_r g_s
\mathcal{F}_{ij}^{rs}. \eea
Explicit expressions for  $\mathcal{T}_{ij}$  and for the matrix elements,
$\mathcal{F}_{ij}^{rs}$, are presented in Appendix
\ref{subsec:Tij}. It is useful to realize that, both in 2D and 3D, the system possesses a rotational symmetry in the $x_1x_2$ plane. This allows us to choose the projection of $\mathbf{q}$ onto the $x_1x_2$ plane as the $x_1$ axis. One more simplification occurs if we note that while performing an integral over $\bk$, reflection about the $x-$axis ($\theta_\bk\rightarrow -\theta_\bk$) implies that $\theta_{\bk+\bq}\rightarrow -\theta_{\bk+\bq}$.
In the subband Green's function $g_{r}$ [Eq.~(\ref{eq:Greens Function_c})],
the angular dependence always enters as
$\cos\theta_\bk$ and/or
$\cos\theta_{\bk+\bq}$. These two points together imply that all the
terms with $\sin\theta_\bk$, $\sin\theta_{\bk+\bq}$, and
$\sin(\theta_\bk\pm\theta_{\bk+\bq})$  that appear in
$\mathcal{T}_{ij}$'s (see Appendix \ref{subsec:Tij}) vanish. This reduces our
consideration to only the following six components:
$\Pi^0_{00}$, $\Pi^0_{02}$, $\Pi^0_{13}$,
$\Pi^0_{11}$, $\Pi^0_{22}$, and $\Pi^0_{33}$. A simple exercise shows
that this also ensures a block-diagonal structure of the matrices
$\hat{\Pi}^0$ and $\hat{\Pi}^U$. It is important to note that, while this property is valid for any rotationally-invariant interaction, it is only guaranteed for linear Rashba SOC.

The six non-zero components of the $4\times 4$ susceptibility tensor (Eq. \ref{eq:RPA_sum}) can then be subdivided into
two decoupled sectors: the 1-3 sector
\bea
\chi_{ij}(Q)&=&-\Pi^U_{ij}~
\mathrm{with}~~\{ij\}\in\{11,13,33\}
\label{eq:susceptibilities3}
\eea
and the 0-2 sector
\bea\label{eq:susceptibilities32}
\chi_{00}(Q)&=&-\frac{\Pi^U_{00}}{1-V\Pi^U_{00}},\nonumber\\
\chi_{02}(Q)&=&-\frac{\Pi^U_{02}}{1-V\Pi^U_{00}},\nonumber\\
\chi_{22}(Q)&=&-\Pi^U_{22}-\frac{\Pi^U_{20}V~\Pi^U_{02}}{1-V\Pi^U_{00}}.
\eea
The remaining $\chi_{ij}$ vanish.
This is precisely the (partial) decoupling of the charge and chiral-spin modes mentioned in Sec.~\ref{sec:intro}: the $22$ susceptibility is coupled to the $00$ (charge) susceptibility, whereas the $11$ susceptibility is coupled to the $33$ susceptibility. The formulation presented above is applicable in both 2D and 3D; the specific results depend on the structure of $\hat{\Pi}^0$. We now apply this general scheme to specific situations, beginning with the 2D case.

\section{Collective modes in a two-dimensional Rashba system}
\label{sec:2D}
The 2D case is obtained by setting $k_3=0$ in Eqs.~(\ref{ham2}) and (\ref{eq:Chiral spectrum}). In Secs.~\ref{subsec:polarization tensor}-\ref{subsec:coupling}, we discuss the analytic results for the collective modes at small $q$, in particular, we derive analytical expressions for the masses of the collective modes in the spin and charge sector, discuss the coupling between the spin and charge modes, and analyze the redistribution of the spectral weight in the conductivity. The numerical results for dispersions of the modes, valid for any $q$, are presented in Sec.~\ref{sec:anyq}.

\subsection{Spin-charge polarization tensor in two dimensions}\label{subsec:polarization tensor}
We begin by discussing the polarization tensor for non-interacting electrons, $\Pi_{ij}^0$, for the case when both Rashba subbands are occupied, i.e., $\mu>0$, as shown in Fig.~\ref{fig:subband}a. The case of only one occupied subband, corresponding to Fig.~\ref{fig:subband}b will be discussed in Sec.~\ref{sec:mu<0}. Since $\Pi^0_{00}(0,\Omega)=0$ by total charge conservation, to capture the physics in the charge
sector we need to preserve the leading order $q$ dependence in
$\Pi^0_{00}$ which, as will be shown below, appears as $q^2$. This requires expanding all the components of $\hat{\Pi}^0$ to
$\mathcal{O}(q^2)$. The diagonal components are expandable in even powers of $q$ while the off-diagonal components are expandable in
odd powers of $q$. Upon analytic continuation $i\Omega_n\to \Omega+i\delta$, we obtain expansions of the six non-zero components of $\hat\Pi^0$ to $\mathcal{O}(q^2)$
\bea\label{eq:NonZeroPi}
\Pi_{00}^0&=&\frac{m_1}{2\pi}\left[\left(\frac{p_0 q}{m_1\Omega}\right)^2-\frac{q^2}{8m_1\Omega}L(\Omega)\right],\nonumber\\
\Pi_{11}^0&=&\frac{m_1}{2\pi}\left[-1-\frac{\Omega}{8m_1\alpha^2}L(\Omega) + A_{11}(\Omega)\left(\frac{q}{2m_1\alpha}\right)^2\right],\nonumber\\
\Pi_{22}^0&=&\frac{m_1}{2\pi}\left[-1-\frac{\Omega}{8m_1\alpha^2}L(\Omega) + A_{22}(\Omega)\left(\frac{q}{2m_1\alpha}\right)^2\right],\nonumber\\
\Pi_{33}^0&=&\frac{m_1}{2\pi}\left[-2-\frac{\Omega}{4m_1\alpha^2}L(\Omega) + A_{33}(\Omega)\left(\frac{q}{2m_1\alpha}\right)^2\right],\nonumber\\
\Pi_{13}^0&=&-i\frac{m_1}{2\pi}A_{13}(\Omega)\left(\frac{q}{2m_1\alpha}\right) + \mathcal{O}(q^3),\nonumber\\
\Pi_{02}^0&=&-\frac{m_1}{2\pi}A_{02}(\Omega)\left(\frac{q}{2m_1\alpha}\right) + \mathcal{O}(q^3).
\eea
Here,
\beq\label{eq:definitions}
p_0\equiv\sqrt{2m_1\mu+m_1^2\alpha^2}
\eeq
and
\beq
\label{eq:p0}
p_\pm\equiv p_0\mp m_1\alpha
\eeq
are the Fermi momenta of the Rashba subbands.
The expressions in Eq.~(\ref{eq:NonZeroPi}) are valid in the limit of
$q\ll m_1\alpha,\Omega/v_F$.

The functions $A_{ij}$ in Eq.~(\ref{eq:NonZeroPi}) are given by
\bea\label{eq:def2} A_{11}(\Omega)&=&
\left(\frac{p_0\alpha}{\Omega}\right)^2 +
\frac{m_1\alpha^2}{4\Omega}L(\Omega) +
\frac{3\Omega}{2\alpha}a_1-\frac38m_1\Omega
a_2,\nonumber\\
A_{22}(\Omega)&=& 3\left(\frac{p_0\alpha}{\Omega}\right)^2 -
\frac{m_1\alpha^2}{4\Omega}L(\Omega) +
\frac{\Omega}{8\alpha}a_1-\frac18m_1\Omega
a_2,\nonumber\\
A_{33}(\Omega)&=& \frac{m_1\alpha^2}{2\Omega}L(\Omega) +
\frac{\Omega}{2\alpha}f_1-\frac12m_1\Omega a_2,\nonumber\\
A_{13}(\Omega)&=& -\frac{\Omega}{4m_1\alpha^2}L(\Omega) +
\frac{\Omega}{4\alpha}
a_1,\nonumber\\
A_{02}(\Omega)&=& -\frac18L(\Omega) +
\frac{2m_1\alpha^2}{\Omega}
\eea
with
\bea\label{eq:f-functions}
a_1&\equiv&\frac{2\Omega(p_-+m_1\alpha)}{m_1(4\alpha^2 p_-^2 -\Omega^2)}-(\alpha\rightarrow -\alpha),\nonumber\\
a_2&\equiv&\frac{2\Omega}{m_1(4\alpha^2 p_-^2-\Omega^2)}-\frac{(p_-+m_1\alpha)^2}{m_1^2}\frac{8p_-\alpha\Omega}{(4\alpha^2 p_-^2-\Omega^2)^2}\nonumber
\\
&&-(\alpha\rightarrow-\alpha).
\eea
The function \beq\label{eq:L}L(\Omega)=
\ln\left[\frac{(\Omega-\Omega_-
+i\delta)(\Omega+
\Omega_++i\delta)}{(\Omega-
\Omega_+
+i\delta)(\Omega+
\Omega_-+i\delta)}\right[
 \eeq
with $\Omega_{\pm}=2\alpha p_{\pm}$ arises from transitions between the Rashba subbands, as shown in Fig.~\ref{fig:subband}. The interval of frequencies $\Omega_{+}\leq \Omega \leq \Omega_-$, where $\I L(\Omega)\neq 0$, corresponds to the Rashba continuum of width $\Omega_--\Omega_+=4m_1\alpha^2$. The logarithmic structure of  $L(\Omega)$ is responsible for most of the interesting properties of the collective modes.

\begin{figure}[htp]
$\begin{array}{cc}
\includegraphics[width=0.5\columnwidth]{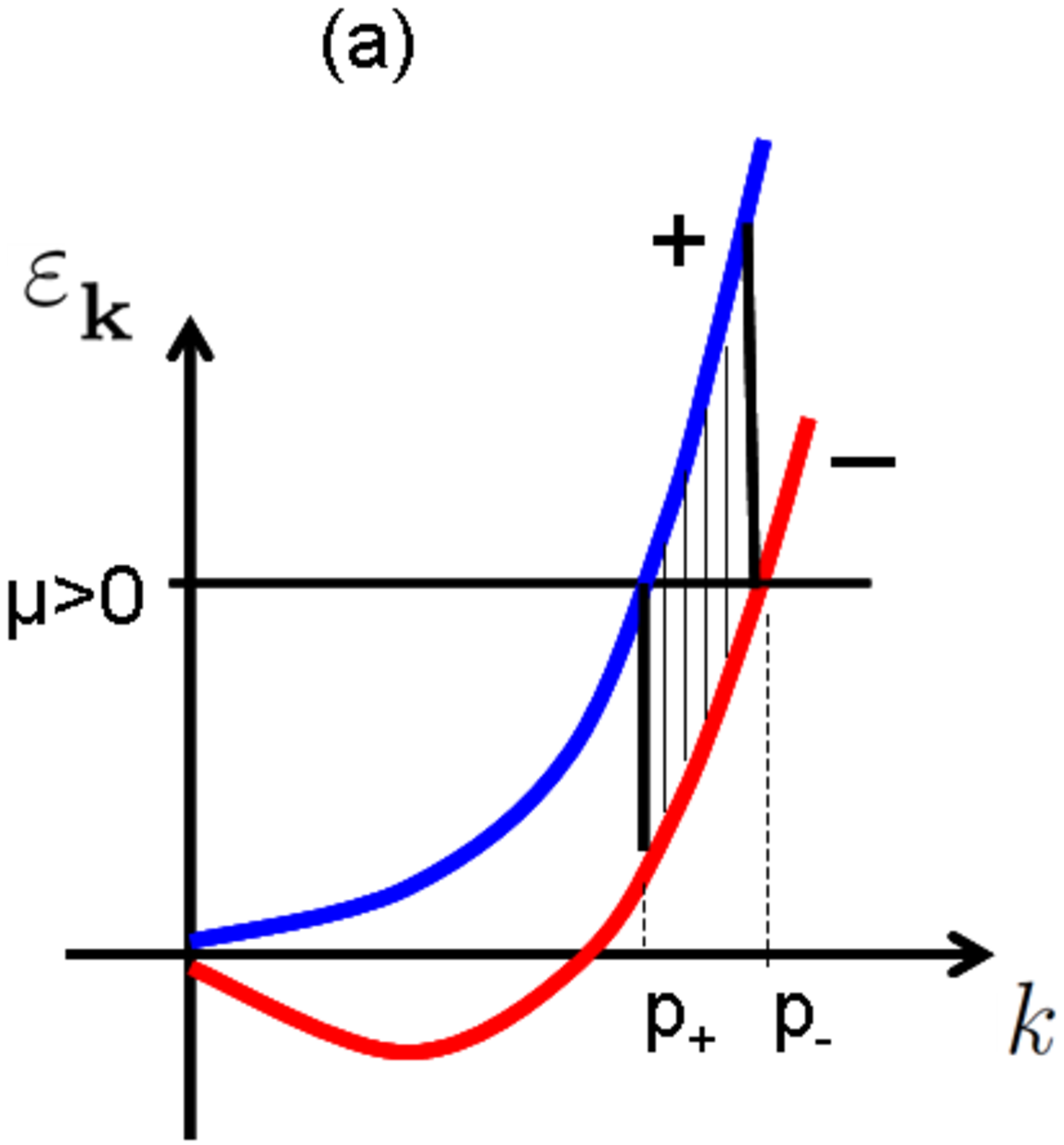}&
\includegraphics[width=0.5\columnwidth]{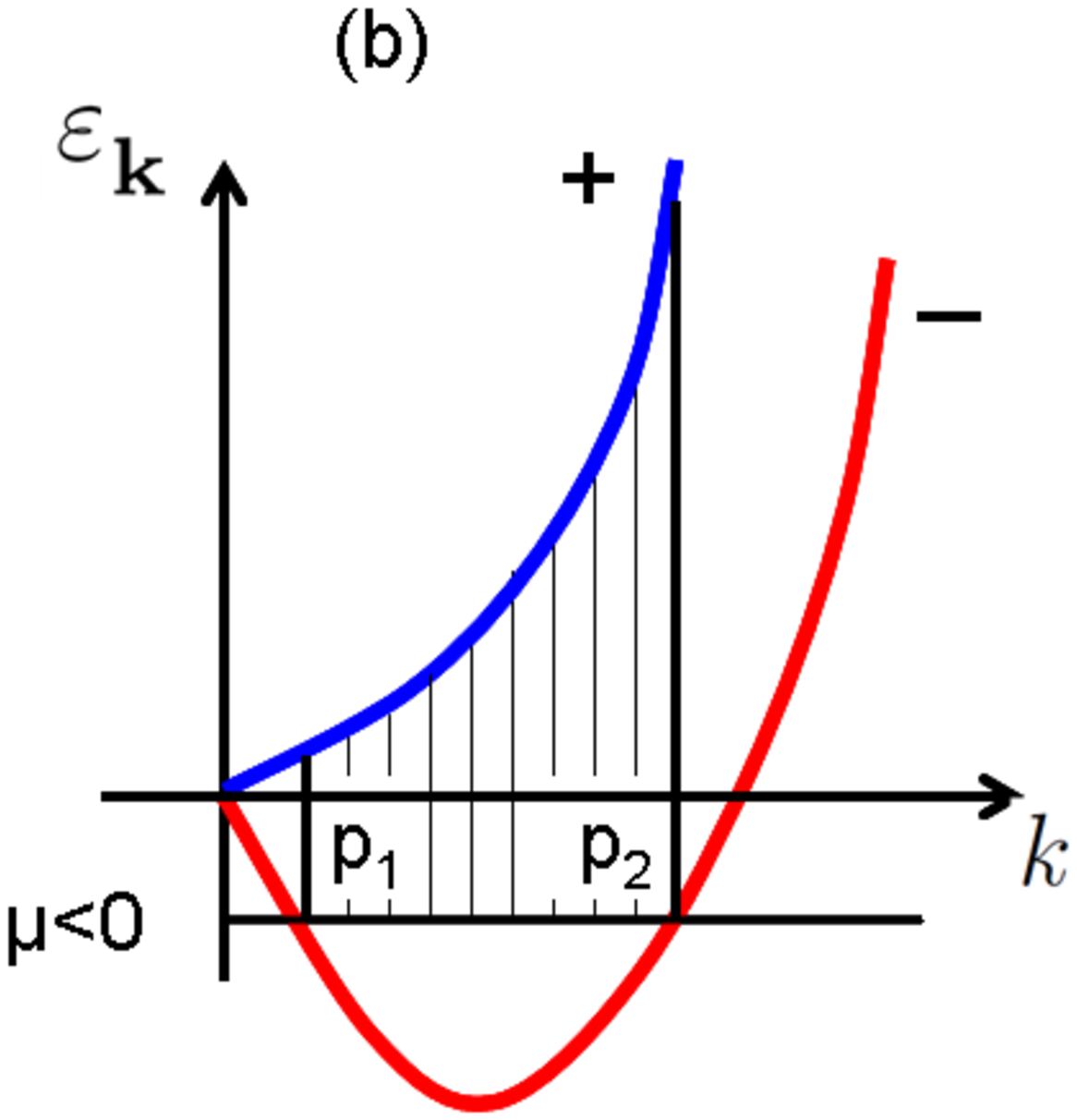}
\end{array}$
\caption{\label{fig:subband}
Intersubband transitions for the cases when a) both the Rashba subbands are occupied and b) only the lowest subband is occupied. Labels $\pm$ denote chiralities of the Rashba subbands.
}
\end{figure}

\emph{A note on a small $q$ expansion:}
While the above expansions are completely straightforward,
the entire derivation is too cumbersome to
be presented here. Nevertheless, we would like to highlight some key
aspects in the behavior of $\Pi_{ij}^0$'s that differ from the case without SOC. It follows from Eq.~(\ref{eq:00}) that, in the limit of small $q$,
the angular factors arising from the matrix elements reduce to $1+\cos(\theta_{\bk}-\theta_{\bk+\bq})\approx 2-\mathcal{O}(q^2)$ and $1-\cos(\theta_\bk-\theta_{\bk+\bq})\approx \frac{q^2}{2k^2}\sin^2\theta_
{\bk}$. Furthermore, the combination $g_+g_++g_-g_-$, which corresponds to intrasubband transitions, gives the same contribution as in the absence of SOC and thus scales as $\propto{q^2}/{\Omega^2}$, while the combination $g_+g_-+g_-g_+$ involves integration in the region between the two Fermi surfaces giving rise to the logarithmic factor $L(\Omega)$ [Eq.~(\ref{eq:L})]. The latter is the intersubband contribution that makes a system with SOC qualitatively different from a 2D Fermi gas. The need to integrate over the momentum interval in between the Rashba subbands, where quasiparticles are  in general not well-defined, is also a roadblock for the development of a FL theory for systems with arbitrary SOC.\cite{ali_maslov_rashba} To evaluate $\Pi^0_{00}$ [Eq.~(\ref{eq:00})] to $\mathcal{O}(q^2)$, we  notice that the contribution of the convolution of the Green's functions  to the intrasubband part  scales as $q^2$, while the matrix element is independent of $q$ in the limit of $q\to 0$. For the intersubband contribution, the matrix element scales as $\propto q^2$ while the Green's functions give a $q$ independent logarithmic factor [$L(\Omega)$ in Eq.~(\ref{eq:L})]. For the $\Pi^0_{33}$ component, however, a similar consideration shows that the intrasubband contribution is absent to order $q^2$ (the first non-zero term occurs at order $q^4$), while one needs to keep $\mathcal{O}(q^2)$ corrections to $L(\Omega)$ in the intersubband part. This is the origin of the function $A_{33}$. All the $\Pi^0_{ij}$'s thus have both intrasubband and intersubband contributions  dressed appropriately by the angular factors arising from the matrix elements.

\subsection{Charge sector:
plasmons, optical conductivity,  and the sum rule}\label{subsec:RPA}
As mentioned in Sec.~\ref{sec:intro}, the plasmon mode of a  two-dimensional electron gas (2DEG) with Rashba SOC has been studied in great detail in earlier work.\cite{Magarill,wu:2003,Wang,Gumbs, Kushwaha} Here, we demonstrate how the plasmon modes are obtained within our formalism and compare our results with
those of prior work. We also study coupling between plasmons and the modes in the spin sector. We show that the existence of two plasmon modes is a 2D Rashba metal follows naturally from the RPA-ladder approach developed in Sec.~\ref{sec:Model}. As a consistency check, we also show in Appendix \ref{subsec:jj} that the same results can also be obtained by calculating the conductivity by using either the quantum Boltzmann equation or the Kubo formula.

\subsubsection{Plasmons from the RPA-approach}
Plasmon modes are manifested by poles in the charge susceptibility, $\chi_{00}$. The poles coincides with the roots of the equation $1-V(q)\Pi_{00}^U=0$, where $\Pi_{00}^U ={\Pi^0_{00}}\lr{1+\frac{U}{2}\Pi^0_{00}}\rr^{-1}$. Neglecting the short-range component of the interaction ($U$), upon which $\Pi^U_{00}=\Pi^0_{00}$,  and using the form of $\Pi^0_{00}$ from Eq.~(\ref{eq:NonZeroPi}), we see that the long-wavelength limit of plasmon modes occur as solutions of a transcendental equation (we choose $\Omega>0$ for convenience)
\bea\label{eq:transcendental2} \frac{\Omega^2}{q}&=&
e^2\left(\frac{p_0^2}{m_1}-\frac{\Omega}{8}\R L(\Omega)\right). \eea
\begin{figure}[htp]
\includegraphics[width=0.9\columnwidth]{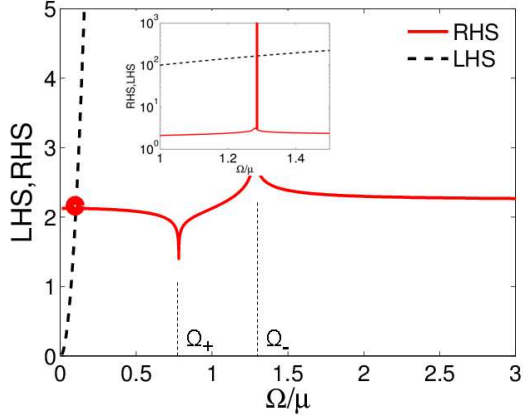}
\caption{\label{fig:trans}
Graphic solution of Eq.~(\ref{eq:transcendental2}). The root marked by a dot corresponds to the 2D plasmon with a  $\sqrt{q}$ dispersion. Inset: A zoom on peak at $\Omega_-$ in logarithmic scale showing the second root which corresponds to the optical plasmon mode arising solely due to SOC.}
\end{figure}

It is obvious that Eq.~(\ref{eq:transcendental2}) has a real solution(s) only outside of the Rashba continuum, i.e., either for $0\leq \Omega<\Omega_+$ or for $\Omega>\Omega_{-}$. The left- and right-hand sides (LHS and RHS, correspondingly) of Eq.~(\ref{eq:transcendental2}) are plotted in Fig.~\ref{fig:trans}. One of the roots (marked by the dot) is prominent.  To obtain its dispersion analytically, one can neglect the second term in the RHS of Eq.~(\ref{eq:transcendental2}), which yields
\bea\label{eq:plasmons_sol1} \Omega_1(q)&=&
\sqrt{\frac{e^2p_0^2}{m_1}}q^{1/2}~~ +~~\mathcal{O}(q).
\eea
This is the usual 2D plasmon with a $\sqrt{q}$ dispersion; the coefficient of the $\sqrt{q}$ term is renormalized by SOC. The second root is more subtle. Since we have already found the mode with dispersion vanishing at $q\to 0$ and
since the state at $q=0$ is non-degenerate, we expect the other mode to have finite frequency at $q\to 0$. At finite $\Omega$ and $q\to 0$, the LHS of Eq.~(\ref{eq:transcendental2}) diverges; therefore, the RHS must diverge too. This is only possible if $\Omega$ approaches the upper boundary of the Rashba continuum ($\Omega_{-}$) from above, such that $\R L(\Omega)$ is negative and diverges as $\ln(\Omega-\Omega_-)$. Neglecting the first term in the RHS of Eq.~(\ref{eq:transcendental2}) and replacing $\Omega$ by $\Omega_{-}$ in all the factors under the logarithm except for the one that vanishes at $\Omega=\Omega_-$, we obtain
\bea
\Omega_2(q) &=& \Omega_- + \frac{\Omega_- +
\Omega_+}{p_0/m_1\alpha}~e^{-\frac{8\Omega_-}{e^2q}}.\eea
This second root is shown in the inset of Fig.~\ref{fig:trans}.

Thus a 2D Rashba system formally has two plasmon modes: one mode is the usual, $\sqrt{q}$ plasmon, expected for any 2D system, and the other mode is split off (exponentially weakly at small $q$) from the upper edge of the Rashba continuum. At larger $q$, the boundaries of the continuum themselves disperse with $q$ (see, e.g., Ref.~\onlinecite{Gritsev}) and the second plasmon tracks the upper boundary of the continuum. At finite but small $q$ ($q\ll m_1\alpha$), the dispersion of the second plasmon can be written as
\beq\label{eq:plasmon2_b} \Omega_2 \approx \Omega_-(q) +
\frac{\Omega_-(q) +
\Omega_+(q)}{p_0/m_1\alpha}~e^{-\frac{8\Omega_-(q)}{e^2q}},
\eeq
where the $q$-dependent boundaries of the Rashba continuum are\cite{Gritsev}

\beq\label{eq:z}\Omega_{\pm}(q)= \pm\frac{(q\pm 2m_1\alpha + p_F)^2-p_F^2}{2m_1}.\eeq

One might wonder if the exponentially weak dispersion of $\Omega_2$ exceeds the accuracy of the small $q$ expansion. We would like to stress this is not the case. To see this, we note that an expansion of $\Pi_{00}^0$ to fourth order in $q$ can be expressed as (see Appendix \ref{subsec:plasmons_rtefact})
\beq\label{eq:3}\Pi_{00}^0
\propto
c_1q^2 + c_2q^4 + (c_3q^2 + c_4q^4)L[\Omega(q)],
\eeq
where $c_i$'s are some coefficients with appropriate dimensions. Dropping the $q^4$ terms  implies the smallness of the $q^4L[\Omega(q)]$
term relative to the $q^2$ term near the second plasmon branch. For our result to be valid, we thus need the distance between $\Omega_2(q)$ and $\Omega_-$ to be larger than $e^{-\mathrm{const}/q^2}$. Our solution suggests that this distance is of order
$\sim e^{-\Omega_-q}\gg e^{-\Omega_-^2/q^2}$, which is well within the region of validity.

At this point we would like to compare our results with the ones obtained previously by different groups, \cite{wu:2003,Wang,Gumbs,Kushwaha,Gritsev,Magarill} not all of which agree with each other on the number and type of plasmons. Our results partially agree with those of Refs.~\onlinecite{wu:2003,Gritsev,Magarill}. Reference~\onlinecite{Magarill} identified the two plasmon branches and also the exponential closeness of the optical branch to the Rashba continuum; however, the dispersion of the continuum boundaries was ignored in this work. Reference~\onlinecite{Gritsev} did not identify the optical plasmon--probably due to its exponential closeness to the continuum. Reference~\onlinecite{wu:2003} correctly identified the two plasmons but also reported a third plasmon at the lower edge of the Rashba continuum, which we do not find. Reference~\onlinecite{Gumbs} also reported the two plasmons, but both the shape of the Rashba continuum and the $\sqrt{q}$-plasmon dispersion  disagree with our results, as well as with that by others. Reference~\onlinecite{Kushwaha} identified two plasmons and noticed correctly that the third plasmon is damped inside the Rashba continuum; however, our result for the optical phonon dispersion disagrees with theirs. We believe that the disagreement is due to the fact that the sign in the equation for the plasmon dispersion in Ref.~\onlinecite{Kushwaha} is opposite to that in ours, as well as to that in other works.

Recall that we had neglected the short-range component of the interaction in arriving at our results for the plasmon modes. It is safe to do so because, to
leading order in $q$, the presence of the short-range interaction does not affect the plasmons. To see this, let us go back to the equation $1-V(q)\Pi^U_{00}=0$ and expand $\Pi^U_{00}$ in $U$ as $\Pi^U_{00}\approx \Pi_{00}^0+\frac{U}{2}(\Pi^0_{00})^2$. From Eq.~(\ref{eq:NonZeroPi}) we see that $(\Pi^0_{00})^2\sim{q^4}/{\Omega^4}$. Hence, the equation for the plasmon mode acquires a correction of order $U{q^3}/{\Omega^4}$. While this provides a subleading, $q^{3/2} U$ correction to the $\sqrt{q}$ plasmon, it leaves the exponential behavior of the second plasmon unchanged.

It is necessary to point out that the exponential proximity of the optical plasmon to the continuum makes it hard to be detected. Any broadening of the continuum due to finite temperature or disorder will smear this mode out.

\subsubsection{Optical conductivity and the sum rule}
In this section, we demonstrate how the optical sum rule is satisfied in the presence of Rashba SOC. The optical conductivity can be found from the Kubo formula as $\sigma_{ij}(\Omega)= i\mathcal{K}_{ij}(\Omega)/{\Omega}$, where $\mathcal{K}_{ij}$ is the current-current correlation function at $T=0$ (as before, $\Omega>0$)
\bea\label{eq:conductivity_bubble}
\mathcal{K}_{11}(\Omega)&=&
e^2\int\frac{d^2k}{(2\pi)^2}
\int^0_{-\Omega}\frac{d\omega}{2\pi}\text{Tr}\left[\hat{v}_1 \hat{G}^R_{\omega}\hat{v}_1\hat{G}^A_{\omega+\Omega}\right]
\eea
where
\beq
\label{eq:vx}
\hat{v}_1=\frac{k_1}{m_1}\sigma_0-\alpha\hat\sigma_2
\eeq
and the superscript $R(A)$ denotes the retarded (advanced) Green's function. Due to in-plane symmetry, we have $\mathcal{ K}_{11}=\mathcal{ K}_{22}\equiv\mathcal{ K}$.  As shown in Ref.~\onlinecite{Magarill} (see also Appendix \ref{subsec:Kubo}),
\bea\label{eq:conductivity_bubble2}
\mathcal{ K}(\Omega)&=&-e^2\alpha^2\frac{m_1}{2\pi}\left(1+\frac{\Omega}{8m_1
\alpha^2}L(\Omega)\right).\eea
\begin{figure}[htp]
\includegraphics[width=0.9\columnwidth]{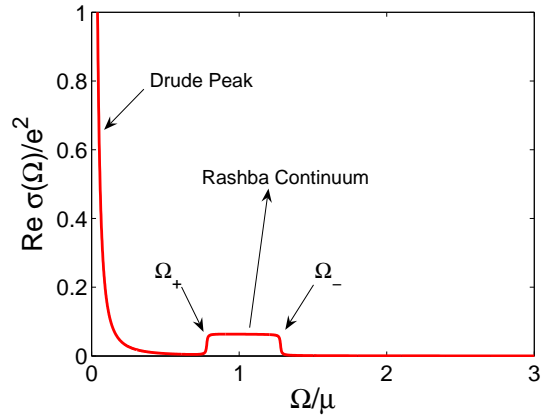}
\caption{\label{fig:Cond2d}
The real part of the conductivity of a non-interacting 2D system with Rashba SOC. A small imaginary part, $1/\tau=0.01\mu $, was added to the denominators of the Green's functions in the Kubo formula to simulate of the effect of disorder.}
\end{figure}

On adding the diamagnetic term ${n_{2D}e^2}/{m_1}$ to the last result, the conductivity can be written as
\bea\label{eq:conductivity_kubo} \sigma(\Omega)&=& i
e^2\left[\frac{\frac{n_{2D}}{m_1}-\frac{m_1\alpha^2}{2\pi}}{\Omega}-
\frac{ L(\Omega)}{16\pi}\right].\eea The real part of $\sigma(\Omega)$
is shown in Fig.~\ref{fig:Cond2d}. We added a small imaginary part ($\frac{i}{2\tau}$) to the denominators of the Green's functions in Eq.~(\ref{eq:conductivity_bubble})
to simulate disorder (under a rather crude assumption that the interband and intraband scattering rates are the same). A detailed description of the effects of disorder is outside the scope of this work.\cite{trushin}

In a clean system, the coefficient of the ${\Omega}^{-1}$ term in the imaginary part of the conductivity is
the Drude weight \beq\label{eq:DrudeWt}\mathcal{D}=e^2\pi\frac{n_{2D}-\frac{m_1^2\alpha^2}{2\pi}}{m_1},\eeq

which can also be defined as
\beq\label{eq:drudewt2} \mathcal{D} =
e^2\pi\lim_{\Omega,q\rightarrow 0} \frac{\Omega^2}{q^2}\Pi_{00},
\eeq
where the limit $q\to 0$ is taken first.  There is thus, a correction $\mathcal{O}(\alpha^2)$ to the Drude weight due to SOC.\cite{Magarill,Agarwal,Mishchenko_Graphene,Mishchenko_Farid1}
In addition, $\R\sigma(\Omega)\neq 0$ in the interval of frequencies corresponding the Rashba continuum,
$\Omega_+\leq \Omega\leq\Omega_-$. This is to be contrasted to the case without SOC when $\R\sigma(\Omega)=0$ for $\Omega>0$. Both
the reduction of the Drude weight and a non-zero $\R\sigma$ at finite $\Omega$ occur because SOC, being a relativistic effect, breaks Galilean (but not translational) invariance.\cite{Agarwal,Mishchenko_Graphene,Mishchenko_Farid1}
While both these features have been discussed in the literature, their consequences for the optical sum rule has not been analyzed.

We now show explicitly that the sum rule is satisfied. The integrated spectral weight should be equal to
\beq
\int_0^\infty ~d\Omega~ \text{Re}\sigma(\Omega)= \frac{\pi}{2}e^2\frac{n_{2D}}{m_1}.
\eeq
The spectral weight of the Drude peak at $\Omega=0$ is ${\mathcal{D}}/{2}$. In the absence of SOC, the area under the Drude peak contains all the spectral weight and thus the sum rule is satisfied automatically. In the presence of SOC, ${\mathcal{D}}$ is reduced from its free-electron value of $e^2\pi{n_{2D}}/{m_1}$.
However, this loss of weight in the Drude peak is exactly recovered in the box-like feature at finite frequency (see Fig.~\ref{fig:Cond2d}), the area under which is exactly $e^2{m_1\alpha^2}/{4}$. Adding up these two contributions, we recover the total spectral weight of $\frac{\pi}{2}e^2{n_{2D}}/{m_1}$.
Electron-electron interaction gives rise two additional effects: (a) a correction to the Drude weight\cite{Agarwal} and (b)
a non-zero value of $\R \sigma(\Omega)$ outside the Rashba continuum. \cite{Mishchenko_Farid1}
Checking the sum rule in the presence of interactions is a more challenging task and we shall not dwell on this point any further.

For lattice systems, the above sum rule needs to be applied with care. Of course, if a sum is performed over all bands, we must recover the spectral sum $e^2\pi n_{2D}/2 m_e$ (the $f$-sum rule), where $m_e$ is the bare electron mass. However, if we model the conduction band by a parabolic spectrum with an effective mass $m_1$, then the sum rule is valid as long as there are no interband transitions. A spectral weight rearrangement due to SOC occurs at the energy scale $2\alpha p_0$. If this redistribution were to be experimentally verified, then it requires the band gap $E_g$ of the semiconductor to be large compared to $2\alpha p_0$ (which guarantees that the energy scales are well separated). In this case, one can formulate a \lq\lq band\rq\rq\/ sum rule (with the band mass $m_1$ as opposed to $m_e$ in the full $f$-sum rule), which stipulates conservation of the spectral weight within a given band. This procedure can then serve as a consistency check between the optical and Hall measurements: the latter provides the value of $n_{2D}$, while the former contains the spectral information. It is important to realize that for stronger SOC, $n_{2D}$ should be deduced not from just the Drude weight but rather from the spectral weight integrated up to $\Omega\sim 2\alpha p_0$.

\subsection{Spin sector:
chiral-spin modes and their coupling to the charge modes
}\label{subsec:coupling}
Thus far, we have analyzed the charge sector. We now investigate the chiral-spin sector and the coupling between the two sectors. Some of the important features can be tracked analytically in the limit of small $q$; those are discussed in Sec.~\ref{sec:q0}.  The quantitative aspects of coupling between the two types of modes  require a full numerical analysis, which will be presented in Sec.~\ref{sec:anyq}.

\subsubsection{Chiral-spin modes at $q=0$}
\label{sec:q0}
The chiral-spin modes were investigated in Refs.~\onlinecite{shekhter} and \onlinecite{ali_maslov} in the limit of weak SOC but for an arbitrarily strong interaction within the FL theory. Here, we relax the constraint of weak SOC and use the RPA+ladder scheme to find the collective modes. We explicitly show that this reproduces correctly the results in the small $\alpha$ limit upon expressing the FL parameters in terms of the short-range coupling constant, $U$.

We have already shown that, in general, the susceptibilities can be grouped into the 0-2 and 1-3 sectors (we remind that 0 stands for the charge component while $1,~2,$ and  $3$ stand for the Cartesian components of magnetization). As is evident from the structure of the polarization tensor discussed in Sec \ref{subsec:polarization tensor},  all the four channels decouple in the limit of $q\rightarrow 0$.  All $\Pi^0_{ij}$s with $i\neq j$ scale as
$q$ and hence vanish at $q=0$. Thus the masses of the modes (which are a $q=0$ feature)  are not affected by the $0-2$ and $1-3$ couplings.

\begin{figure}[htp]
\includegraphics[width=0.9\columnwidth]{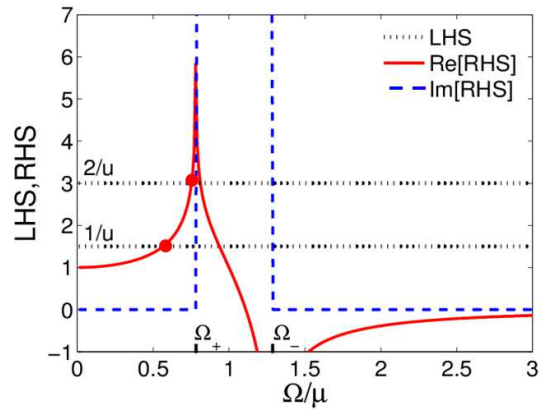}
\caption{\label{fig:trans2}Graphic solution of Eq.~(\ref{eq:modal mass}). The solid and dashed lines are the real and imaginary parts of the RHS of Eq.~(\ref{eq:modal mass}), respectively. The dash-dotted lines are the LHSs for $u=0.66$. The analytical forms of the weak coupling solutions are discussed in the text.}
\end{figure}

In the $q\rightarrow 0$ limit, Eqs.~(\ref{eq:vertex chi}), (\ref{eq:susceptibilities3}) and (\ref{eq:NonZeroPi}) suggest that the only non-zero components are $\chi_{jj}$ with $j=1,2,3$ which, in this limit, are given by $-\Pi^U_{jj}$. The collective modes correspond to poles of $\Pi^U_{jj}$ and are thus are given by the roots of the equations $1+\frac{U}{2}\Pi^0_{jj}=0$. This leads to the following transcendental equations for the masses of the modes
\bea\label{eq:modal mass}
\frac{2}{u}&=1+\frac{\Omega}{8m_1\alpha^2}L(\Omega),&~~~\text{for 11 and 22 modes};\nonumber\\
\frac{1}{u}&=1+\frac{\Omega}{8m_1\alpha^2}L(\Omega),&~~~\text{for 33 mode},
\eea
where $u=m_1 U/2\pi$ is the dimensionless interaction and $L(\Omega)$ is defined in Eq.~(\ref{eq:L}). The $11$ and $22$ modes are degenerate at $q=0$: this is guaranteed by the in-plane rotational symmetry of Rashba SOC. The LHS and RHS of Eq.~(\ref{eq:modal mass}) are plotted in Fig.~\ref{fig:trans2}.
Due to a logarithmic singularity in the real part of $L(\Omega)$, a solution exists for any value of $u$. At weak coupling ($u\ll 1$), the solution is exponentially close to the lower boundary of the Rashba continuum (marked by the dashed line). Searching for a solution of the form $\Omega= \Omega_+ -\delta$ with $|\delta|\ll \Omega_+$, we find for the masses of the $11$ and $22$ modes
\bea\Omega_{11}(0)=\Omega_{22}(0)&=&\Omega_+\left( 1- \frac{2m_1\alpha}{p_0} e^{-\left( \frac{2}{u}-1\right)\frac{8m_1\alpha^2}{\Omega_+}} \right).\nn\\
\label{eq:mode mass2}
\eea
The mass of the $33$ mode is obtained by replacing $u\rightarrow 2u$; therefore, $\Omega_{33}(0)$ is the smallest mass. Equation (\ref{eq:mode mass2}) is valid if the argument of the exponential is much larger than unity in magnitude.
For weak SOC ($\alpha\ll v_F$, where $v_F$ is the Fermi velocity at $\alpha=0$), the last condition implies that $u\ll \alpha/ v_F\ll1$. Therefore, the found solution corresponds to the regime when the electron-electron interaction is weaker than SOC (when measured in appropriate units). To find the solution in the opposite regime of weak SOC ($\alpha/ v_F\ll u$), we expand $L(\Omega)$ in $2m_1\alpha^2$. A straightforward calculation yields
\bea\label{eq:mode mass2_2}
\Omega_{11}(0)=\Omega_{22}(0)&=&2\alpha p_F\sqrt{1-\frac{u}{2}}.
\eea
[Again, $\Omega_{33}(0)$ is obtained from Eq. (\ref{eq:mode mass2_2}) above by replacing $u\to 2u$.]
Equation (\ref{eq:mode mass2_2}) is valid if $2m_1\alpha^2\ll |\Omega-2\alpha p_F|$, where one should substitute the masses of the modes for $\Omega$. Doing so and expanding in $u$, we indeed see that the condition for validity of Eq.~(\ref{eq:mode mass2_2}) is $\alpha/v_F\ll u$. The crossover between regimes described by Eqs.~(\ref{eq:mode mass2}) and (\ref{eq:mode mass2_2}) occurs at $u\sim \alpha/v_F$. In Fig. \ref{fig:ap}, we show the exact solution of Eq.~(\ref{eq:modal mass}) for the $11$ and $22$ modes as a function of $u$ along with the asymptotic solutions given by Eqs.~(\ref{eq:mode mass2}) and (\ref{eq:mode mass2_2}). We see here that for weaker SOC ($\alpha/v_F=0.1$), the solution is well approximated by Eq. (\ref{eq:mode mass2_2}). But, for the stronger SOC ($\alpha/v_F=0.5$), the crossover between Eqs.~(\ref{eq:mode mass2}) and (\ref{eq:mode mass2_2}) is clearly apparent.

\begin{figure}[htp]
$\begin{array}{c}
\includegraphics[width=0.9\columnwidth]{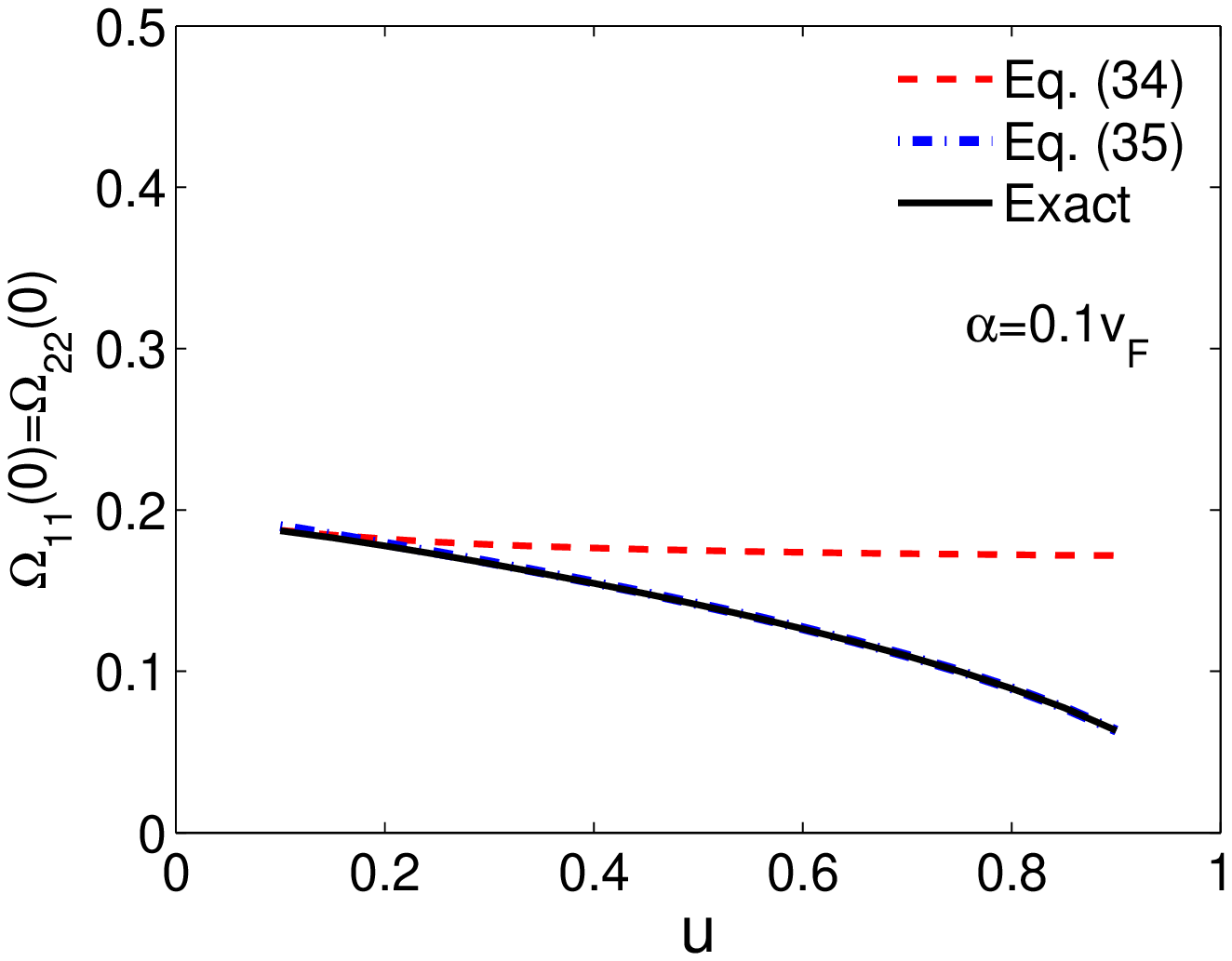}\\
\includegraphics[width=0.9\columnwidth]{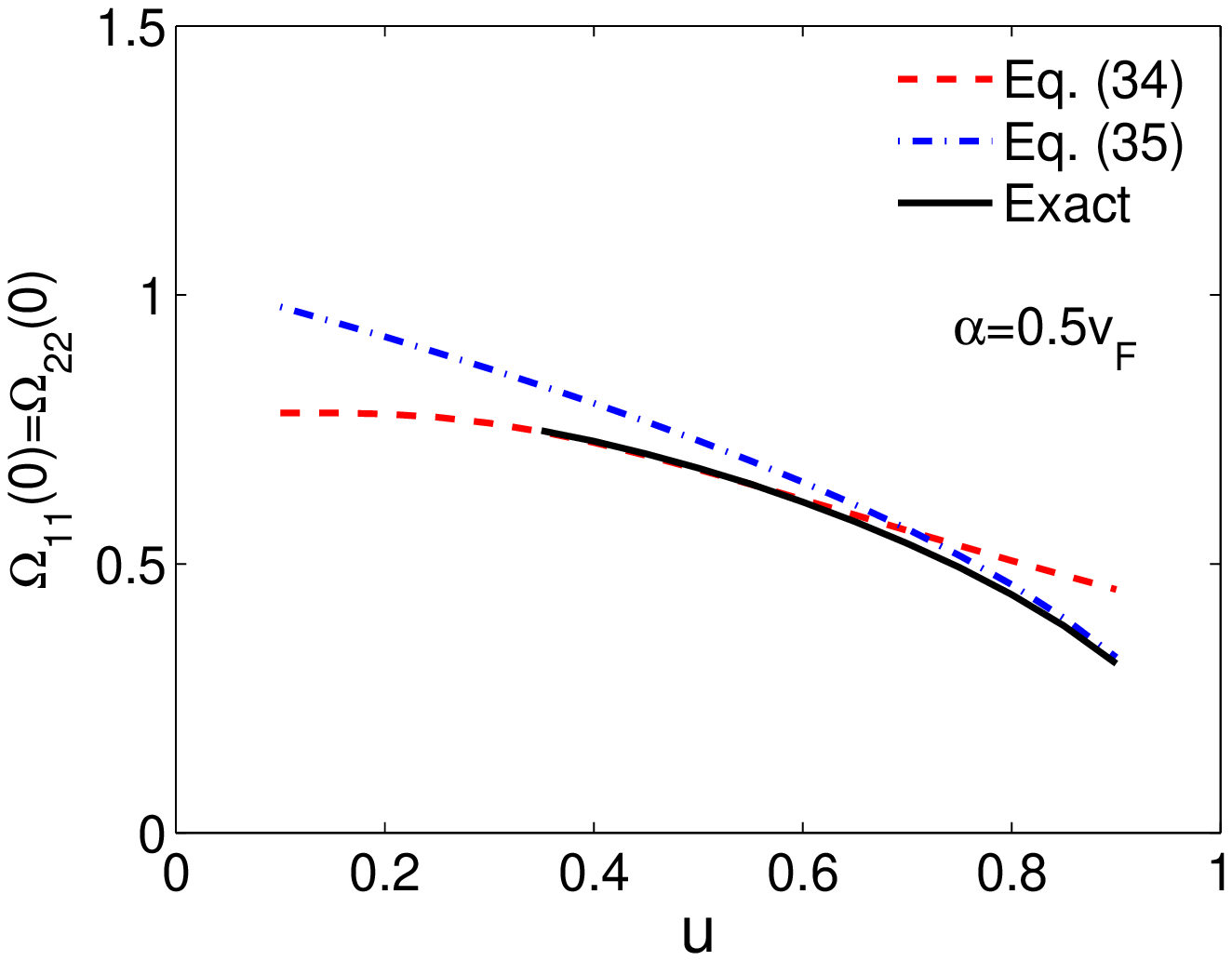}
\end{array}$
\caption{\label{fig:ap}
Masses of the $11$ and $22$ modes as a function of the dimensional electron-electron coupling $u$. Solid: an exact solution of Eq.~(\ref{eq:modal mass}). Dashed: Eq.~(\ref{eq:mode mass2}) valid for $u\ll \alpha/v_F$. Dash-Dotted: Eq.~(\ref{eq:mode mass2_2}) valid for $u\gg \alpha/v_F$. Top: $\alpha=0.1 v_F$; Bottom: $\alpha=0.5 v_F$. The crossover between Eqs.~(\ref{eq:mode mass2}) and (\ref{eq:mode mass2_2}) occurs approximately at $u\approx 0.7$.
}
\end{figure}

Within the FL formalism, which assumes that SOC is weak, the mode masses can be expressed via the FL parameters evaluated in the absence of SOC.~\cite{shekhter,ali_maslov} If one further adopts the $s$-wave approximation, in which all but the zeroth angular harmonic of the Landau function are absent, the FL results for masses of the modes reduce to:\cite{ali_maslov}

\beq
\Omega_{11}(0)=\Omega_{22}(0)=2\alpha p_F \sqrt{1+F_0^a/2},
\label{eq:FL}
\eeq
where $F_0^a$ is the zeroth harmonic of the spin-asymmetric part of the Landau function. For the $33$ mode, one replaces $F_0^a\rightarrow 2 F_0^a$.
Since  $F_0^a=-u$ to first order in the short-range interaction,\cite{ali_maslov_rashba} the FL result [Eq.~(\ref{eq:FL})] corresponds to the weak-SOC  limit of the RPA result, Eq.~(\ref{eq:mode mass2_2}). We see, however, that the form of the RPA result, Eq.~(\ref{eq:mode mass2_2}), has no correspondence in the FL theory. This implies that the assumption of weak SOC of Refs.~\onlinecite{shekhter} and \onlinecite{ali_maslov} is quite stringent: SOC must be smaller not only compared to the Fermi energy but also to the electron-electron coupling.

\subsubsection{Dispersion and coupling of charge and spin sectors for arbitrary $q$}
\label{sec:anyq}
\begin{figure*}[htp]
\includegraphics[width=1.8\columnwidth]{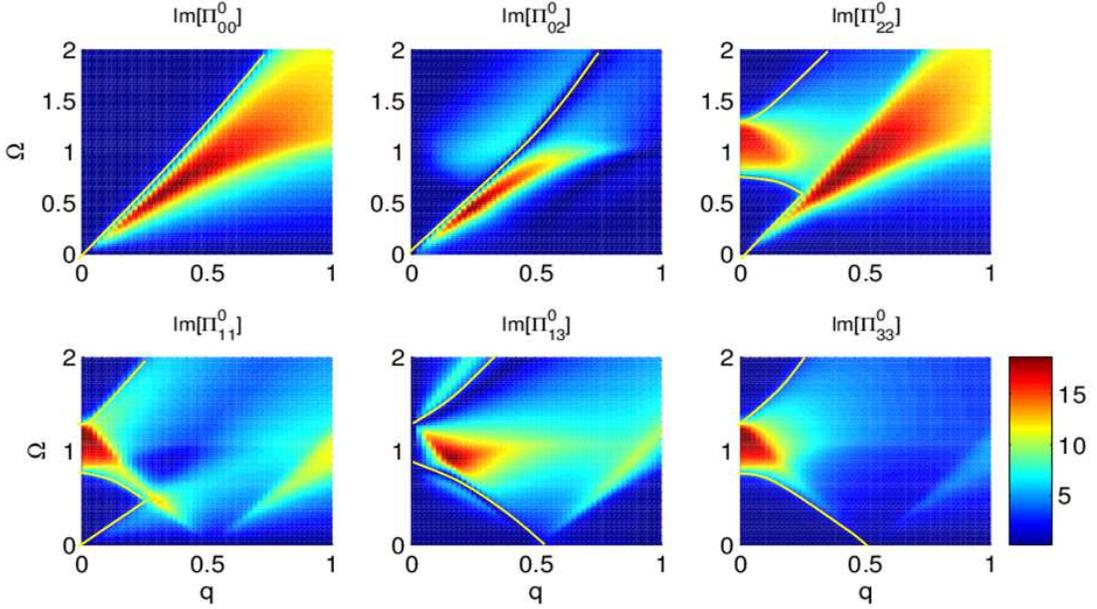}
\caption{\label{fig:cont}
Imaginary parts of the six non-zero components of the spin-charge polarization tensor $\Pi^0_{ij}$
[Eq.~(\ref{eq:susceptibilities2}), arbitrary units]. $\Omega$ is in units of $\mu$ and $q$ is in units of $\sqrt{2m_1\mu}$. The solid lines are guides to they eye that mark the boundaries of the single-particle continua.
}
\end{figure*}
In this section, we present numerical results for the dispersions of the charge and chiral-spin modes, supplemented by the analytical treatment
of limiting cases. The collective modes are manifested by the poles in the components of the susceptibility tensor given by
Eqs.~(\ref{eq:susceptibilities3}) and (\ref{eq:susceptibilities32}).
A collective mode is not Landau-damped if it lies outside (regions where Im$\Pi^0_{ij}\neq 0$). Since the various components of the spin-charge polarization tensor are coupled to each other, it is important to determine the boundaries of the continua for all the six $\Pi^0_{ij}$. These are shown in Fig. \ref{fig:cont} the lines are guides to the eye. The top row corresponds the $0-2$ sector, where the continua present in all $\Pi^0_{ij}$'s have both the charge and Rashba
regions. The charge continuum is the region starts below the diagonal line, while the Rashba continuum occupies a finite segment of the vertical axis and disperses into the $\Omega-q$ plane. [The Rashba continuum in $\Pi^0_{00}$ is not easily seen on the color scale as $\I\Pi_{00}^0\propto q^2$ but is nevertheless present--see Eq.~(\ref{eq:NonZeroPi})]. This suggests that the collective modes in this sector are affected by both charge and Rashba continua. In the $1-3$ sector, $\Pi^0_{11}$ has both charge and Rashba continua, but $\Pi^0_{13}$ and $\Pi^0_{33}$ only have a Rashba continuum.

The absence of the charge continuum in these two last cases can be seen analytically, at least in the limit of $\alpha\ll v_F$. Since the gap in the Rashba continuum is present only for $0\leq q\leq 2m_1\alpha \ll p_0$,  we can also look at the small-$q$ case, assuming that $q\ll p_F$. In this limit, the first term in Eq.~(\ref{eq:ZZ}) is equal to zero to $\mathcal{O}(q^2)$ since $\cos(\theta_{\bk}-\cos\theta_{\bk+\bq}) = 1-\mathcal{O}(q^2)$ and $g_+g_+\sim \mathcal{O}(q^2)$. The leading term in $\Pi_{33}^0(q,\Omega)$ is thus given by
\bea\label{eq:pzi3} \Pi_{33}^0(q,\Omega)&=&\int_K \left(g_+g_-+g_-g_+\right)\\
&=&\int \frac{d^2k}{(2\pi)^2}\frac{n_F(\e^+)-n_F(\e^-)}{\Omega + i\delta + (2m_1\alpha-q\cos\theta)v_F}.\nn
\eea
The imaginary part of this integral gives only the Rashba continuum.  Thus there is no charge continuum in $\Pi_{33}$ to $\mathcal{O}\left[(q/p_F)^2\right]$ and $\mathcal{O}\left[\left(\frac{m_1\alpha}{p_0}\right)^2\right]$.

Since the $33$ mode is coupled to the $11$ mode through $\Pi^0_{13}$, we also need to look at Im$\Pi_{13}^0$. We focus here only on the diagonal, $\int\left(g_+g_+-g_-g_-\right)$ term in Eq.~(\ref{eq:XZ}), as the charge continuum can arise only from this term. Integrating the product $g_+g_+$, we obtain $\frac{n_F(\e^+)-n_F(\e^+)}{\Omega + i\delta + v_F q\cos\theta)}(\cos\theta_\bk-\cos\theta_{\bk+\bq})$.
Furthermore, $(\cos\theta_\bk-\cos\theta_{\bk+\bq})\approx -\frac{q}{p_F}\sin^2\theta$. Integrating over $\bk$, we obtain for the imaginary part
\bea\label{eq:pzi13}
\text{Im}\int_K~g_+g_+(\cos\theta_\bk-\cos\theta_{\bk+\bq})&=&\frac{m_1}{2\pi}\frac{\sqrt{v_F^2q^2-\Omega^2}}{v_Fk_F}\nn\\
\eea
for $\Omega<v_F q $
and $0$ otherwise. This is region where the charge continuum should be. However, the combination of $g_-g_-$ also yields the same expression which cancels the contribution of $g_+g_+$. This leads to complete cancellation of the charge part
in $\I\Pi_{13}^0$ to $\mathcal{O}\left[(q/p_F)^2\right]$  and $\mathcal{O}\left[\left(\frac{m_1\alpha}{p_0}\right)^2\right]$.
While we have shown this explicitly for weak SOC, we have also checked numerically that this remains true for larger $\alpha$ as well.

The absence of the charge continuum has an important consequence for the damping of the $33$ mode.  To see this, we recall that, according to Eqs.~(\ref{eq:vertex chi}) and (\ref{eq:susceptibilities3}),
\beq
\chi_{33}=-\lr\Pi^0_{33}M_{33}+\Pi^0_{31}M_{13}\rr.
\eeq
Since we have shown that in the charge-continuum region both Im$\Pi^0_{33}$ and Im$\Pi^0_{13}$ are equal to zero, a trivial exercise in matrix inversion suggests that Im$\chi_{33}$ is also zero in that region (even though Im$\Pi^0_{11}$ is finite). Thus the 33-mode does not \lq\lq see\rq\rq\/ the charge continuum and hence is not damped.

The dispersions of the collective modes are obtained by numerically evaluating $\chi_{ij}$ defined in Eqs.~(\ref{eq:susceptibilities3}) and (\ref{eq:susceptibilities32}) (see Appendix \ref{sec:appA} for details).\cite{footnote} The results of these calculations are shown in Fig.~\ref{fig:velocity}.
The imaginary part of the charge susceptibility ($\chi_{00}$) is shown in the top panel. The $\sqrt{q}$-plasmon approaches the charge continuum for larger $\Omega$ and $q$. The optical charge plasmon is not seen here because of a weak damping added to improve numerical convergence. The boundaries of the Rashba continuum, which is also not visible on the color plot, are marked by the yellow dotted lines. The charge plasmon is damped within the Rashba continuum.

The chiral-spin modes are manifested by the poles in the three components of the spin susceptibility ($\chi_{11}$, $\chi_{22}$, and $\chi_{33}$).
Their dispersions are shown in the bottom panel of Fig.~\ref{fig:velocity}. The dashed line is the boundary of the Rashba continuum. The insets show the calculated Im$\chi_{ii}$ ($i=1,~2,~3$) separately.
The $11$ and $22$ modes start out degenerate at $q=0$ but split off at finite $q$ and eventually run  into the Rashba continuum. The fact that the two modes split at finite $q$ can already be seen analytically from the Eq.~(\ref{eq:NonZeroPi}); this analysis is presented in Appendix \ref{subsec:split}. The $33$ mode disperses downward and approaches the Rashba continuum but at larger value of $q$ (in the FL theory,\cite{ali_maslov} the merging point coincides with the end point of the continuum).  The $33$ mode is not damped by the charge continuum, but will be damped due to broadening of the Rashba continuum by disorder and thermal fluctuations, as well as by quasiparticle scattering.
\begin{figure}[htp]
$\begin{array}{c}
\includegraphics[width=0.7\columnwidth]{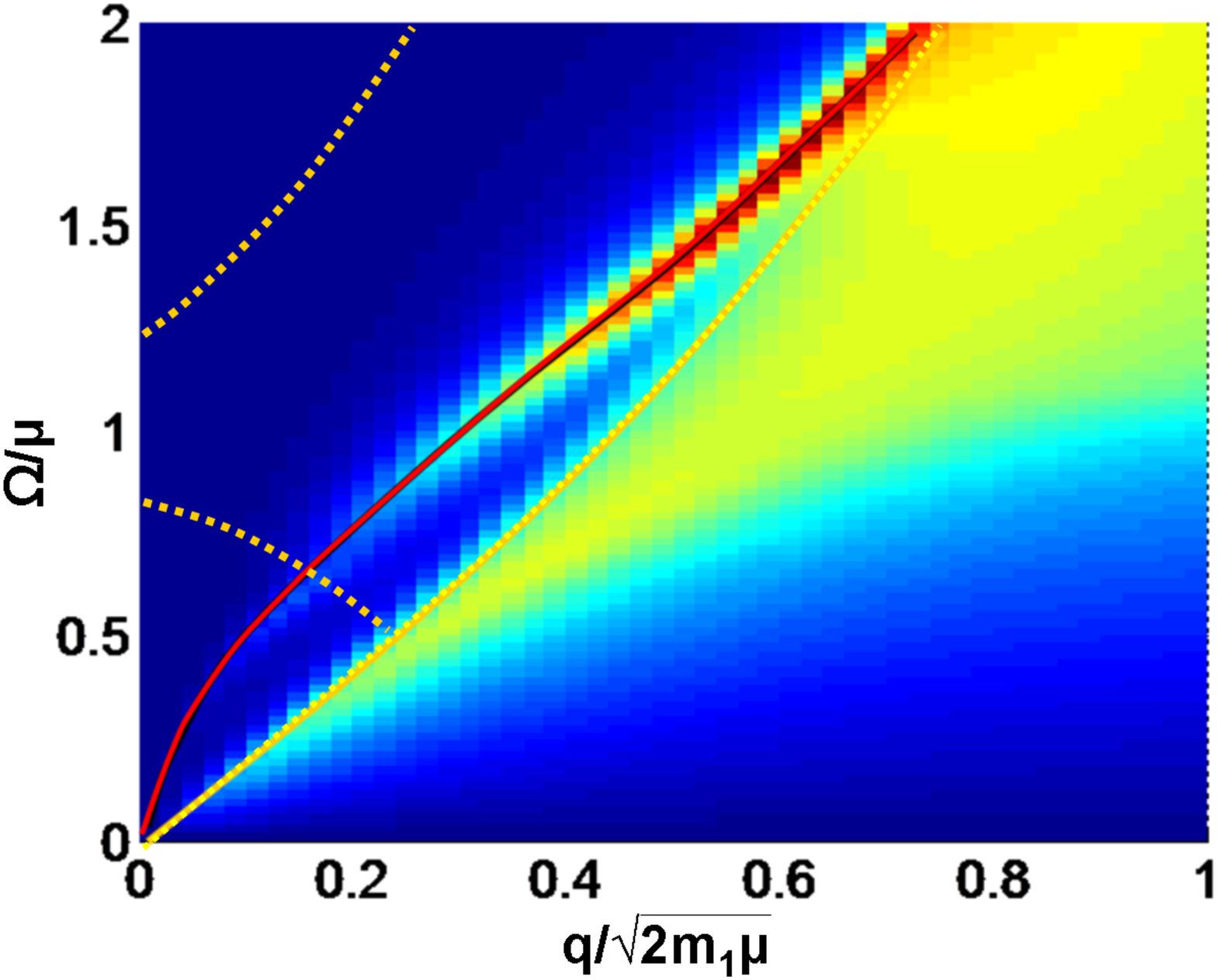}\\
\includegraphics[width=0.7\columnwidth]{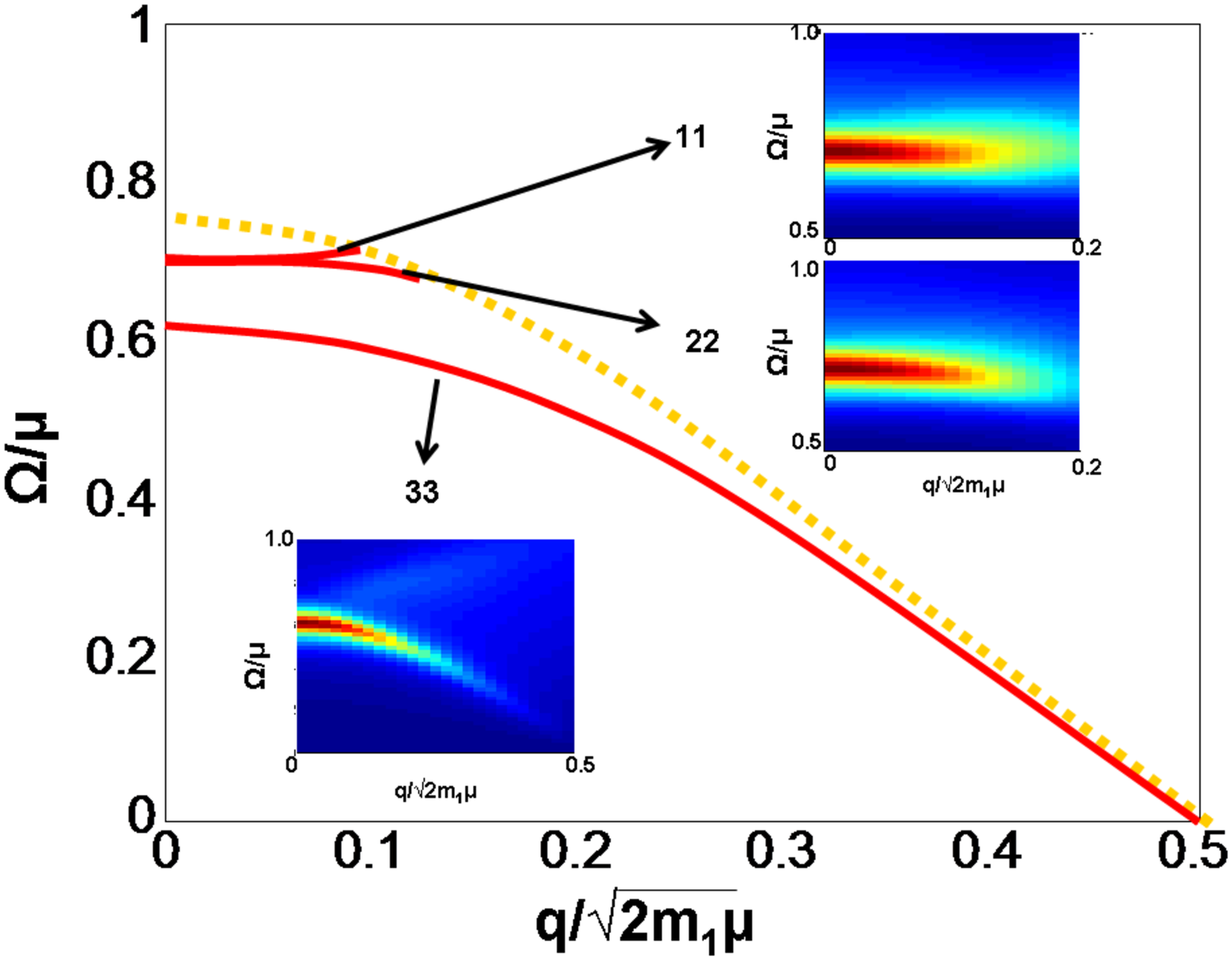}
\end{array}$
\caption{\label{fig:velocity}Top: Imaginary part of the charge susceptibility
$\I\chi_{00}$ in the $\Omega$-$q$ plane.
The $\sqrt{q}$ plasmon is shown by the solid red line. For larger $\Omega$ and $q$, the plasmon merges with the charge continuum (whose boundary is marked by the solid yellow line). The second (optical) charge plasmon is smeared out by weak damping, added to improve numerical convergence,
and is not visible in the plot.
Bottom: The dispersions of the spin modes shown are shown by solid red lines.
The $11$ and $22$ modes are degenerate at $q=0$ but split off at finite $q$ and run into the Rashba continuum. The $33$ mode disperses downward and merges with the Rashba continuum. This mode does not feel the charge continuum, as discussed in the text. The boundary of the Rashba continuum is marked by the dashed yellow line. The insets show individual Im$\chi_{ii}$ ($i=1,~2,~3$), from which the dispersions are extracted. Here, $\alpha=0.25\sqrt{2\mu/m_1}$, $u=0.2$, and $1/\tau=0.02\mu$.}
\end{figure}

Figure~\ref{fig:velocity2} shows the effect of coupling of the plasmon to the $22$ mode and the coupling of the $33$ mode to the $11$ mode. The dashed lines are the dispersions obtained by ignoring the coupling between the respective modes; the solid lines are the actual dispersions of the modes. We note two important features: 1)The masses of the modes are not affected by the inter-mode coupling but the dispersions are. 2) The $33$ mode is pushed away from the continuum but the plasmon is pulled towards the continuum.

\begin{figure}[htp]
$\begin{array}{c}
\includegraphics[width=0.7\columnwidth]{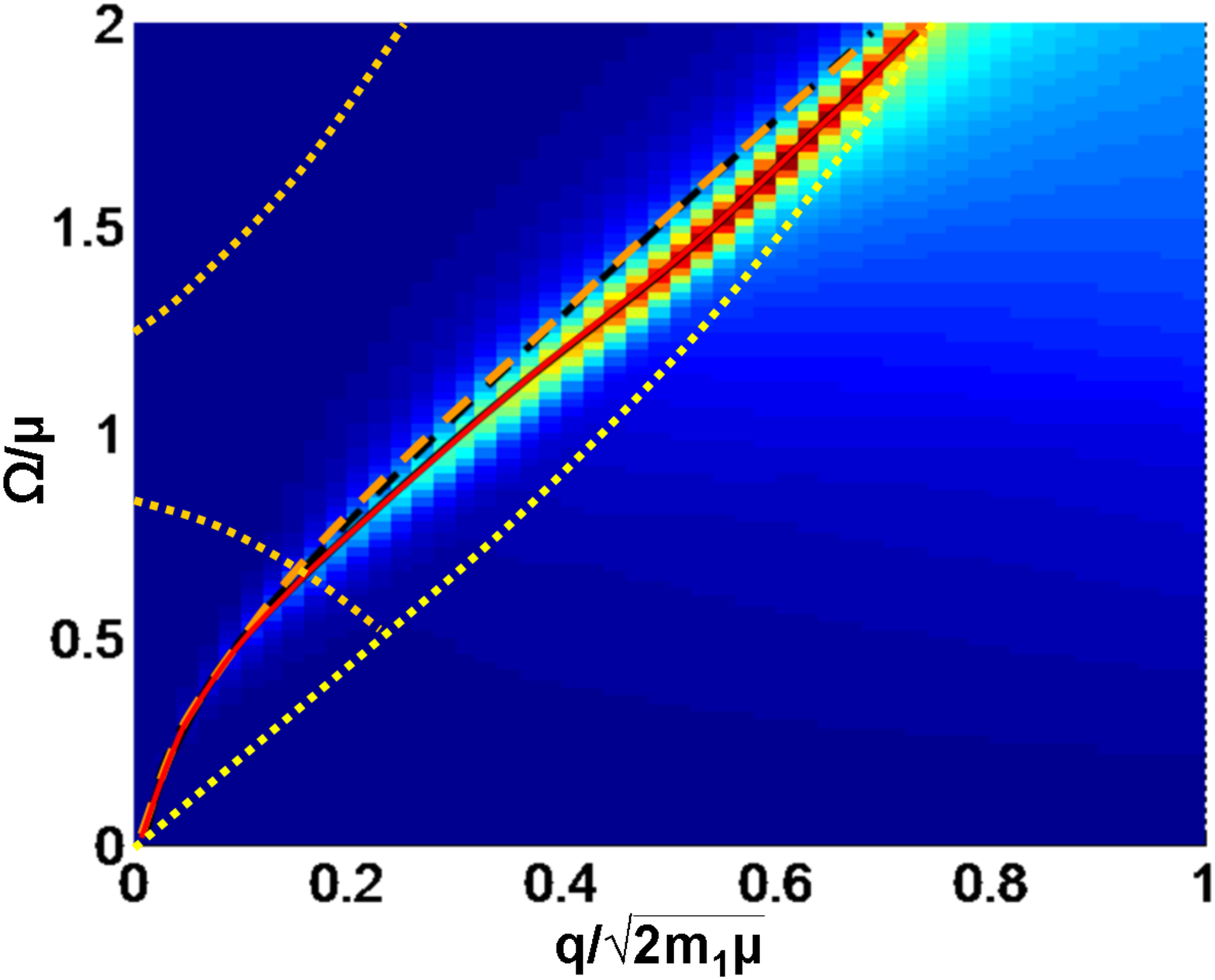}\\
\includegraphics[width=0.7\columnwidth]{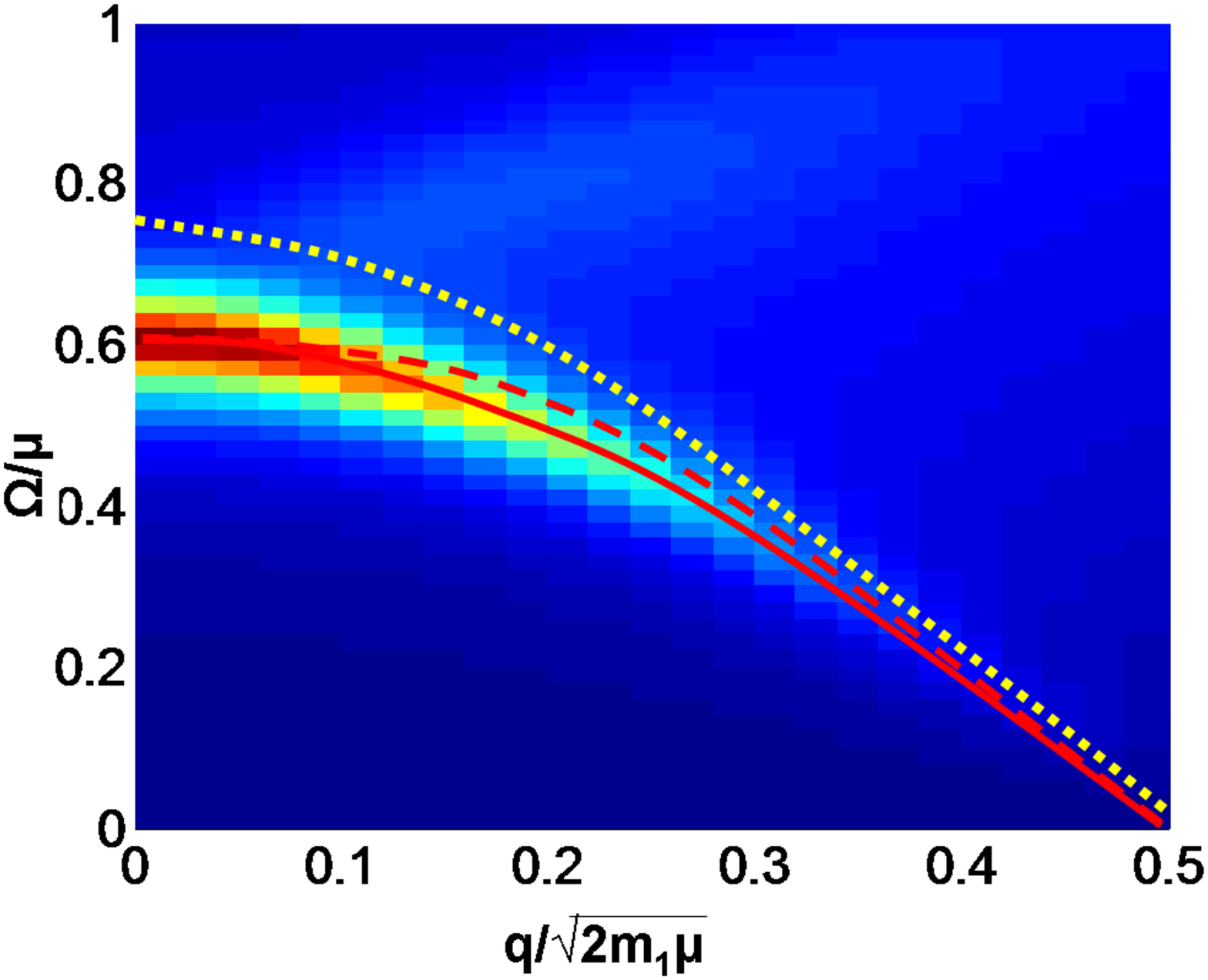}
\end{array}$
\caption{\label{fig:velocity2} (Top) Effect of spin-orbit coupling on the dispersion of plasmon. The red line is the true dispersion in a system with Rashba SOC (where $\Pi_{02}\neq0$) The dashed line is the dispersion ignoring the coupling to chiral sector ($\Pi_{02}=0$). (Bottom) Change in dispersion of the $33$ mode as a result of coupling to the $11$ mode via $\Pi_{13}$ (red line). The dashed line is the dispersion ignoring this coupling. Here $\frac{m\alpha}{\sqrt{2m\mu}}=0.25$ and $u=0.2$.}
\end{figure}

\subsubsection{Spin-chiral modes for $\mu<0$}
\label{sec:mu<0}
In this section, we show that chiral-spin modes exists even when only the lower Rashba subband is occupied, i.e., $\mu<0$ To see this,  let's look at $\Pi^0_{33}(0,\Omega)= \frac12\int_K\left(g_+g_- + g_-g_+\right)$ for $\mu<0$. Carrying out the $\omega$ and angle integrations, we find
\bea\label{eq:pitilde4} \Pi^0_{33}(0,\Omega)&=&\frac12\int\frac{kdk}{2\pi}\left[\frac{n_F(\e^+)-n_F(\e^-)}{i\Omega + 2\alpha k}+~~~\alpha\rightarrow -\alpha\right].\nonumber\\
\eea
For $\mu<0$, $n_F(\e^+)=0$ and $n_F(\e^-)=1$ in the interval $p_1<p<p_2$
where $p_{1,2}= m_1\alpha\mp\sqrt{-2m_1|\mu|+m_1\alpha^2}$ are the inner and outer radii of the annular Fermi surface, see Fig.~\ref{fig:subband} b.
Integrating over $k$, we arrive at
\bea\label{eq:pitilde5} \Pi^0_{33}(0,\Omega)&=&-\frac{m_1}{2\pi}\left[ \frac{2p_0}{m_1\alpha}+\frac{\Omega}{4m_1\alpha^2}L'(\Omega) \right],
\eea
where $L'(\Omega)$ is the same as $L(\Omega)$ in Eq.~(\ref{eq:L}) but with $p_+\rightarrow p_1$ and $p_-\rightarrow p_2$.
Intersubband transitions that give rise to the function $L'(\Omega)$ are shown by the hatched region in Fig.~\ref{fig:subband} b. Since $L(\Omega)$ and $L'(\Omega)$ are qualitatively the same, the structure of the poles and of the continua is the same as for $\mu>0$.

\subsection{Manifestations of the collective modes in the observable quantities}
\label{subsec:2D Exp consqnce}
In this section,
we discuss the relation between the collective modes and observable quantities.  Due to decoupling of the $0-2$ and $1-3$ sectors, the $11$- and $33$ chiral-spin modes in the dipole approximation can only be excited magnetically and can be seen in the spin susceptibility measurements or using the mode confinement proposal of Ref.~\onlinecite{ali_maslov}. (At the next, quadrupole, order, the $11$ and $33$ modes couple to the electric field as well.)
The coupling between the charge ($00$) and the in-plane, transverse chiral-spin mode ($22$) occurs already at dipole order,\cite{shekhter} and we will focus on this channel. In this section, we study only the theoretical aspects of the relation between the collective modes and various observables, hence the parameters chosen for the plots do not necessarily correspond to any real system. Our predictions for specific materials are given in Sec.~\ref{sec:other effects}.
\subsubsection{Probing the modes at $q=0$}
It was shown in Ref.~\onlinecite{shekhter} that the part of the optical conductivity arising from the intersubband transitions is proportional to the in-plane spin susceptibility  at $q=0$. This suggests a possibility to observe the chiral-spin mode at $q=0$ directly in the optical conductivity, measured either via absorption or reflectivity. For completeness, we show how the result of
Ref.~\onlinecite{shekhter} is reproduced within our approach.

Recall that  the velocity operator in the presence of SOC [Eq.~(\ref{eq:vx})] contains an off-diagonal part proportional
to $\alpha$.
In the non-interacting case, the corresponding off-diagonal contribution to the current-current correlation function, $\mathcal{K}_{\mathrm{off}}$,
is directly proportional to the $22$ component of the spin susceptibility at $q=0$:
\bea\label{eq:pitilde3}
{\cal K}_{\mathrm{off}}(\Omega)&=&e^2\alpha^2\int\frac{kdk}{2\pi}\int\frac{d\theta}{2\pi}\int\frac{d\omega}{2\pi}\text{Tr}
\left[\hat\sigma_2\hat{G}_{\omega}\hat\sigma_2\hat{G}_{\omega+\Omega}\right]\nonumber\\
&=&e^2\alpha^2\Pi_{22}^0(0,\Omega).
\eea
Within our RPA+ladder formalism, taking into account the electron-electron interaction amounts to calculating vertex corrections to the conductivity.
This changes ${\cal K}_{\mathrm{off}}$
to ${\cal K}^U_{\mathrm{off}}$, where \bea\label{eq:cond_corr_U1}
{\cal K}^U_{\mathrm{off}}(\Omega)&=& e^2\int_K\text{Tr}\left[\hat{v}_1\hat{G}_K \hat\beta\hat{G}_{K+Q}\right],\nonumber\\
\hat\beta&=&\hat v_1 -U\int_P \hat G(P)\hat\beta \hat G(P+Q),
\eea
and $\hat v_1$ is defined in Eq.~(\ref{eq:vx}).
We represent $\hat\beta$ as $\hat\beta=N_a \hat\sigma_a$ with $a\in{0,1,2,3}$. Substituting this form into Eq.~(\ref{eq:cond_corr_U1}), multiplying by $\hat\sigma_0$ and taking trace, we find: $N_{1,3}=0$, $N_0=k_1/m_1$, and $N_2=\frac{\alpha}{1+\frac U2\Pi^0_{22}}$. This results in
\bea\label{eq:cond_corr_U2}
{\cal K}^U_{\mathrm{off}}(\Omega)&=& e^2\alpha^2\frac{\Pi_{22}^0(0,\Omega)}{1+\frac U2\Pi_{22}^0(0,\Omega)}.
\eea
Thus the $22$ mode at $q=0$ (the \lq\lq chiral-spin resonance\rq\rq\/ in the terminology of Ref.~\onlinecite{shekhter}) shows up as a pole in the conductivity.
The real part of the total conductivity
 \bea
 \sigma(\Omega)=i\frac{e^2}{\Omega}\left[\frac{n_{2D}}{m_1}-\frac{m_1\alpha^2}{2\pi}+\alpha^2
 \frac{\Pi_{22}^0(0,\Omega)}{1+\frac U2\Pi_{22}^0(0,\Omega)}\right]
 \eea
is shown in Fig.~\ref{fig:cond_2D_u}, were again we added a small $1/\tau$ to mimic the effect of disorder.  The new feature, compared to the non-interacting case (Fig.~\ref{fig:Cond2d}), is a sharp peak below the Rashba continuum. In 2D systems, $\R\sigma(\Omega)$ is measured via absorption.

 \begin{figure}[htp]
$\begin{array}{c}
\includegraphics[width=0.8\columnwidth]{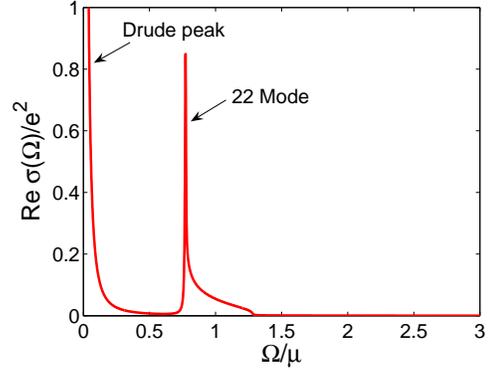}\\
\end{array}$
\caption{\label{fig:cond_2D_u}  The real part of the conductivity of an interacting 2D electron system with Rashba SOC. A sharp peak
is due to the $22$ chiral-spin mode at $q=0$.
Here, $\alpha=0.25\sqrt{2\mu/m_1}$ and $u=0.2$.}
\end{figure}

The optical conductivity can be also measured via reflectivity. The reflectance of a single 2D sheet is related to its conductivity via
\beq\label{eq:RR}
R=\left|\frac{2\pi\sigma(\Omega)}{c+2\pi\sigma(\Omega)} \right|^2,
\eeq
where $c$ is the speed of light. The reflectance is plotted in Fig.~\ref{fig:Refl} for the non-interacting case (left panel) and in the presence of the interactions (right panel).

\begin{figure*}[htp]
$\begin{array}{cc}
\includegraphics[width=0.9\columnwidth]{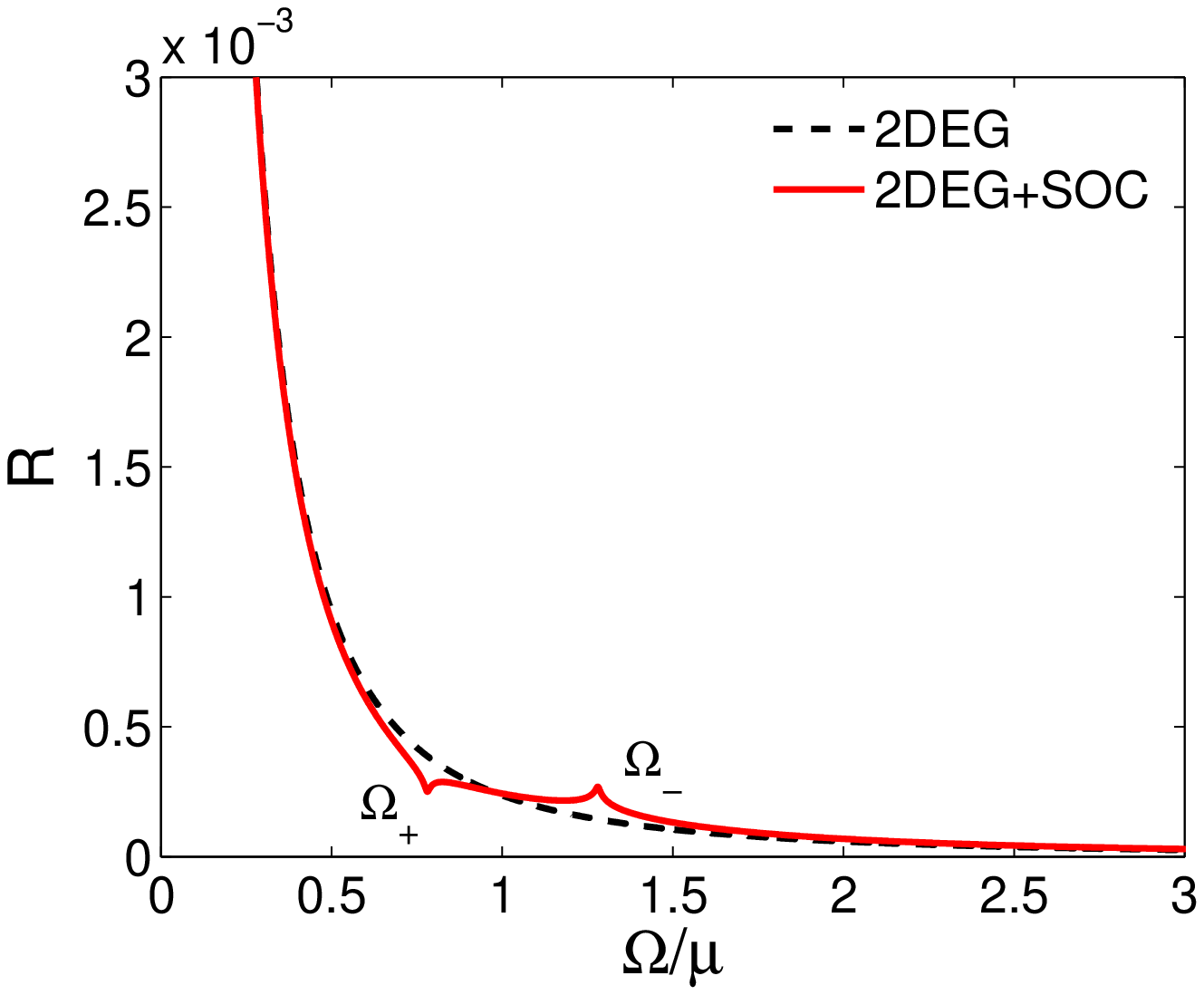}&
\includegraphics[width=0.9\columnwidth]{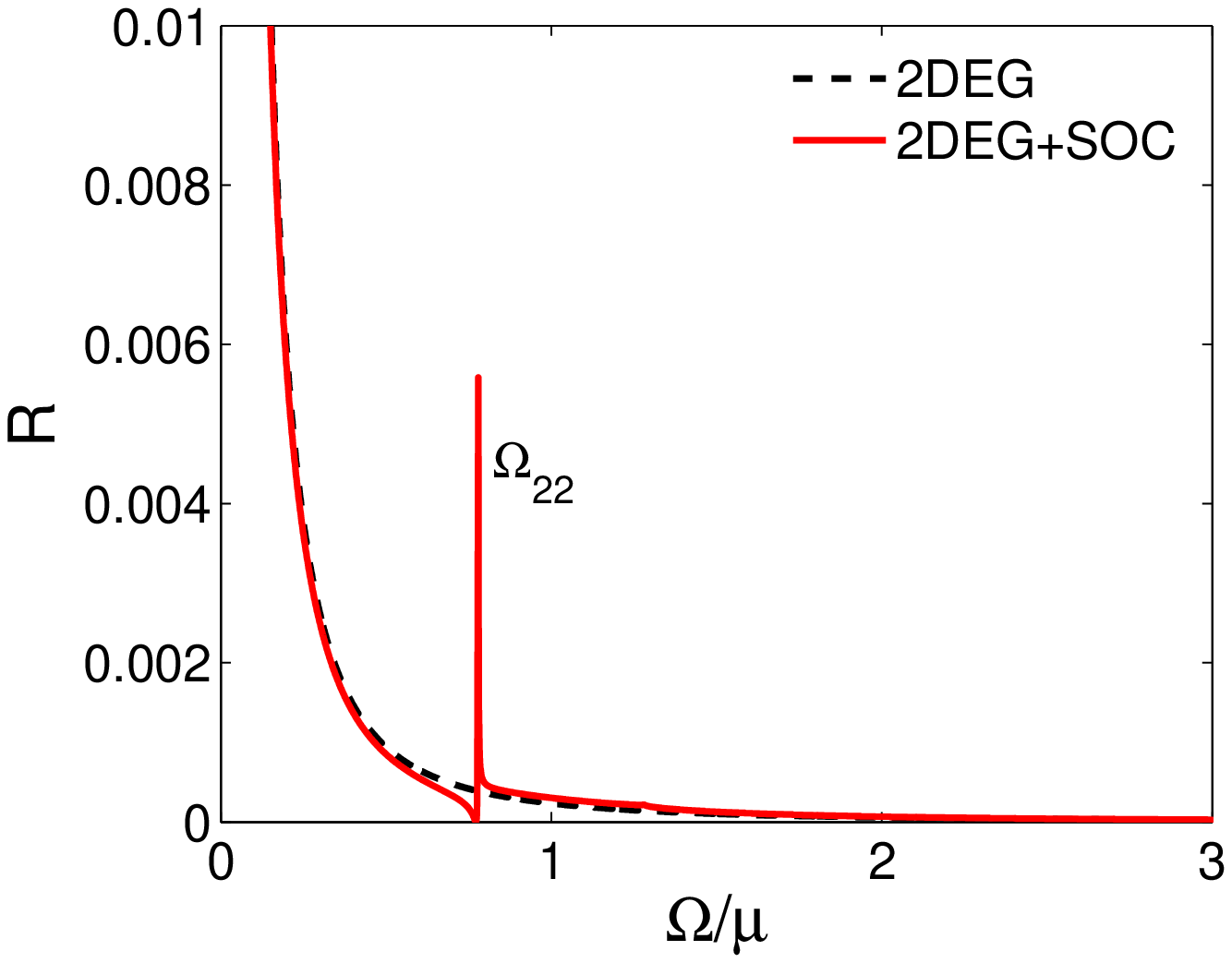}
\end{array}$
\caption{\label{fig:Refl} Left:
Reflectance of a single 2D layer of non-interacting electron gas with Rashba SOC (solid) and without SOC (dashed). The two features at $\Omega=\Omega_-$ and $\Omega=\Omega_+$ are due to the logarithmic singularities at the boundaries of
the Rashba continuum. Right: The same but with the electron-electron interaction taken into account. The $22$ mode shows up as a strong peak in the reflectance. Since the 22 mode is close to the lower boundary of the Rashba continuum, the logarithmic feature at $\Omega_+$ is washed out by the peak at $\Omega_{22}$. Here, $\alpha = 0.25\sqrt{2 \mu/m_1}$ and $u=0.2$.}
\end{figure*}

\subsubsection{Probing the modes at finite $q$}
The dispersion of the $22$ mode can be accessed via measuring
the nonuniform conductivity, i.e., $\sigma(q,\Omega)$ with $q\neq 0$.
There is a well-developed technique of measuring $\sigma(q,\Omega)$ via absorption of the
incident power by a 2DEG with a grating structure imposed on it. \cite{ando} The absorbed power is proportional to
\ $(1/2)E_0^2\R\sigma_{\text{eff}}$, where $E_0$ is
the the amplitude of the incident electric field,
\bea\label{eq:eff_cond}
\sigma_{\text{eff}}(q,\Omega)&=&\frac{\sigma_{11}(q,\Omega)}{1+\frac{2\pi i q}{\Omega\epsilon_{\mathrm{eff}}}\sigma_{11}(q,\Omega)}
\eea
$2\pi/q$ is the grating period and $\epsilon_{\mathrm{eff}}=\left[\epsilon_2+\epsilon_1\coth (qd)\right]/2$ is the effective dielectric constant
of a structure vacuum/insulator$_1$/insulator$_ 2$ with $\epsilon_{1,2}$ being the dielectric constants of insulator$_{1/2}$, correspondingly, and $d$ being the thickness of insulator$_1$. Both ${\bf E}$ and
$\bq$ are in the $x_1$-direction. A well-established feature is a peak in $\sigma_{\mathrm{eff}}$ corresponding to the 2D plasmon.\cite{ando} SOC modifies the plasmon dispersion; more importantly, however, it brings in a qualitatively new effect: a dispersing chiral-spin mode.
Therefore, one should expect to see two peaks: one from the plasmon and another one from the chiral-spin mode.

Our goal now is to find
$\sigma_{11}(q,\Omega)$ given by
\bea\label{eq:full_cond}
\sigma_{11}(q,\Omega)&=&\frac{i}{\Omega}{\cal K}^U_{11}(q,\Omega).
\eea
We notice that, in contrast to the $q=0$ case, ${\cal K}^U_{11}(q,\Omega)$ at finite $q$ is not simply related to the $22$ component of the spin susceptibility. This is already evident for the non-interacting case, when
\bea\label{eq:cond_corr}
\mathcal{K}_{11}(q,\Omega)&=& e^2\int_K\text{Tr}\left[\hat{v}_1(q)\hat{G}_K \hat{v}_1(-q)\hat{G}_{K+Q}\right]
\eea
with
\beq
\hat{v}_1(q)\equiv \frac{k_1+q/2}{m}\hat\sigma_0-\alpha\hat\sigma_2.\label{vq}
\eeq  Carrying out the trace, we find
\bea\label{eq:cond_corr2a}
\frac{
\mathcal{K}_{11}(q,\Omega)
}{e^2}
&=&\int_K \left( \frac{k_1^2-\frac{q^2}{4}}{m^2}\frac12\mathcal{T}_{00}-2\alpha\frac{k_1}{m} \frac12\mathcal{T}_{02}\right)\nn\\
&&+ \alpha^2\Pi^0_{22}(q,\Omega),
\eea
where $\mathcal{T}_{00}$ and $\mathcal{T}_{02}$ are given by Eqs.~(\ref{eq:00}) and (\ref{eq:0Y}), correspondingly. The last
term in Eq.~(\ref{eq:cond_corr2a}) is proportional to the spin susceptibility while the first two terms vanish at $q=0$ because in this case
$\int_\omega \mathcal{T}_{00}=0$ and $\int_\omega \mathcal{T}_{02}=0$.
These terms is an extra contribution which distinguishes between $\mathcal{K}_{11}$ and $\Pi_{22}$ at finite $q$.

The current-current correlation function for interacting electrons is evaluated in Appendix \ref{subsec:correction}. The final result is that the
the dispersion of the mode probed by the conductivity at finite $q$ is different from the dispersion probed by the spin susceptibility: the difference is in a $q$-dependent term that scales as
$u^2 q^2$ at small $q$. At arbitrary $q$, $\mathcal{K}_{11}(q,\Omega)$ needs to be computed numerically.
The full result for $\sigma_{\mathrm{eff}}$ in shown in Fig.~\ref{fig:cond_log}, where for simplicity we set $\epsilon_{\mathrm{eff}}=1$. In addition to the plasmon peak at lower energies, there is also a (much weaker dispersing) peak at higher energies from the $22$ chiral-spin mode. As larger $q$, the $22$ modes merges with the Rashba continuum. For the parameters chosen for Fig.~\ref{fig:cond_log}, this happens at $q\approx 0.32\sqrt{2m_1\mu}$, which is why the there is no 22-peak at $q=0.32\sqrt{2m_1\mu}$ in this figure.

\begin{figure*}[htp]
$\begin{array}{cc}
\includegraphics[width=0.9\columnwidth]{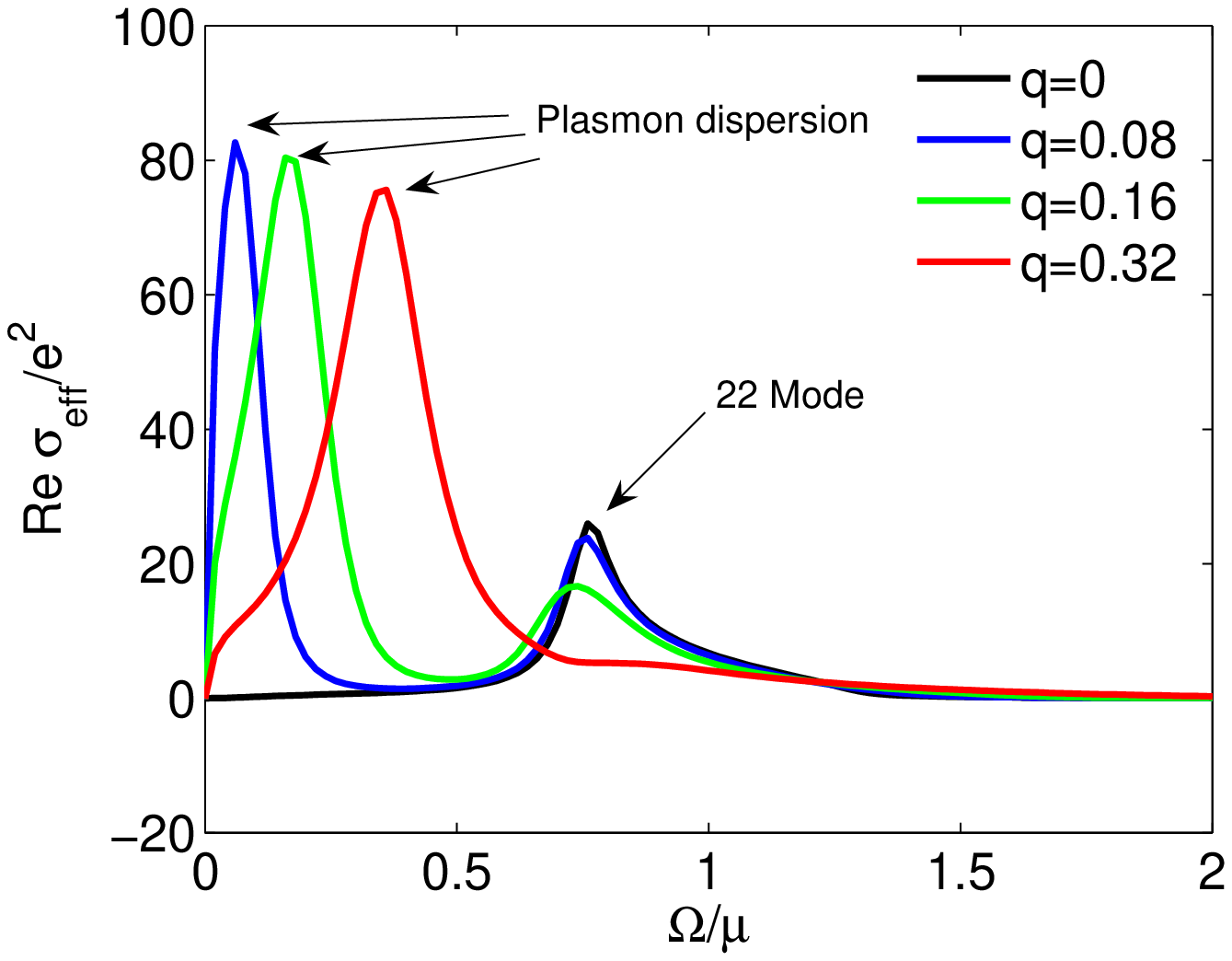}&
\includegraphics[width=0.9\columnwidth]{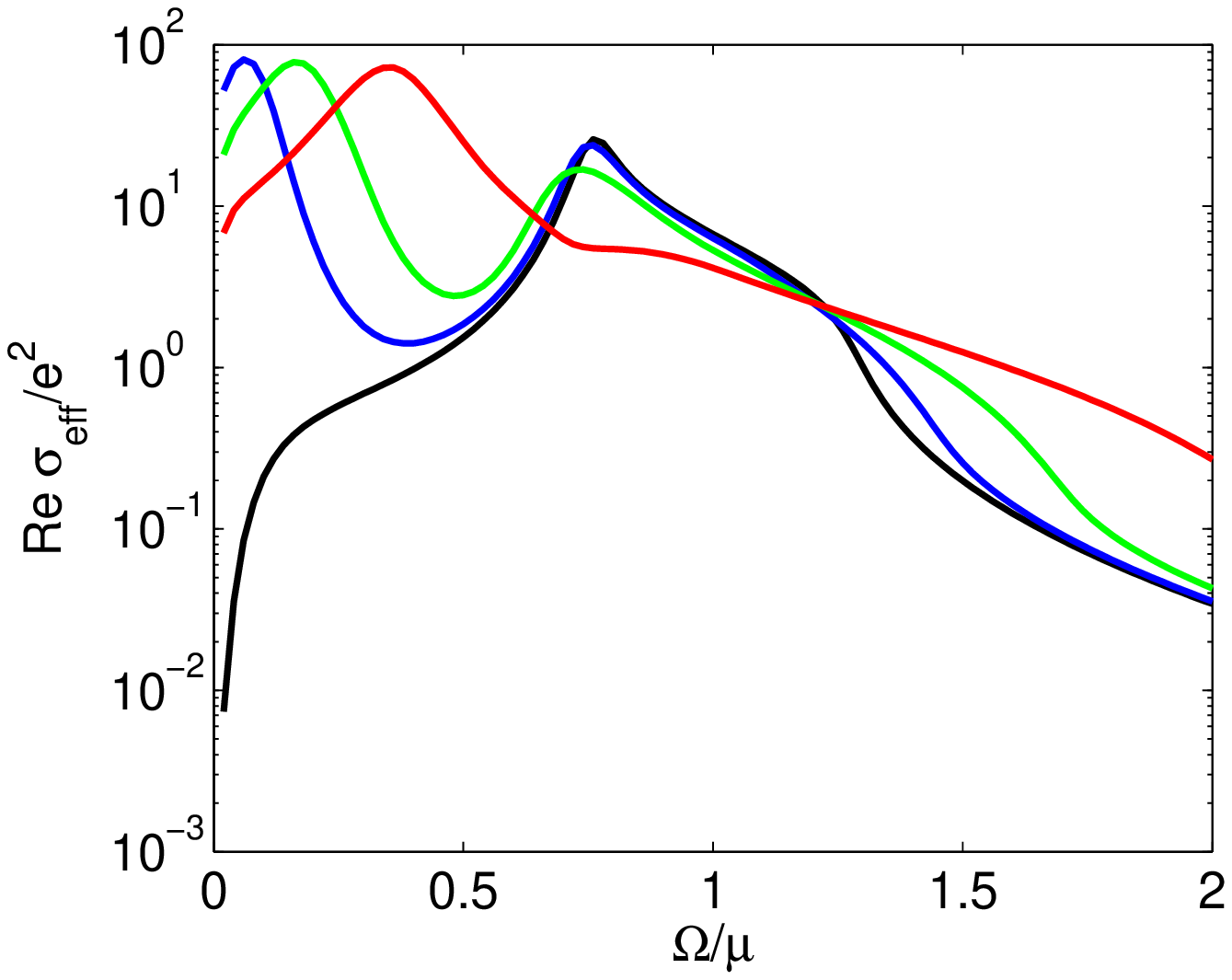}
\end{array}$
\caption{\label{fig:cond_log}
Left: The real part of the effective conductivity $\sigma_{\text{eff}}(q,\Omega)$ [Eq.~\ref{eq:eff_cond}] for a 2D Rashba system with electron-electron interactions. The Coulomb interaction gives rise a plasmon peak at lower energies while the short-range part of the interaction gives rises to a chiral-spin peak at higher energies. The wavenumber $q$ is measured in units of $\sqrt{2m_1\mu}$, $\alpha=0.25\sqrt{2\mu/m_1}$, and the dimensionless coupling constant of the short-range interaction $u=0.2$. Right: The same as on the left but on a log scale.}
\end{figure*}

\section{Collective modes in a three-dimensional Rashba system}\label{sec:3D}

In this section, we address the collective modes in a 3D Rashba system. At the non-interacting level, our model is the Hamiltonian in Eq.~(\ref{eq:Hamiltonian}) with finite $k_3$. The 3D Fermi surface is of the toroidal shape. Its projection onto the $x_1x_2$ plane is the same two circles as in a 2D system (cf. Fig.~\ref{fig:FS}, top), while the $x_1x_3$ projections are two overlapping ellipses (cf. Fig.~\ref{fig:FS}, bottom). It is convenient to define the following dimensionless parameters
\beq\label{eq:3D ratios}
s_1= \frac{m_1\alpha}{p_0};~~~~~s_2=\frac{\sqrt{2m_1\mu}}{p_0}.
\eeq
\begin{figure*}[htp]
$\begin{array}{cc}
\includegraphics[width=0.7\columnwidth]{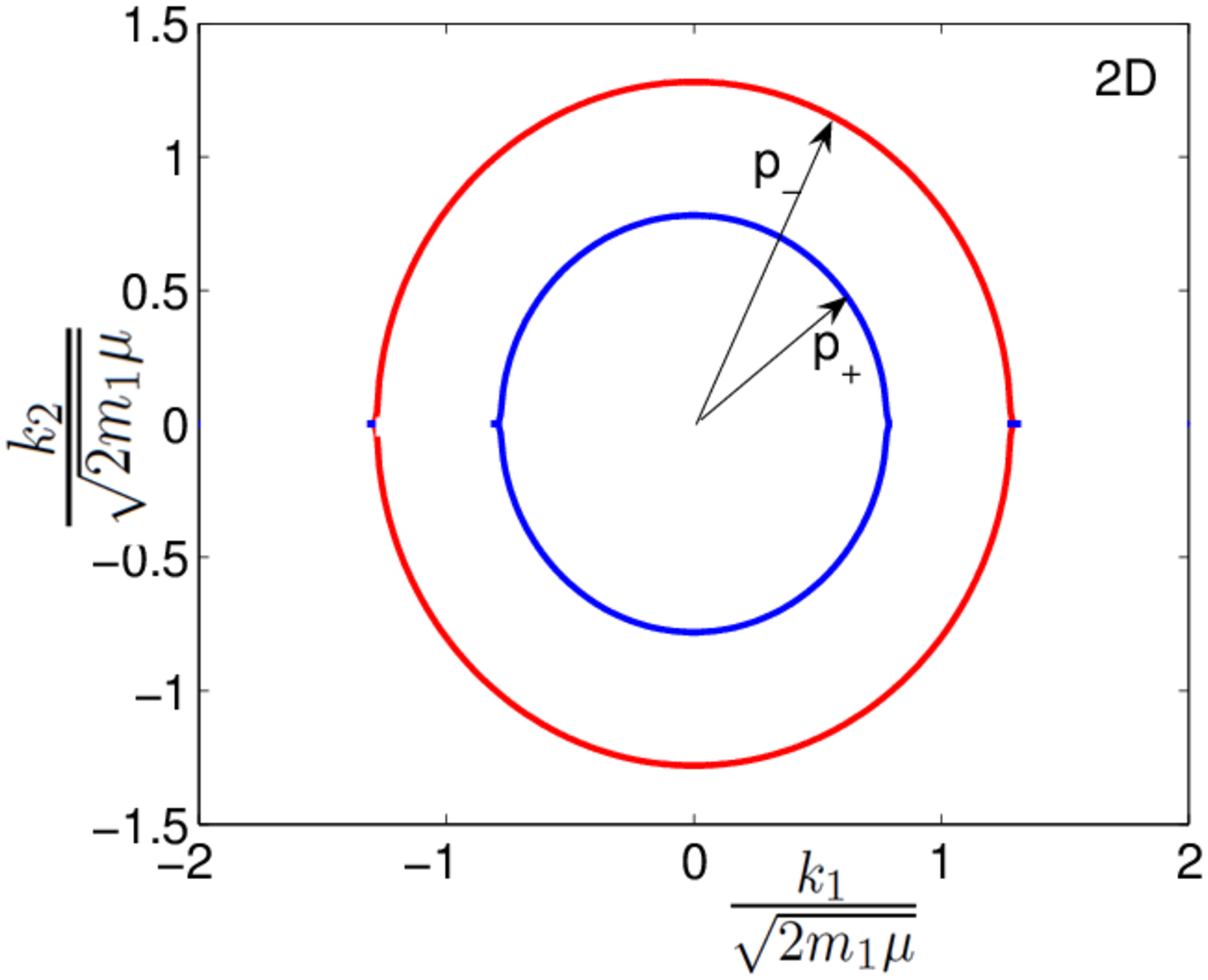}&
\includegraphics[width=0.7\columnwidth]{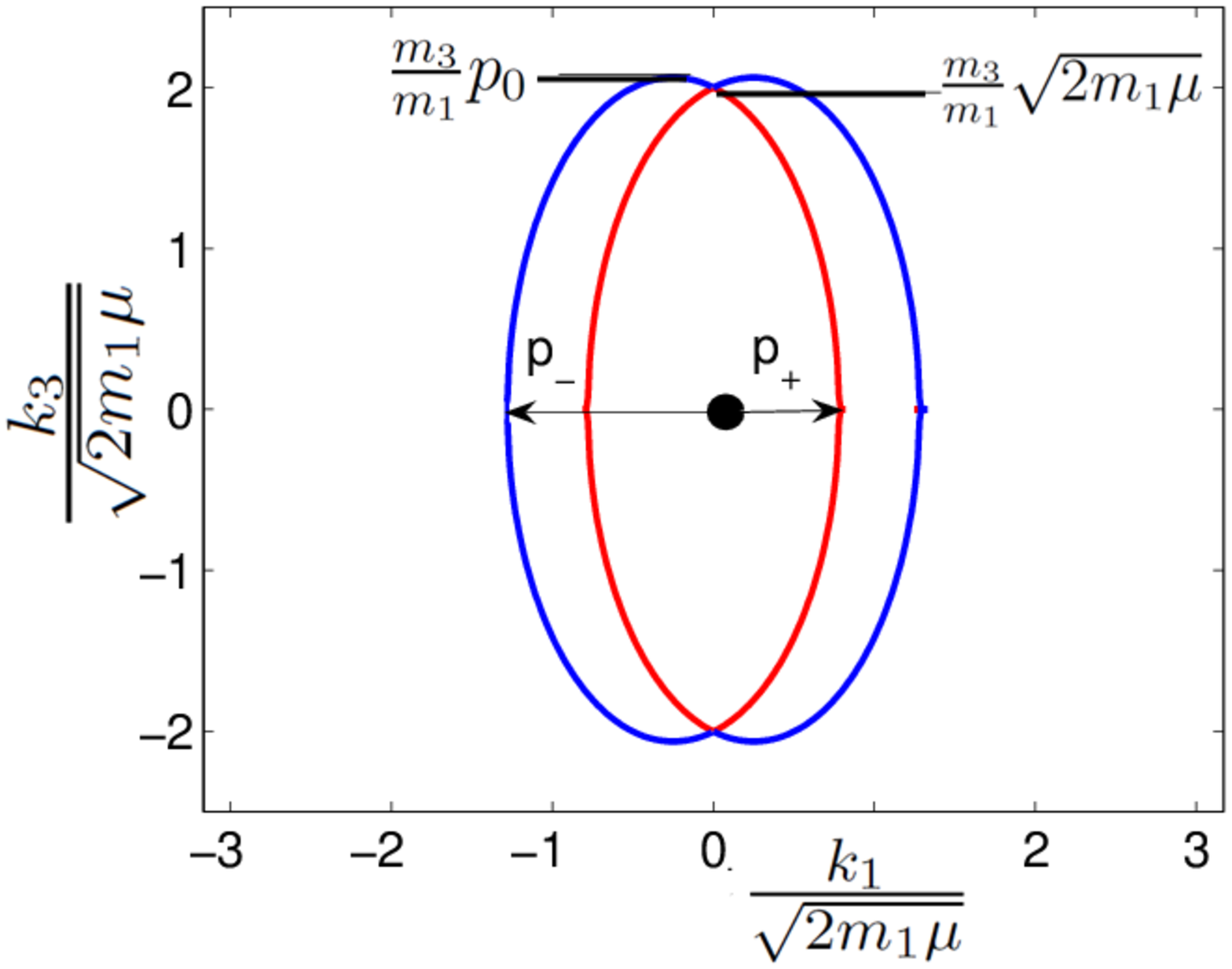}
\end{array}$
\caption{\label{fig:FS}
Left: Fermi surface of a 2D electron system with Rashba spin-orbit coupling for $\mu>0$. Right: A cut along $k_2=0$ plane of the 3D Fermi surface corresponding to the spectrum (see \ref{eq:Chiral spectrum}),  which is isotropic in the $x-y$ plane. $\mu>0$. The 3D Fermi surface is obtained by rotating the 2D contours about the $k_3$ axis. Here, $\alpha=0.25\sqrt{2\mu/m_1}$ and ${m_3}/{m_1}=2$.}
\end{figure*}

The general formalism for studying the collective modes is the same in 2D (Sec.~\ref{sec:Model}); an important difference, however, is $k_3$ integration which will lead to qualitative differences between the 2D and 3D cases.

\subsection{Charge sector: plasmons}\label{subsec:plasmons3D}
Since the 3D system is anisotropic (the $x_3$ direction is different from any direction in the $x_1x_2$ plane ), we would expect the in-plane and out-of-plane responses to be different. Indeed, the system does support an anisotropic plasmon mode whose frequency, given by the equation $1-V(q)\Pi^0_{00}=0$, depends on the direction of  $\bq$.
Although the exact solution for generic $\bq$ is a quite involved,  one can readily work out the cases of the out-of-plane ($\bq=q\hat{x}_1$) and in-plane ($\bq\cdot\hat{x}_3=0$) plasmons.

\subsubsection{Out-of-plane plasmon ($\bq=q\hat{x}_3$)}
In this case, the calculation is quite simple because angular integration  is rendered trivial. According to Eq.~(\ref{eq:pi}), $\Pi^0_{00}$ is given by an integral of $\mathcal{T}_{00}$ [Eq.~(\ref{eq:00})]. Because $\bq$ is out of plane, $\cos(\theta_{\bk}-\theta_{\bk+\bq})=1$, and thus one needs to evaluate only the convolution$\int g_+g_+ + g_-g_-$. Expanding $\mathcal{T}_{00}$ to order $q_3^2$ and integrating over $K$, we obtain
\bea
\Pi^0_{00}(\bq,\Omega)
=\frac{q^2}{\Omega^2}\frac{p_0^3}{\pi^2\sqrt{m_1m_3}}\left(\frac13 s_2^3+\frac12 s_1^2s_2+ \frac12 s_1 \sin^{-1}s_1\right),
\nonumber\\
\label{eq:pi_00_3d}
\eea
where $q=q_3$ and $s_{1,2}$ are defined in Eq.~(\ref{eq:3D ratios}). [As a check, one can show that if $\alpha=0$ and $m_1=m_3=m$, then $p_0\rightarrow p_F=\sqrt{2m\mu}$ and we reproduce the isotropic 3D limit: $\Pi^0_{00}= ({p_F^3/3\pi^2}){q^2}/m{\Omega^2}= nq^2/m{\Omega^2}$.]
The prefactor in Eq.~(\ref{eq:pi_00_3d}) is related to the total number density. Indeed,
a rather straightforward calculation shows that the number density for the spectrum (\ref{eq:Chiral spectrum}) is given by

\bea\label{eq:volume2}
n_{3D}&=& m_3\frac{p_0^3}{\pi^2\sqrt{m_1m_3}}\left(\frac13 s_2^3+\frac12 s_1^2s_2+ \frac12 s_1 \sin^{-1}s_1\right),\nonumber\\
\eea
and thus Eq.~(\ref{eq:pi_00_3d}) can be re-written as
\bea\label{eq:pi_00_3d_v2}
\Pi^0_{00}(q,\Omega)&=& \frac{n_{3D}}{m_3}\frac{q^2}{\Omega^2}.
\eea
Consequently, the plasmon frequency at $q=0$ is $\Omega^2={4\pi n_{3D}e^2}/{m_3}$, which is the same as in the absence of SOC.

\subsubsection{In-plane plasmon ($\bq\cdot\hat{x}_3=0$)}
\label{sec:inplane}
This case is more involved because of angular integration. In the limit of small $q_{\parallel}\equiv \sqrt{q_1^2+q_2^2}$, one can expand the angular factors in Eq.~(\ref{eq:00}) as:
\bea\label{eq:xpansion}
1+\cos(\theta_{\bk}-\theta_{\bk+\bq})&\approx & 2-\frac{q^2_{\parallel}}{2k^2_{\parallel}}\sin^2\theta,\nonumber\\
1-\cos(\theta_{\bk}-\theta_{\bk+\bq})&\approx & \frac{q^2_{\parallel}}{2k^2_{\parallel}}\sin^2\theta.
\eea
After some algebra, we obtain

\bwt
\bea\label{eq:inplane_pi00} \Pi^0_{00}(\bq,\Omega)=
\frac{q^2}{\Omega^2}\frac{m_3}{m_1}\frac{p_0^3}{\pi^2\sqrt{m_1m_3}}\left(  \frac13 s_2^3 + \frac14 s_1^2s_2+ \frac14 s_1 \sin^{-1}s_1 + \frac{m_1\Omega}{16 p_0^2}\left[ L_1(\Omega)+ L_2(\Omega)\right]\right),
\eea
\ewt
where $q=q_{||}$ and
\bea\label{eq:L1L2}
L_1(\Omega)&=&\int_0^{s_2} dk~\text{ln}\left[\frac{\lr\frac{\Omega}{2\alpha p_0}+s_1+i\delta\rr^2-1+k^2}{\lr
\frac{\Omega}{2\alpha p_0}-s_1+i\delta\rr^2-1+k^2}\right],\nonumber\\
L_2(\Omega)&=&\int_{s_2}^1 dk~\text{ln}\left[\frac{\lr\frac{\Omega}{2\alpha p_0}+\sqrt{1-k^2}+i\delta\rr^2-s_1^2}{\lr\frac{\Omega}{2\alpha p_0}-
\sqrt{1-k^2}+i\delta\rr^2-s_1^2}\right].
\nn\\
\eea
The integrals in the expressions above can be performed analytically but the final results are not very insightful, and we  refrain from presenting them. As a consistency check, one can verify that in the limit $\Omega\gg 2\alpha p_0$, when SOC becomes irrelevant, Eq.~(\ref{eq:inplane_pi00}) reproduces correctly the $\alpha=0$ result, i.e.,
\bea\label{eq:inplane_pi00_v2}
\Pi^0_{00}(q,\Omega\gg 2\alpha p_0)&=& \frac{q^2}{\Omega^2}\frac{n_{3D}}{m_1}.\eea
[Indeed,  in this limit $L_1(\Omega)\rightarrow \frac{p_0^2}{16m_1\Omega} \left(  \frac12 s_1^2s_2\right)$ and $L_2\rightarrow\frac{p_0^2}{16m_1\Omega} \left(  -\frac14 s_1^2s_2+ \frac14 s_1 \sin^{-1}s_1\right)$, which, with the help of Eq.~(\ref{eq:volume2}) for the number density, yields Eq.~(\ref{eq:inplane_pi00_v2}).]

Substituting $\Pi_{00}^0$ from Eq.~(\ref{eq:inplane_pi00}) into $1-V(q)\Pi^0_{00}=0$, we re-write the plasmon equation as

\begin{widetext}
\bea\label{eq:plasmon_3D}
\Omega^2&=&\frac{4\pi e^2}{m_1}\tilde{n}(m_3,\alpha,\Omega),\nonumber\\
\text{with}~~~\tilde{n}(m_3,\alpha,\Omega)&=&\frac{m_3p_0^3}{\pi^2\sqrt{m_1m_3}}\left( \frac13 s_2^3 + \frac14 s_1^2s_2+ \frac14 s_1 \sin^{-1}s_1 + \frac{m_1\Omega}{16 p_0^2}\left[L_1(\Omega)+L_2(\Omega)\right] \right),
\eea
\end{widetext}
where the function $\tilde{n}(m_3,\alpha,\Omega)$  reduces to $n_{3D}$ in the limit $\Omega\gg 2\alpha p_0$. An important feature of the current case, as compared to the case without SOC, is that  $\tilde n$ has an imaginary part (due to imaginary parts of $L_1$ and $L_2$) and thus the plasmon can be Landau-damped by particle-hole excitations within the Rashba continuum. On its turn, the Rashba continuum in the 3D case is different from that in 2D: along the frequency axis, the 3D continuum starts right at $\Omega=0$ and goes up to frequency $\Omega_-$ (see Fig.~\ref{fig:cont_diff}). Therefore, if the plasma frequency happens to lie below $\Omega_-$, the plasmon is damped even at $q=0$.
\begin{figure}[htp]
\includegraphics[width=0.8\columnwidth]{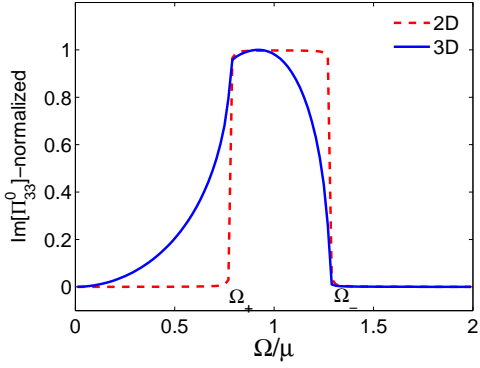}
\caption{\label{fig:cont_diff} The Rashba continuum--the region where Im$\Pi^0_{33}
$ (and other Im$\Pi^0_{ij}$'s) is non-zero--in 2D (dashed) and 3D (solid). In 2D, the continuum is bounded above and below by $\Omega_{-}$ and $\Omega_+$, correspondingly. In 3D, the upper limit is at $\Omega_-$ while the lower limit extends all the way to $\Omega=0$ with a cusp at $\Omega_+$. The plots have been normalized to their respective maximum values for the sake of comparison. Here $\alpha=0.25\sqrt{2\mu/m_1}$ and $m_3=2m_1$.}
\end{figure}

The boundaries of the Rashba continuum in 3D can be found analytically. To this end, we note that the imaginary part comes from the off-diagonal convolution of the Greens functions, $\int g_+g_-+g_-g_+$, rather than from the angular factors. The same convolution occurs in $\Pi_{00}^0$ and $\Pi_{33}^0$ [cf. Eqs.~(\ref{eq:00}) and (\ref{eq:ZZ})]. It is more convenient then to determine the continuum boundaries from Im$\Pi_{33}^0$, which is finite at $q=0$, than from Im$\Pi_{00}^0$, which vanishes at $q=0$ and hence needs to be expanded to order $q^2$. The expression for  $\Pi^0_{33}$ at $q=0$ reduces to
\bea
\Pi^0_{33}(0,\Omega)&=&\int_K (g_+g_- + g_-g_+)\label{eq:im}\\
&=&\int\frac{d^3k}{(2\pi)^3} \left[\frac{n_F\left(\varepsilon^+)-n_F(\varepsilon^-\right)}{i\Omega+2\alpha k_{\parallel}}+~~\alpha\rightarrow-\alpha\right].\nn
\eea
For $\Omega>0$, the imaginary part of Eq.~(\ref{eq:im}) is given by
\bea
\I\Pi^0_{33}(0,\Omega)&=&-\frac{1}{2\pi}\int_0^{\sqrt{2m_3\mu}} dk_3 \int_{\tilde{p}_+}^{\tilde{p}_-} k_{\parallel}dk_{\parallel}
\delta\left(\Omega-2\alpha k_{\parallel}\right),\nn\label{eq:im2}\\
\eea
where
\beq
\label{eq:tildep}
\tilde{p}_{\pm}(k_3)=\sqrt{2m_1\left(\mu-\frac{k_3^2}{2m_3}\right) +m_1^2\alpha^2}\mp m_1\alpha.
\eeq
In 2D, when the $k_3$ integral is absent and $\tilde p_\pm(k_3=0)= p_\pm$,
$\I\Pi^0_{33}(0,\Omega)\neq 0$
is non-zero only for $2\alpha p_+<\Omega<2\alpha p_-$. In 3D, the spectral weight is further integrated over $k_3$:
\bea\label{eq:im3}
\I\Pi^0_{33}(0,\Omega)&=&-\frac{1}{2\pi}\frac{\Omega}{(2\alpha)^2}\int_0^{\sqrt{2m_3\mu}} dk_3~~\Theta\left(\frac{\Omega}{2\alpha}-\tilde p_+\right)\nn\\
&\times&\Theta\left(\tilde p_--\frac{\Omega}{2\alpha}\right).
\nonumber\\
\eea
Since $\tilde p_+(k_3=\sqrt{2 m_3\mu})=0$ and $\tilde p_{-}(k_3=0)=p_-$,
the integral is non-zero for all frequencies in the interval
$0<\Omega<2\alpha p_-$. For $\Omega\ll2\alpha p_0$, $\I\Pi^0_{00}(0,\Omega)\propto |\Omega|\Omega$.
The profiles of $\I\Pi_{33}^0$ as a function of $\Omega$ in 2D and 3D are shown in Fig.~\ref{fig:cont_diff}.

A graphic solution of Eq.~(\ref{eq:plasmon_3D}) is shown in Fig.~\ref{fig:trans3}. The solid and dashed-dotted lines depict the real and imaginary parts of the RHS, correspondingly. In agreement with the argument given in the preceding paragraph, the imaginary part is non-zero in the interval of frequencies
$0\leq \Omega\leq \Omega_-$.
The root of the equation is shown by the dot. We chose the material parameters appropriate for BiTeI:
$\alpha=4.0$ eV$\times$\AA,\cite{Hamlin,Tokura}$m_1=0.1~m_e$, \cite{Hamlin} $m_3=1.0~m_e$,\cite{optics_w_abinitio}, and the backround dielectric constant $\epsilon_\infty\approx 20$.\cite{Tanner_PRL} Note that in the case of a semiconductor with the background dielectric constant $\epsilon_{\infty}$, one needs to replace $e^2$ in Eq.~(\ref{eq:plasmon_3D}) by $e^2/\epsilon_\infty$ to obtain the \lq\lq screened plasma frequency\rq\rq\/; the plasma frequencies in Figs.~\ref{fig:trans3} and \ref{fig:p3d} are the screened ones. In the top panel of Fig.~\ref{fig:trans3}, $\mu=E_R$ (which is the right order of magnitude for
a typical BiTeI sample). The plasmon lies within the Rashba continuum and is hence Landau-damped even at $q=0$. In the bottom panel, we show a hypothetical case of $\mu=500 E_R$, when the plasmon is above the continuum.
\begin{figure}[htp]
$\begin{array}{c}
\includegraphics[width=0.9\columnwidth]{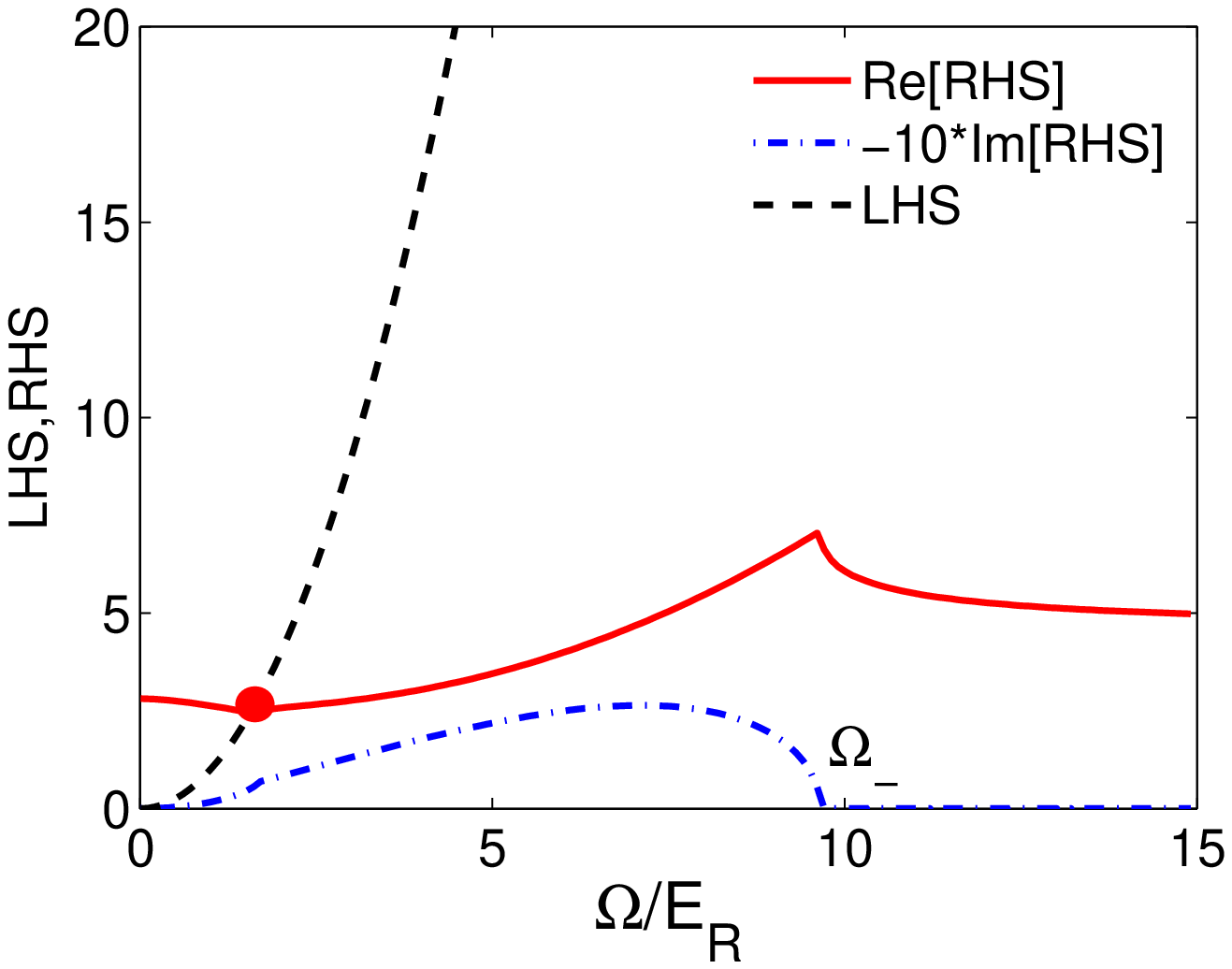}\\
\includegraphics[width=0.9\columnwidth]{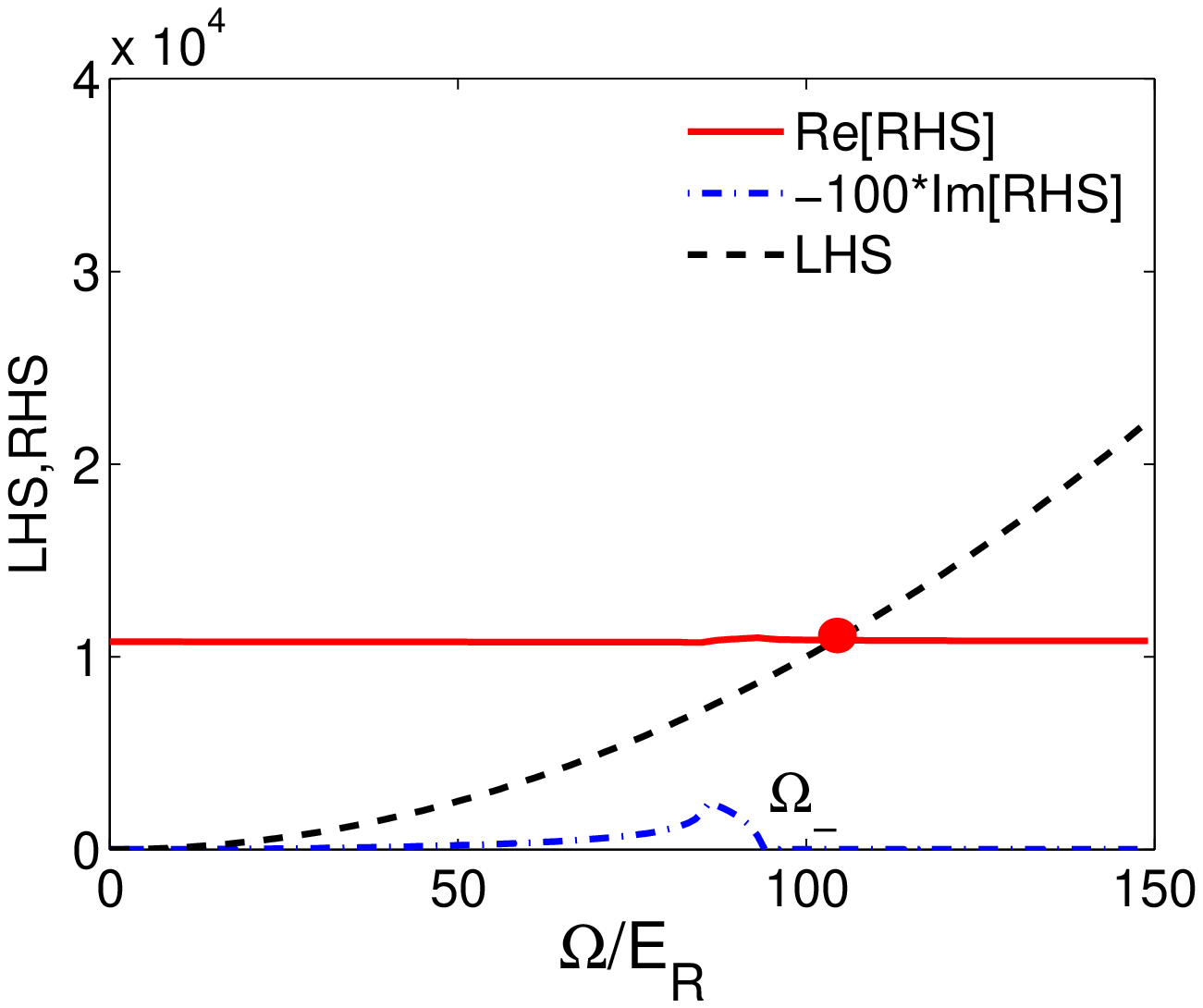}
\end{array}$
\caption{\label{fig:trans3} Graphic solution of Eq.~(\ref{eq:plasmon_3D}) for a 3D plasmon at $q=0$. Solid: real part of RHS, dashed-dotted: imaginary part of RHS, dashed: LHS. The frequency is measured in units of the Rashba energy, $E_R=m_1\alpha^2/2$. The top is for $\mu=E_R$, which is a realistic value for BiTeI, and the bottom is for $\mu=500E_R$. Other material parameters are chosen for BTeI, as specified in the main text.
}
\end{figure}

Figure~\ref{fig:p3d} shows the plasma frequency and the upper edge of the Rashba continuum, $\Omega_-$, as functions of the chemical potential, $\mu$; all frequencies in units of the Rashba energy, $E_R=m_1\alpha^2/2$; the rest of the material parameters is the same as in Fig.~\ref{fig:trans3} and corresponds to BiTeI. In regard to real materials, we note that $E_R\approx 140$ meV while $\mu$ varies substantially from sample to sample. The highest value of $\mu$ reported in the literature is $66$ meV (above the Dirac point)\cite{Tokura} which is more than twice smaller than $E_R$. Therefore,
BiTeI (and other materials from this family) appear to be squarely in the regime of damped plasmons.
For our choice of the material parameters, undamped plasmons exist only when $\mu \gtrsim
330 E_R$, which implies an unrealistically high doping level.

\begin{figure}[htp]
\includegraphics[width=.9\columnwidth]{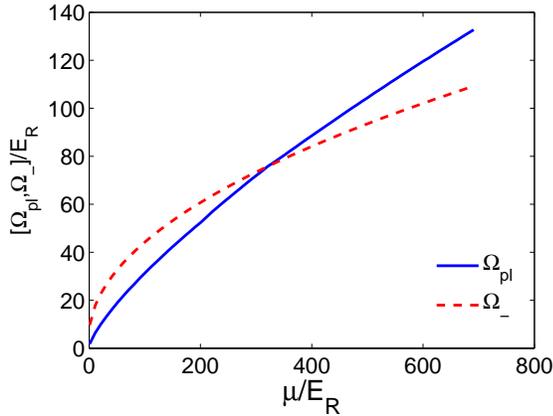}
\caption{\label{fig:p3d}
The plasma frequency ($\Omega_{\mathrm{pl}}$) and the upper boundary of the Rashba continuum ($\Omega_-$) as functions of the chemical potential. All energies are measured in units of $E_R=m_1\alpha^2/2$. Material parameters for BiTeI are specified in the main text.
}
\end{figure}

Although we looked at the limiting case of the in- and out-of-plane plasmons, in reality, there is just one plasmon branch with anisotropic dispersion. This is fundamentally different from the 2D case, where there are two distinct plasmon branches. Furthermore, as long as $\bf q$ has an in-plane component, the plasmon is modified by Rashba SOC and, if the number density of carriers is sufficiently small, it may be damped by particle-hole excitations with the Rashba continuum. We emphasize that Landau damping of plasmons at $q=0$ is a unique feature of a 3D Rashba system.

\subsection{Optical conductivity and the sum rule}\label{subsec:conductivity3D}
As in the 2D case, intersubband transitions give rise to a non-zero real part of the optical conductivity even in the absence of scattering. In 2D, the Rashba continuum occupies an interval of frequencies $\Omega_{+}<\Omega<\Omega_{-}$ and, consequently, $\R\sigma(\Omega)\neq 0$ within the \lq\lq box\rq\rq\/ bounded by these frequencies; outside this box $\R\sigma(\Omega)=0$ in the absence of scattering (see Fig.~
\ref{fig:Cond2d}). In 3D, the Rashba continuum occupies the interval $0<\Omega<\Omega_{-}$ and $\R\sigma(\Omega)\neq 0$ within this range; thus the gap between the Drude $\delta$-function peak at $\Omega=0$ and $\Omega_-$ is filled (see Fig.~\ref{fig:cond_3d}). The frequency $\Omega_+$ in 3D marks a pronounced feature in $\R\sigma(\Omega)\neq 0$ but not the lower edge of absorption, as it does in 2D.  The overall shape of $\R\sigma(\Omega)$ in Fig.~\ref{fig:mu} agrees qualitatively with the optical data on BiTeI.\cite{optics_w_abinitio,optics2}

Following the standard procedure, one can relate $\sigma(\Omega)$ to an experimentally observable reflectance spectrum, $R(\Omega)$. The resulting $R$, along with real and imaginary parts of the dielectric function, $\epsilon(\Omega)$, are shown in Fig.~\ref{fig:mu}. The top panel corresponds to the case when the plasmon is damped by the Rashba continuum ($\mu=E_R$), while the bottom one to the case when the plasmon is above the Rashba continuum ($\mu=500E_R$). Note a weak feature in $R(\Omega)$ below the plasma edge for $\Omega<\Omega_-$ in the second case. The material parameters are the same as chosen for Fig.~\ref{fig:trans3}. For comparison, we also show the reflectance of a system without SOC (the curve labeled $R_0$).
\begin{figure}[htp]
\includegraphics[width=0.9\columnwidth]{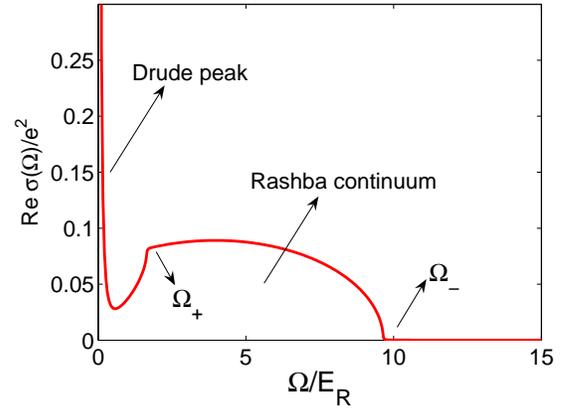}
\caption{\label{fig:cond_3d}The real part of the conductivity of a 3D system with Rashba SOC for $\mu=E_R$. To simulate the effect of disorder, the levels were broadened by $1/\tau=0.01\mu$. This plot is to be contrasted to the 2D case in Fig. \ref{fig:Cond2d}.}
\end{figure}

\begin{figure}[htp]
$\begin{array}{c}
\includegraphics[width=0.9\columnwidth]{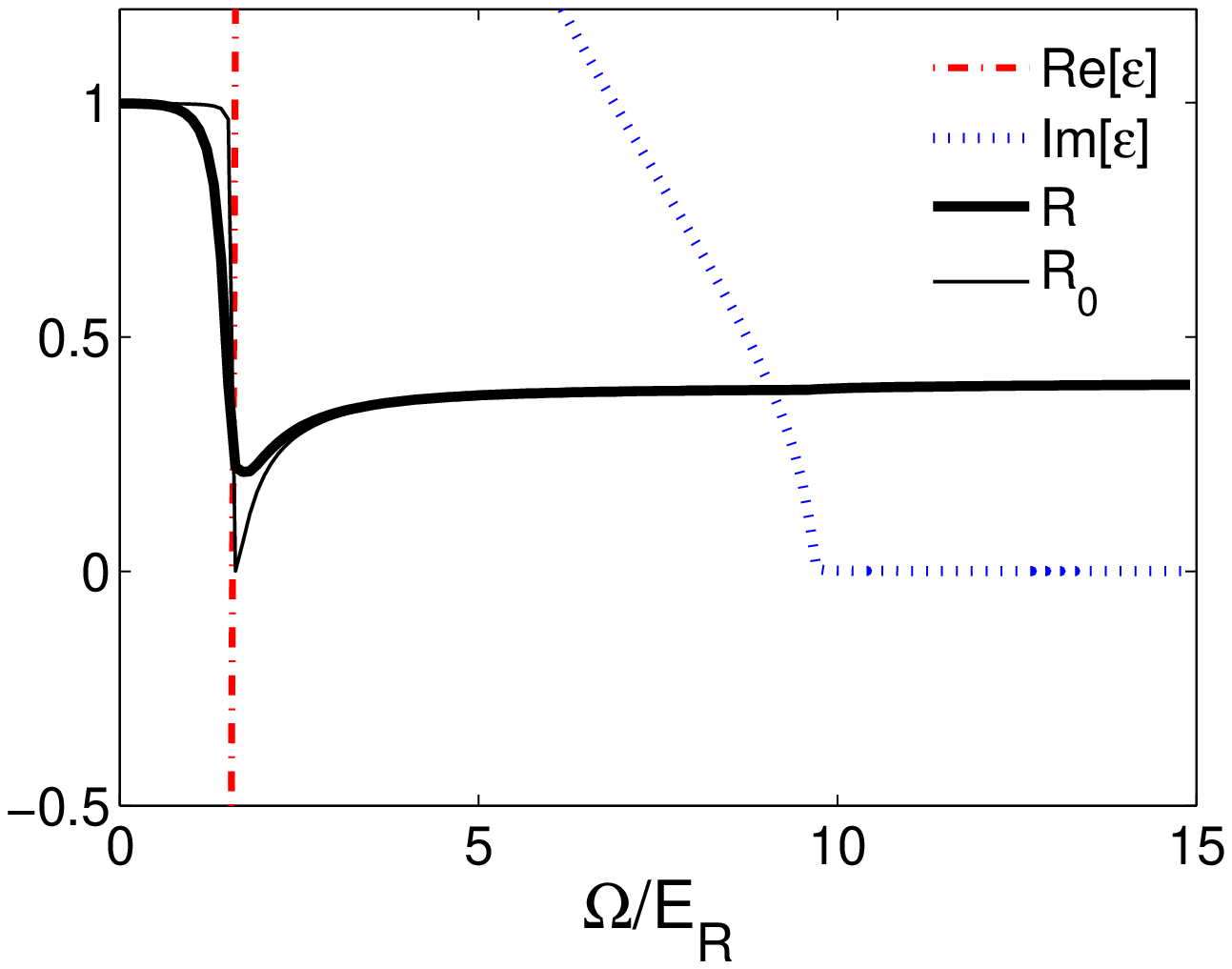}\\
\includegraphics[width=0.9\columnwidth]{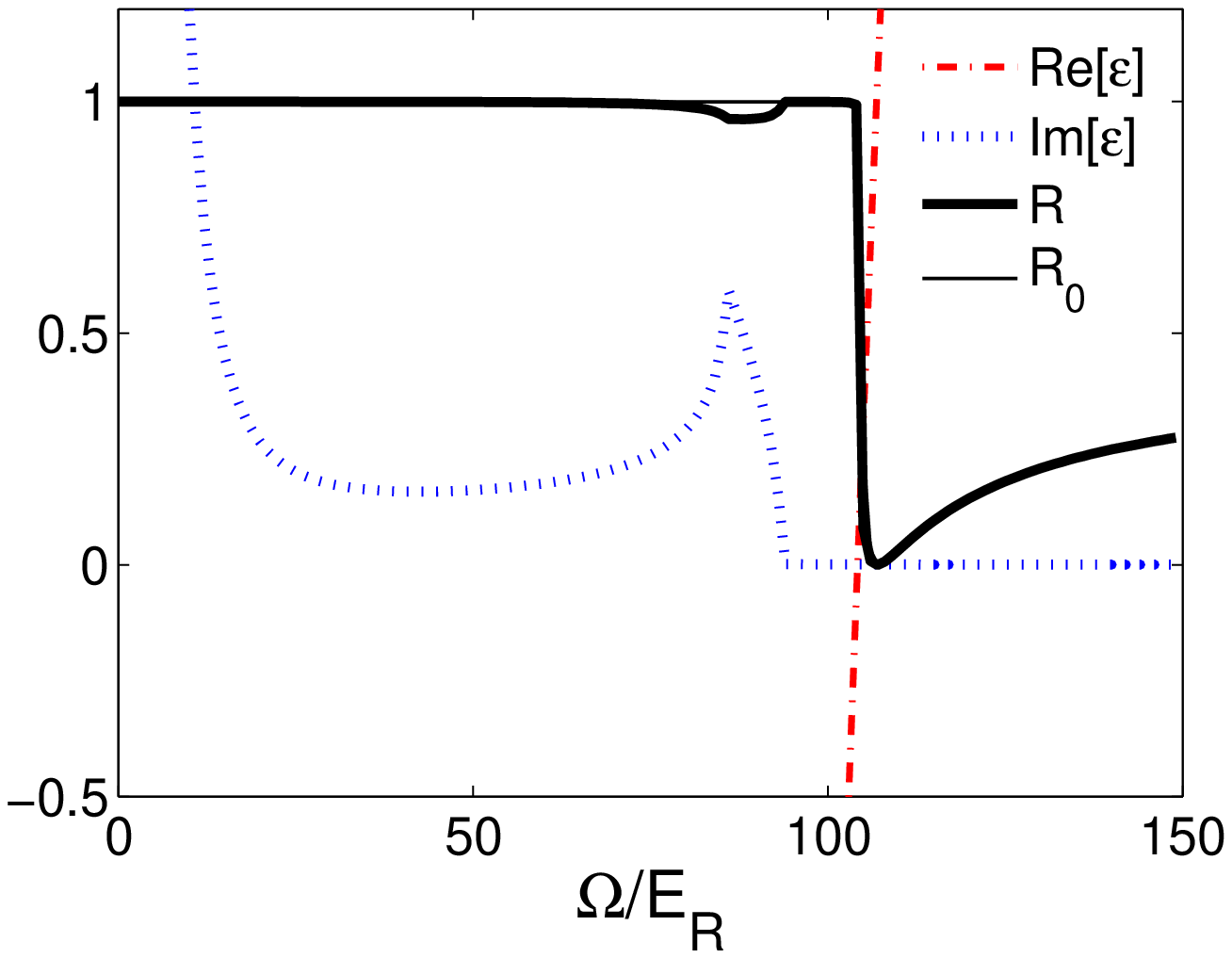}
\end{array}$
\caption{\label{fig:mu}
Real and imaginary parts of the dielectric function $\varepsilon(\Omega)$ and reflectance $R$ of a 3D material with Rashba SOC. $R_0$ is the reflectance without SOC. The material parameters are chosen for BiTeI. Top: $\mu=E_R$. In this case the plasma edge is broadened. Bottom: $\mu=500 E_R$. There is a weak feature in $R$ for $\Omega<\Omega_-$ and a sharp plasma edge at the plasmon frequency.}
\end{figure}

We see from Eq.~(\ref{eq:pi_00_3d}) that the Drude weight ($\sim \frac{\Omega^2}{q^2}\Pi^0_{00}$) is affected by SOC. Indeed, the Drude weight is proportional to $\tilde n(m_3,\alpha,\Omega=0)$ which, in general, does not coincide with the actual number density, $n_{3D}$. Just as in 2D, this indicates a spectral weight redistribution between the Drude and Rashba-continuum parts of the conductivity. To check if the sum rule is satisfied, we calculate the in-plane conductivity of a 3D system using the Kubo formula [Eq.~(\ref{eq:conductivity_kubo})].
On adding the diamagnetic term, we find
\bea\label{eq:cond_3D}
\sigma(\Omega)&=&\frac{i{\cal K}(\Omega)}{\Omega},\nn\\
{\cal K}(\Omega)&=&\frac{e^2}{m_1}\left[n_{3D} -\sqrt\frac{m_3}{m_1}\frac{p_0^3}{\pi^2}\left(\frac14s_1^2s_2 +\frac14s_1\sin^{-1}s_1\right)\right.\nonumber\\
&&\left.+\sqrt\frac{m_3}{m_1}\frac{p_0^3}{\pi^2} \frac{m_1\Omega}{16p_0^2} \left(L_1(\Omega) + L_2(\Omega)\right)\right].
\eea
The frequency-independent term in $\R{\cal K}$ represents the Drude weight reduced by intersubband transitions. Just as in 2D, the loss of the spectral weight
at $\Omega=0$ is compensated by the contribution from the Rashba continuum at finite frequencies. Indeed, it can easily be checked that the integral
\beq\label{eq:redistribution}
\int_0^{\Omega_-} d\Omega ~\text{Im}\left[L_1(\Omega)+L_2(\Omega)\right]=\frac{2\pi p_0^2}{m_1} \left(s_1^2s_2 +s_1\sin^{-1}s_1\right)\eeq
gives exactly the weight missing from the Drude term.

\subsection{Spin-chiral modes in three dimensions}\label{subsec:spin modes3D}
The situation with the chiral-spin modes in 3D differs qualitatively from that in 2D. First, the chiral-spin modes in 3D occur only
if the strength of the electron-electron interaction exceeds a threshold value--this is in contrast to the 2D case, where the modes occur even for an infinitesimally weak interaction. Second, even the modes do occur, they lie within the Rashba continuum and are thus Landau-damped by intersubband particle-hole excitations. Generically, the width of a mode is comparable to its frequency and thus one can expect
to see only broad resonances rather than well-defined excitations.

We begin the analysis of the 3D case with the limit of weak SOC and consider the modes only at $q=0$. In this case, the off-diagonal components  of
the spin-charge susceptibility tensor vanish and the three chiral-spin modes decouple. The masses of the modes are determined from the equation $1+(U/2)\Pi^0_{ii}(0,\Omega)=0$ with $i=1, 2, 3$. Noticing also that $\Pi^0_{11}(0,\Omega)=\Pi^0_{22}(0,\Omega)=(1/2)\Pi^{0}_{33}(0,\Omega)$,
we focus on the $33$ mode. The frequency of the mode at $q=0$ is given by the equation $1+(U/2)\R\Pi^0_{33}(0,\Omega)=0$
with
\bea
\R\Pi^0_{33}(0,\Omega)&=&\frac{1}{\pi^2}{\cal P}\int^{\sqrt{2m_3\mu}}_0 dk_3\int^{\tilde p_-}_{\tilde p_+} dk_{||}\nn\\
&\times&
\frac{2\alpha k_{||}^2}{\Omega^2-(2\alpha k_{||})^2},\label{sc3d1}
\eea
where  $\tilde p_{\pm}$ are defined in Eq.~(\ref{eq:tildep}) and  ${\cal P}\int$ denotes the Cauchy principal value of an integral. For weak SOC, momenta $\tilde p_+$ and
$\tilde p_-$ are close to each other and thus $k_{||}$ under the integral can be replaced by $\sqrt{2m_1\left(\mu-k_3^2/2m_3\right)}$.
After some re-arrangements, Eq.~(\ref{sc3d2}) is reduced to
\bea
\R\Pi^0_{33}(0,\Omega)&=&-\nu
\lr 1-\frac{\Omega^2}{\Omega_0^2}{\cal P}\int^{1}_0 \frac{dx}{x^2-1+
\lr\Omega/\Omega_0\rr^2}\rr,\nn\\
\label{sc3d2}
\eea
where $\nu=m_1\sqrt{2m_3\mu}/2\pi^2$ is the density of states at the Fermi level (per one spin projection), $x\equiv k_3/\sqrt{2m_3\mu}$,  and $\Omega_0=2\alpha \sqrt{2m_1\mu}$ in the limit of small $\alpha$. Solving the integral, we obtain

\bea
  &&\R\Pi^0_{33}(0,\Omega)=-2 \nu P\lr \frac{\Omega}{\Omega_0}\rr\nn\\
  &&P(y)=1+
  \left\{
  \begin{array}{llc}
  \frac{y^2}{2\sqrt{1-y^2}}\ln\frac{1+\sqrt{1-y^2}}{1-\sqrt{1-y^2}},\;\mathrm{for}\;0<y<1;\\
  -\frac{y^2}{\sqrt{y^2-1}}\arctan\frac{1}{\sqrt{y^2-1}},\;\mathrm{for}\;y>1.
  \end{array}
  \right.
  \label{sun1}
 \eea
Notice that $P(0)=1$, $P(y\to 1^-)=2$, and $P(y\to 1^+)=-\infty$. A plot of $\R\Pi^0_{33}(0,\Omega)$ is shown in Fig.~\ref{fig:new}.

\begin{figure*}[htp]
$\begin{array}{ccc}
\includegraphics[width=0.7\columnwidth]{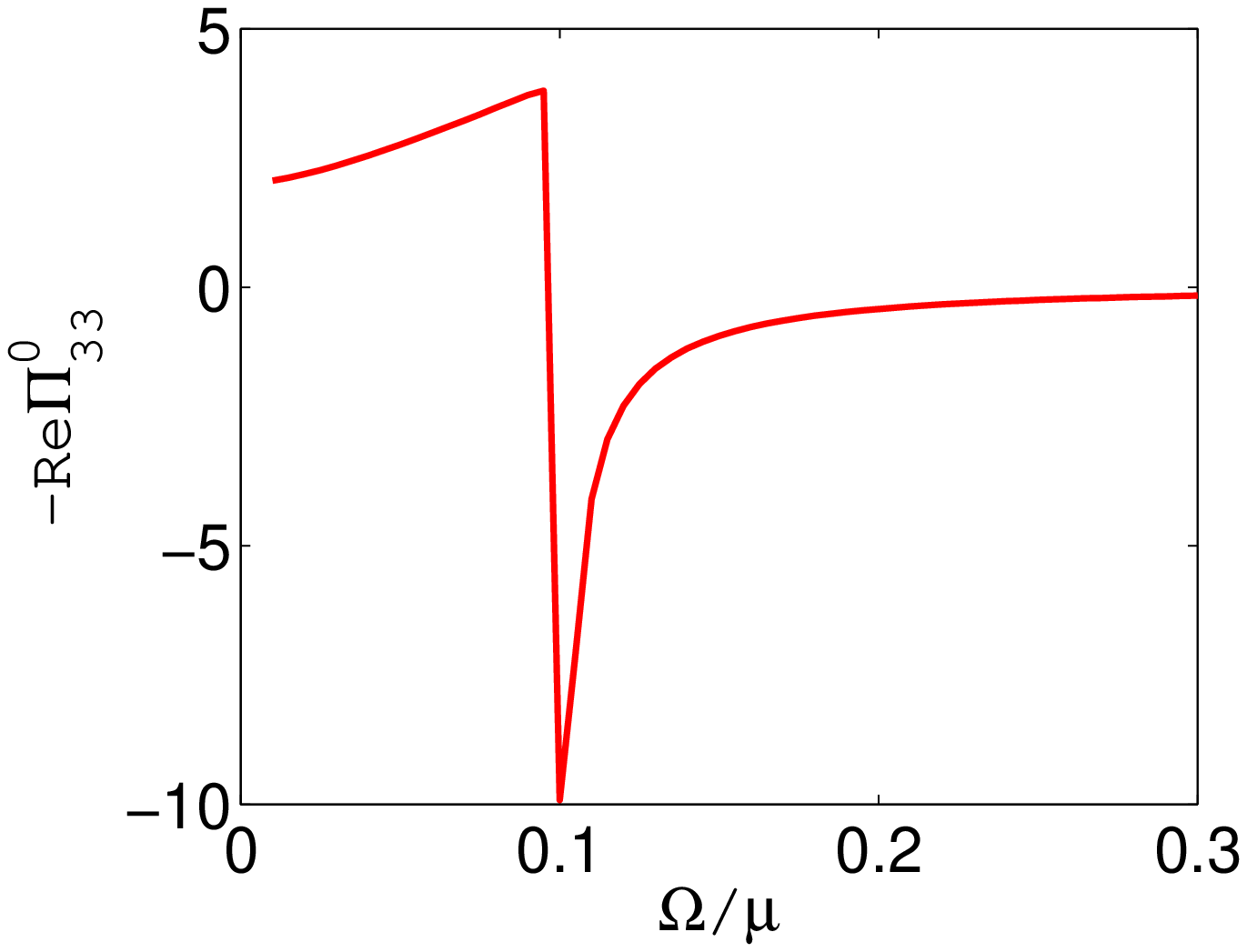}&
\includegraphics[width=0.7\columnwidth]{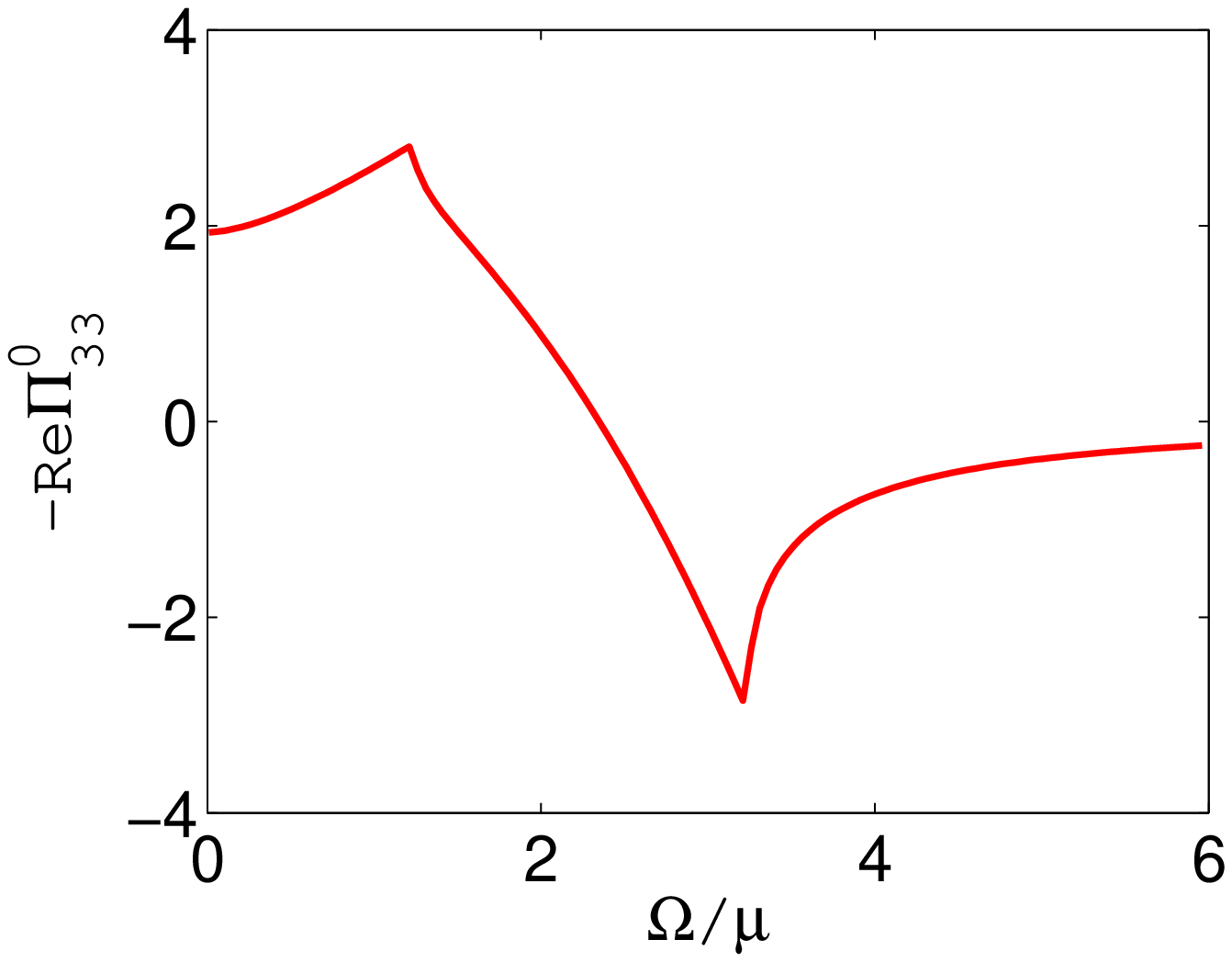}&
\includegraphics[width=0.7\columnwidth]{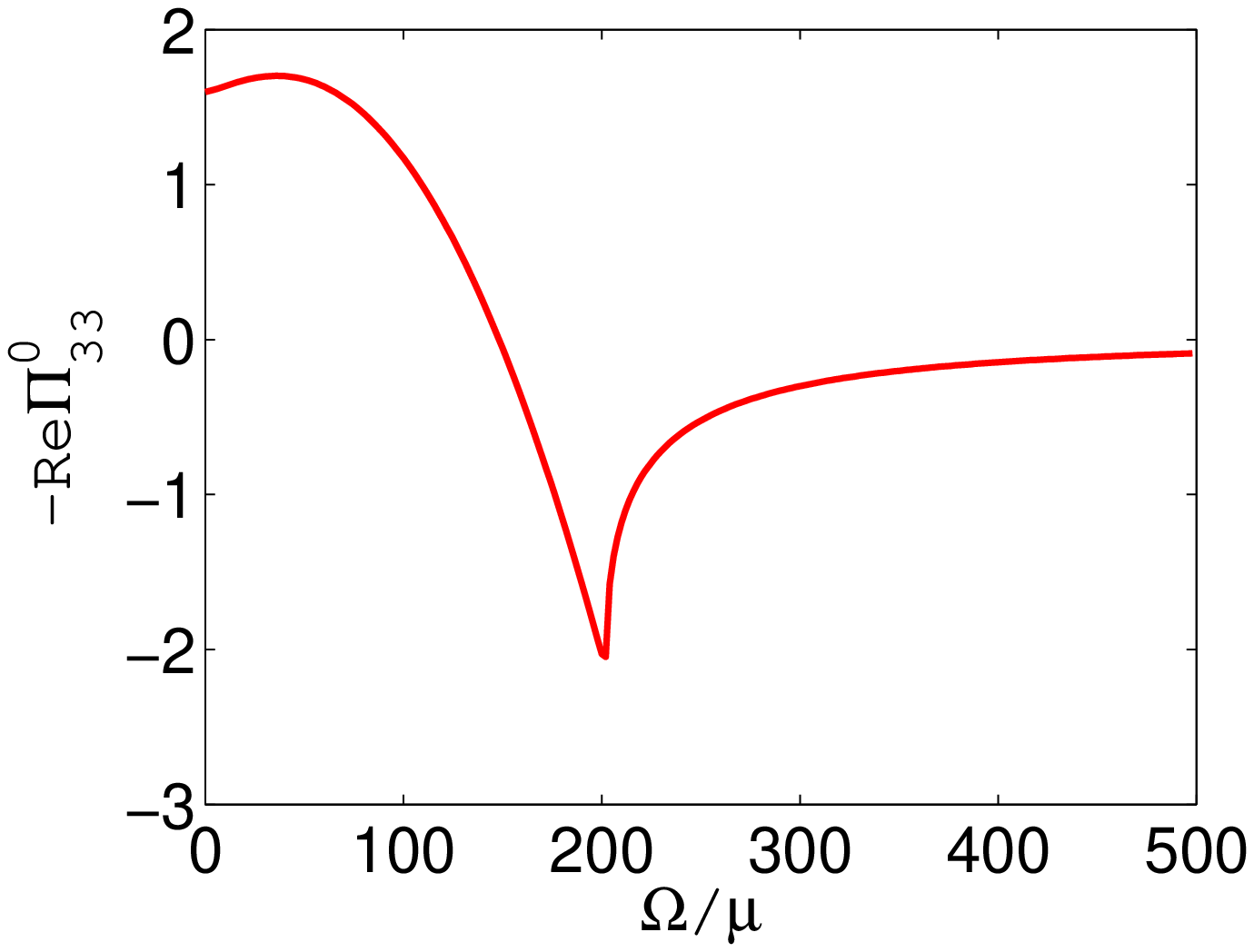}
\end{array}$
\caption{
\label{fig:new} Re$\Pi^0_{33}(0,\Omega)$ for
increasing
values of $\alpha$:
$\alpha/\sqrt{2\mu/m_1}=0.025$ (left); $0.5$ (center); $5.0$ (right). Here, $m_3=2m_1$.}
\end{figure*}

In a dimensionless form, the equation for the mass of the mode reads
\beq\label{eq:r} \frac{2}{u}=P\lr\frac{\Omega}{\Omega_0}\rr,\eeq
where $u\equiv U\nu$. At $u>0$, the solution is possible only for $\Omega<\Omega_0$, where $P$ is positive. In contrast to the 2D case, however, the LHS of the equation is finite within this interval and thus there are no solutions at weak coupling ($u\ll 1$). For the 33 mode, a solution exists if $1/2<u<1$. Notice that $u=1$ corresponds to a ferromagnetic (Stoner) instability, at least  in the mean-field approximation. For the $11$ and $22$ modes, one simply has to replace $u\to u/2$; consequently,  the minimum value of $u$ moves up to $u=1$. Within this approximation, therefore, there are no $11$ and $22$ collective mode in the paramagnetic phase.    Still, the mean-field criterion for ferromagnetism may not be accurate. Assuming for a moment that all the modes do occur already in the paramagnetic phase, we proceed with estimating their damping. Since the mode frequency is below $\Omega_0$, it is within the Rashba continuum and thus damped by particle-hole excitations. The imaginary part of $\Pi^0_{33}$ has already been evaluated in Sec.~\ref{sec:inplane} [see Eqs.~(\ref{eq:im}-\ref{eq:im3})].
In the small-$\alpha$ limit, we find \beq
\I\Pi^{0}_{33}(0,\Omega)=-2\nu\frac{\Omega|\Omega|}{\Omega_0^2}
\eeq
for $\Omega\ll \Omega_0$. Extrapolating this formula with an order-of-magnitude accuracy to the region $\Omega\sim \Omega_0$ and recalling that $u\sim 1$, we conclude that the width of the mode is on the order of $\Omega_0$ and thus comparable to the frequency of the mode itself. Therefore, the mode is actually a rather broad resonance.
\begin{figure*}[htp]
$\begin{array}{cc}
\includegraphics[width=0.9\columnwidth]{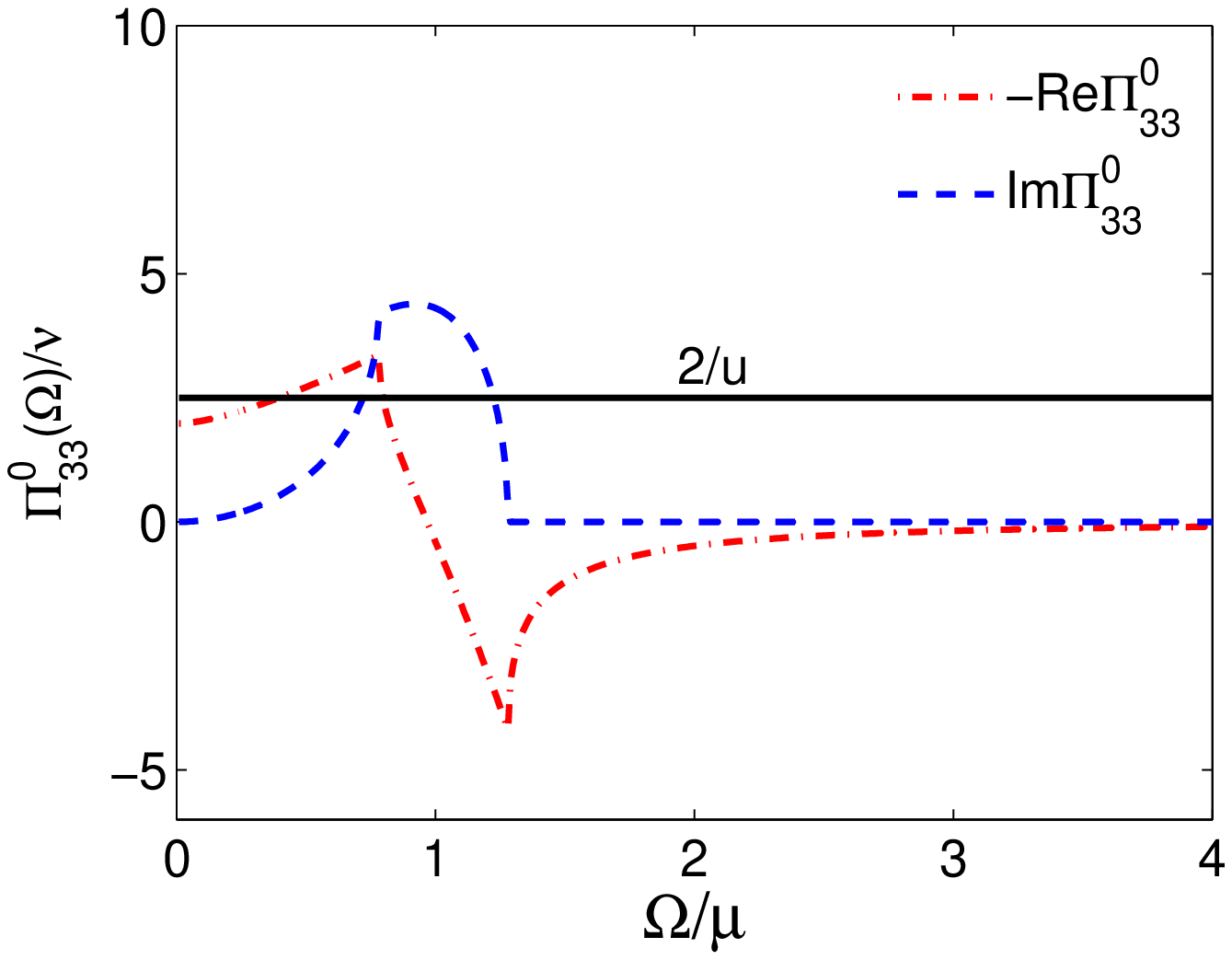}&
\includegraphics[width=0.9\columnwidth]{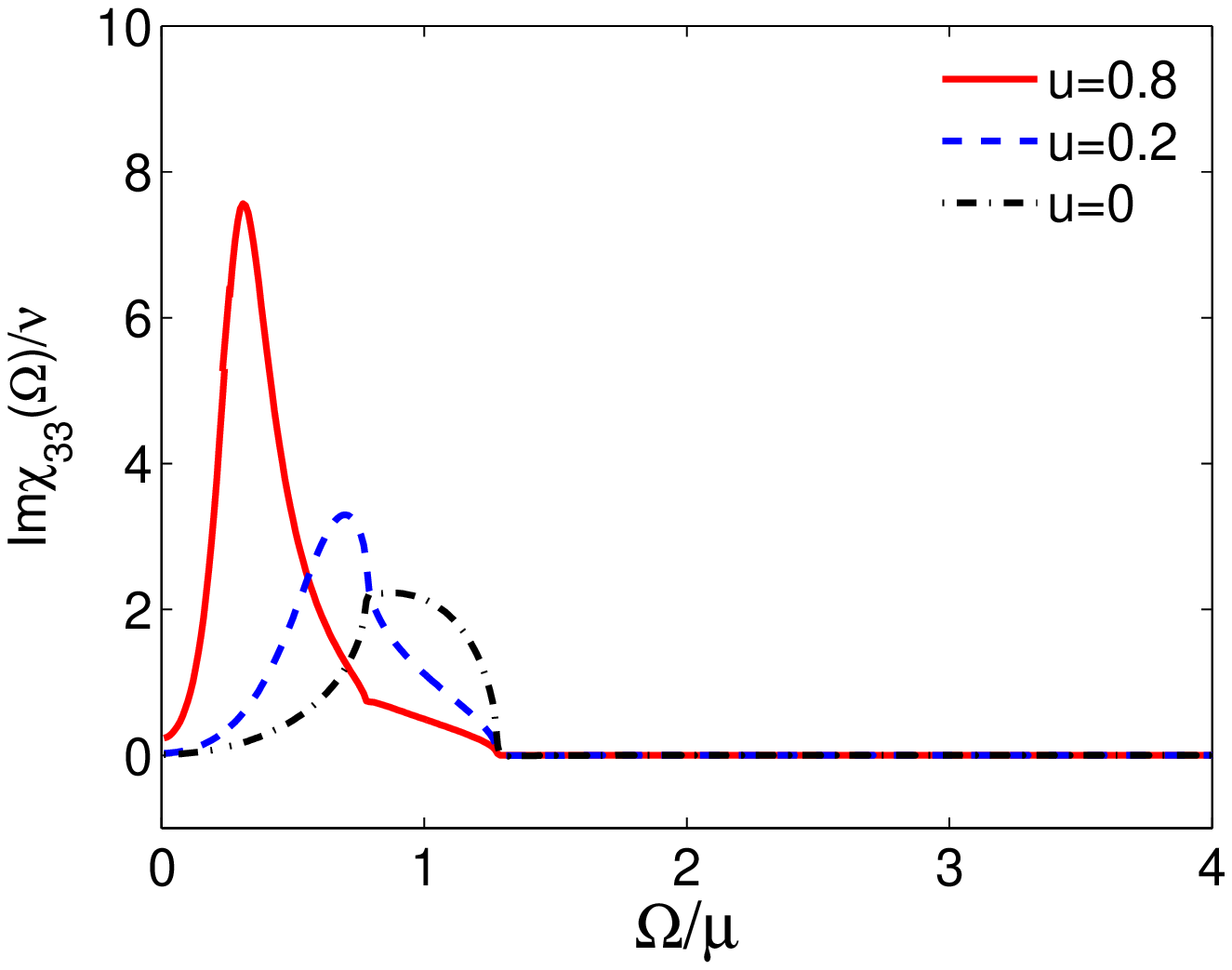}
\end{array}$
\caption{
\label{fig:modes_3d}
Left: Graphic solution of the equation $1+(U/2)\Pi^0_{33}(0,\Omega)=0$ defining the frequency of the mode in the $33$ channel. The dimensionless electron-electron coupling $u\equiv U\sqrt{m_1m_3}p_0/2\pi^2$ is chosen to be 0.8. Right: evolution of $\I\chi_{33}(q,\Omega)$ with the strength of the electron-electron interaction. A resonance peak develops for stronger $u$.}
\end{figure*}

Thus far, we have considered the case of weak SOC, which is not relevant to 3D materials with giant Rashba splitting. However, relaxing the assumption of weak SOC does not change qualitatively the conclusions obtained above:
the modes occur only if the interaction is above a threshold value and are damped; the quantitative change is the the threshold value for $u$ gets larger as $\alpha$ increases. For arbitrary $\alpha$, we find

\begin{widetext}
\bea\label{eq:chi_zz 3D}
\Pi^0_{33}(0,\Omega)=-\sqrt{m_1m_3}\frac{p_0}{2\pi^2}\frac{1}{s_1^2}\left(s_1^2s_2 + s_1\sin^{-1}s_1- \frac{m_1\Omega}{4 p_0^2}\left[L_1(\Omega)+L_2(\Omega)\right]\right),
\eea
\end{widetext}
where $p_0$ is given in Eq.~(\ref{eq:p0}), $s_{1,2}$ are defined in Eq.~(\ref{eq:3D ratios}), $L_{1,2}(\Omega)$ are given in Eq.~(\ref{eq:L1L2}),
and, as before, $\Pi^0_{11}(0,\Omega)=\Pi^0_{22}(0,\Omega)=(1/2)\Pi^0_{33}(0,\Omega)$. The behavior of Re$\Pi^0_{33}(0,\Omega)$ for a range of $\alpha$ is shown in Fig.~\ref{fig:new}. For very small $\alpha$ (left panel), $-\R\Pi^0_{33}$ reaches a maximum value of 4, in agreement with Eq.~(\ref{sun1}). As $\alpha$ increases, the maximum value of $-\R\Pi^0_{33}$  decreases and, consequently, the threshold value of $u$ (defined now as $u=U\sqrt{m_1m_3}p_0/2\pi^2$) increases too: $u_{\text{min}}=0.52,~0.71,~1.18$ from the left to right, correspondingly. For stronger SOC, therefore, a stronger interaction is needed to see the resonances in the susceptibility.

As before, the masses of the modes are given by $1+\frac U2 \Pi_{ii}^0(0,\Omega)=0$ with $i=1,~2,~ 3$. For arbitrary $\alpha$, Eq.~(\ref{eq:r}) for the mass of the $33$ mode is replaced by
\beq\label{eq:r2}
\frac{2}{u}=s_2 + \frac{\sin^{-1}s_1}{s_1} -\frac{\Omega}{4m_1\alpha^2}\left[L_1(\Omega)+L_2(\Omega)\right].
\eeq
This equation is solved graphically in the left panel of Fig.~\ref{fig:modes_3d} for $u=0.8$. In contrast to the weak-SOC case, there are now two solutions of $1+(U/2) \R\Pi_{ii}^0(0,\Omega)=0$ both of which are, however, damped by the continuum. The right panel of Fig. \ref{fig:modes_3d} depicts  $\I\chi_{33}(0,\Omega)$ for various values of $u$. For $u$ above the threshold value, $\I\chi_{33}(0,\Omega)$ exhibits a peak whose width (relative to its center) can be shown to be equal to $
u\pi{(1-s_1)s_2}/{4s_1}$. The peak becomes more prominent for stronger electron-electron interactions.

As in 2D,  the inter-subband part of the in-plane optical conductivity, $\sigma_{11}(\Omega)$, is related to the $22$ component of the spin susceptibility. Subtracting the Drude peak, we obtain for the remainder \beq\label{eq:deltasigma}
\Delta\sigma_{22}(\Omega)=\frac{ie^2\alpha^2}{\Omega}\Pi^U_{22}(0,\Omega).
\eeq
The real part of $\Delta\sigma_{22}(\Omega)$ is plotted in Fig.~\ref{fig:con} for material parameters corresponding to BiTeI and $1/\tau=0.01\mu$. The broad feature just below $\Omega_+$ is due to a damped chiral-spin mode. Comparing Fig.~\ref{fig:con} with the experimental data,\cite{optics_w_abinitio,optics2} we see that the observed conductivity resembles more the theoretical prediction for small $u$, in which case the chiral-spin collective modes do not exist.
\begin{figure}[htp]
\includegraphics[width=0.9\columnwidth]{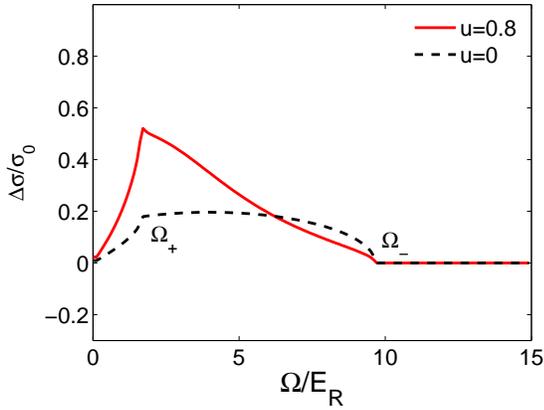}
\caption{
\label{fig:con} Real part of $\Delta \sigma$, as defined by Eq.~(\ref{eq:deltasigma}) and normalized to $\sigma_0\equiv e^2\alpha^2\sqrt{m_1m_3}p_0^2/2\pi^2$. Here $\mu=E_R$, $1/\tau=0.01\mu$ and the rest of the material parameters correspond to BiTeI, as specified in the main text.}
\end{figure}

\section{Prospects for experimental observation of the chiral-spin modes in 2D}\label{sec:other effects}

In the section, we discuss conditions under which the chiral-spin modes can be observed in semiconductor heterostructures via optical probes, which include absorption spectroscopy, both without \cite{absorb} and with a grating structure,\cite{allen} inelastic light scattering,\cite{Oleg,Babu1} and ultrafast pump-and-probe techniques.\cite{pump} We focus here on detecting the in-plane transverse ($22$) chiral-spin mode. The out-of-plane transverse ($33$) mode can be detected using the technique suggested in Ref.~\onlinecite{ali_maslov}. A detailed discussion of experimental conditions for observing the $22$ mode at $q=0$ is given in Ref.~\onlinecite{shekhter}. For completeness, we first re-visit the $q=0$ case and extend the analysis by including the Dresselhaus mechanism of SOC, and then discuss the $q\neq 0$ case.

Since disorder smears out the chiral-spin modes via the Dyakonov-Perel mechanism, the candidate material must have as little disorder as possible. Various effects arising from SOC were studied in 2D electron gases in high-mobility GaAs/AlGaAs and InGaAs/InAlAs quantum wells, and in what follows we limit our discussion to these two systems.

Rashba SOC is characterized by two energy scales: the spin-orbit splitting $\Delta=2\alpha k_F$ and the Rashba energy $E_R=m^*\alpha^2/2\hbar^2$. (In the section, we restore $\hbar$ and measure $\alpha$ in units of energy$\times$length.) Since  $\alpha/\hbar$ is much smaller than the Fermi velocity in both systems, it is permissible to use the value of $k_F$ in the absence of SOC. For future reference,
\bse
\bea
&&k_F\left[\mu\mathrm{m}^{-1}\right]=0.80\times 10^{2} \sqrt{n\left[10^{11}\mathrm{cm}^{-2}\right]},\\
&&\mu[\mathrm{meV}]=\frac{\hbar^2 k_F^2}{2m_1}=0.2\frac{m_e}{m_1} n\left[10^{11}\mathrm{cm}^{-2}\right],
\eea
\ese
where $m_e$ is the free electron mass.

The strength of SOC relative to disorder may be characterized by three ``quality factors" (two of which are related to each other)
\bse
\bea Q_1&\equiv& \Delta\tau_{\mathrm{tr}}/\hbar,\\
Q_2&\equiv& E_R\tau_{\mathrm{tr}}/\hbar,\\
Q_3&\equiv& 16Q_2^2,\eea
\ese
where $\tau_{\mathrm{tr}}$ is the transport mean free time. [Strictly speaking, the width of the chiral-spin resonance is determined  by two characteristic times: $\tau^{(1)}\equiv \tau_{\mathrm{tr}}$ and $\tau^{(2)}$, where $1/\tau^{(l)}=n_iv_F\int d\phi\left[1-\cos(l\phi)\right]d\Sigma/d\phi$, $n_i$ is the impurity number density and $d\Sigma/d\phi$ is the differential scattering cross-section of  a single impurity.\cite{shekhter} We neglect this detail here.]

The $Q_1$-factor characterizes the sharpness of the chiral-spin resonance. The condition $Q_1\gg 1$ coincides with the requirement for the Dyakonov-Perel mechanism to be in the ballistic regime and is readily achieved in high-mobility systems. The $Q_2$-factor determines whether the Rashba continuum can be detected in an optical measurement because the width of the continuum is given by $8E_R$. Also, $Q_2$ defines the height of the peak in the optical conductivity at the chiral-spin resonance frequency $\Omega_0$
\beq
\sigma_{\mathrm{peak}}=\frac{e^2}{h}Q_2.
\eeq
Finally, the ratio of $\sigma_{\mathrm{peak}}$ to the Drude conductivity at the resonance frequency $\Omega_0$  determines the contrast of the peak against the Drude background. Setting $\Omega_0=\Delta$ for an estimate, one obtains $\sigma_D\equiv\R\sigma(\Omega_0)\approx(e^2/h)\mu/\Delta^2\tau_{\mathrm{tr}}$; thus
 \beq
 \frac{\sigma_{\mathrm{peak}}}{\sigma_D}=Q_3.
 \eeq
Since $E_R\ll \Delta$ for weak SOC, both $Q_2$ and $Q_3$ are much smaller than  $Q_1$.
If $Q_3\ll 1$, the resonance is masked by the Drude tail.

In conventional units, $Q_1$, $Q_2$, and $Q_3$ can be written as
\bse
\bea
Q_1&=&1.4 \alpha[\mathrm{meV\times\mathrm{\AA}}]\sqrt{n[10^{11}\mathrm{cm}^{-2}]}\left(\frac{m_1}{m_e}\right)\mu\left[\frac{\mathrm{cm}^2}{\mathrm{V}\cdot \mathrm{s}}\right]\nn\\&&\times 10^{-5},\\
Q_2&=&6.5\left(\frac{m_1}{m_e} \alpha[\mathrm{meV\times\mathrm{\AA}}]\right)^2\mu\left[\frac{\mathrm{cm}^2}{\mathrm{V}\cdot \mathrm{s}}\right]\times 10^{-8},\nn\\
\\
Q_3&=&6.8\left(\frac{m_1}{m_e} \alpha[\mathrm{meV\times\mathrm{\AA}}]\right)^4\left(\mu\left[\frac{\mathrm{cm}^2}{\mathrm{V}\cdot \mathrm{s}}\right]\times 10^{-7}\right)^2.\nn\\
\eea
\ese

The highest reported value of $\alpha$ for a GaAs/AlGaAs heterostructure is $\alpha=5\; \mathrm{meV}\times\mathrm{\AA}$ (Ref.~\onlinecite{miller:2003}). Using $m_1=0.067m_e$, $n=10\times 10^{11}\mathrm{cm}^{-2}$, and $\mu=10^7\;\mathrm{cm}^2/\mathrm{V}\cdot\mathrm{s}$, which is available in the best samples,\cite{umansky} we obtain  $Q_1=148.0$, $Q_2=0.07$, and $Q_3=0.09$. Although a large value of $Q_1$ guarantees that the peak in $\R\sigma$ is sharp, small values of $Q_2$ and $Q_3$ make the amplitude of the peak to be very small. Also, the Rashba continuum is smeared out. The total conductivity (the sum of the Drude and resonance parts) for parameters specified above shows no discernible features associated either with the chiral-spin resonance or with the Rashba continuum. The peak becomes visible on subtracting the Drude tail; however, it is likely that this procedure will not be accurate enough when applied to real data.

It needs to be pointed out that a typical GaAs/AlGaAs quantum well has both Rashba and (linear) Dresselhaus types of SOC with comparable coupling constants. For example, $\alpha=5\; \mathrm{meV}\times\mathrm{\AA}$ and $\beta=4\; \mathrm{meV}\times\mathrm{\AA}$ (Ref.~\onlinecite{miller:2003}) or, according to a different study,\cite{ensslin} $\alpha=1.5\;\mathrm{meV}\times\mathrm{\AA}$ and $\beta=-1.4\;\mathrm{meV}\times\mathrm{\AA}$.  Without repeating all the derivations of the previous sections with Dresselhaus SOC taken account, we consider the (relevant) case when both $\alpha$ and $\beta$ are small (compared to the Fermi velocity). For a $(001)$ quantum well with both  Rashba and Dresselhaus couplings, the energy spectrum of spin-split subbands is given by
\beq
\e^{\pm}_{\bk}=\frac{\hbar^2 k^2}{2m_1}\pm k\sqrt{\alpha^2+\beta^2+2\alpha\beta\sin(2\theta_{\bk})}.
\eeq
An optical measurement probes direct transitions between the subbands. The width of the region of allowed transitions--the intersubband part of the particle-hole continuum--depends on $\alpha$ and $\beta$. At $\beta=0$ and to linear order in $\alpha$, this region reduces to a point: $\Omega=2|\alpha| k_F$.  If both $\alpha$ and $\beta$ are present, the width of the region is determined by the maximum and minimal values of the energy splitting $\Delta(\bk)=\e^{+}_{\bk}-\e^{-}_{\bk}=2 k\sqrt{\alpha^2+\beta^2+2\alpha\beta\sin(2\theta_{\bk})}$, evaluated at $k=k_F$. This gives for the width of the particle-hole continuum
\beq
2k_F\big\vert|\alpha|-|\beta|\big\vert\leq \hbar\Omega\leq 2 k_F \left(|\alpha|+|\beta|\right).
\eeq
The lower boundary of the continuum is always smaller compared to the case when only one of the two mechanisms is present. Since the chiral-spin modes are located below the continuum, their frequencies will also be reduced correspondingly. We thus see that a
competition between the Rashba and Dresselhaus mechanisms is quite detrimental for chiral-spin waves. As the chiral-spin resonance is practically invisible in the optical conductivity for parameters relevant for a GaAs/AlGaAs heterostructure  even if only the Rashba mechanism is taken into account, we conclude that this system is not an optimal material for the purpose of observing chiral-spin waves, at least at $q=0$. At finite $q$, one can maximize the combined effect of the Rashba and Dresselhaus couplings by choosing $\bq$ along the direction in which the energy splitting is maximal ($\theta=\pm \pi/4$ for the same and opposite signs of $\alpha$ and $\beta$, correspondingly). We defer a detailed analysis of this situation to another occasion.

In InGaAs/InAlAs quantum wells,  SOC is much stronger, e.g., $\alpha=100\; \mathrm{meV}\times\mathrm{\AA}$ at $n=16\times 10^{11}\;\mathrm{cm}^{-2}$ (Ref.~\onlinecite{InGaAs}), which helps to compensate for smaller mobilities typical for these structures; the highest mobilities reported for InGaAs/InAlAs samples are in the range $\mu=(2-5)\times 10^5 \;\mathrm{cm}^2/\mathrm{V}\cdot\mathrm{s}$ (Refs.~\onlinecite{sato:2001} and \onlinecite{yamada:2003}). Also, SOC in these structures is predominantly of the Rashba type,\cite{InGaAs,luo:1990} which alleviates the problem with a competition between the Rashba and Dresselhaus mechanisms.    Choosing $\alpha=100\; \mathrm{meV}\times\mathrm{\AA}$, $\mu=2\times10^5\;\mathrm{cm}^2/\mathrm{V}\cdot \mathrm{s}$, $n=16\times 10^{11}\; \mathrm{cm}^{-2}$, and\;$m_1=0.042m_e$, we obtain $Q_1=47.0$, $Q_2=0.23$, and $Q_3=0.84$. Notice that $\mu=83\;$meV at this value of $n$.
Both the Rashba continuum for non-interacting electrons and the chiral-spin resonance for interacting once are now visible in the total optical conductivity (Fig.~\ref{fig:cond_gaas}, top); on subtracting the Drude tail, the peak becomes very much pronounced  (Fig.~\ref{fig:cond_gaas}, bottom). The only free parameter in this estimate is the value of the dimensionless coupling constant for short-range interaction, which we chose as $u=0.5$.
In this case, the frequency of the $22$ mode is $\approx 0.066\mu$.

The chiral-spin mode can also be detected by measuring reflectivity rather than absorption. The disadvantage of this method in 2D is that the reflectance is proportional to the absolute value of the conductivity rather than to its real part [at least as long as $|\sigma(\Omega)|\ll c$, cf. Eq.~(\ref{eq:RR})]; therefore a large imaginary part of Drude tail also serves as a background for the resonance.
The reflectance spectrum for parameters corresponding to a InGaAs/AlGaAs  is plotted in Fig.~\ref{fig:f cond ingaas}.
The resonance feature in bare reflectance (inset) is quite faint but becomes quite pronounced on subtracting the reflectance of a quantum well without SOC (main panel).

We now turn to probing the dispersion relations of the chiral-spin modes. A conventional way to measure the dispersion of a collective mode in 2D is via imposing a grating structure, as discussed in Sec.~\ref{subsec:2D Exp consqnce}. If the thickness of the insulating layer, $d\gg1/q$, then $\epsilon_{\text{eff}}$, in Eq. (\ref{eq:eff_cond}) for the effective conductivity is reduced to $\epsilon_{\text{eff}}=(\epsilon_1+\epsilon_2)/2=12$ [from Ref. \onlinecite{dielectric}]. Using this, we calculate the effective conductivity for a GaAsIn/InGaAs quantum well for the same set of material parameters as used in calculating $\R\sigma(\Omega)$ in the bottom panel, except for $n=10^{12}$cm$^{-1}$. This is shown in the bottom panel of Fig.~\ref{fig:cond_gaas}. For clarity, the corresponding quantity without SOC was subtracted. The evolution of the peak with $q$ follows the dispersion of the in-plane transverse ($22$) mode.

\begin{figure}[htp]
$\begin{array}{c}
\includegraphics[width=0.7\columnwidth]{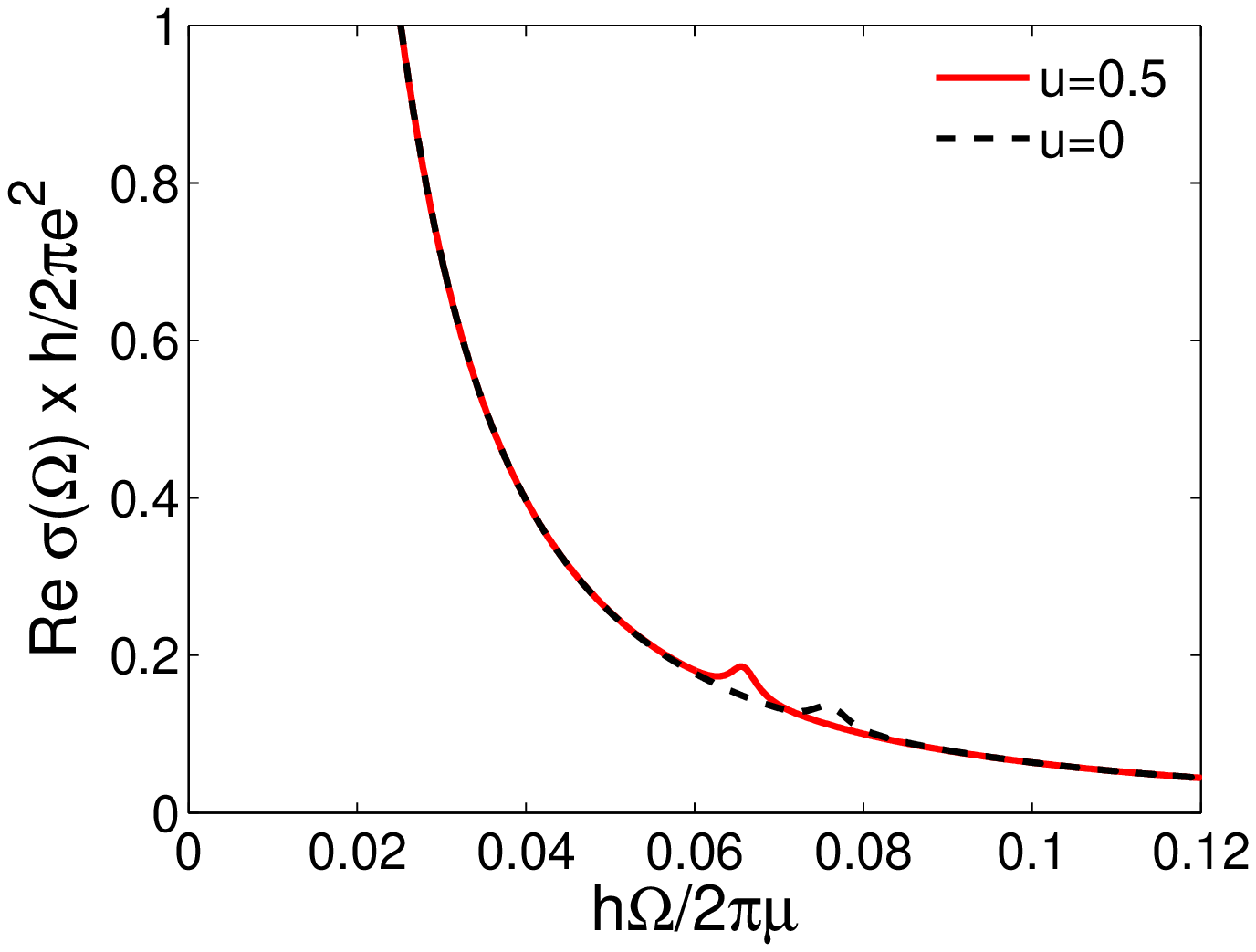}\\
\includegraphics[width=0.7\columnwidth]{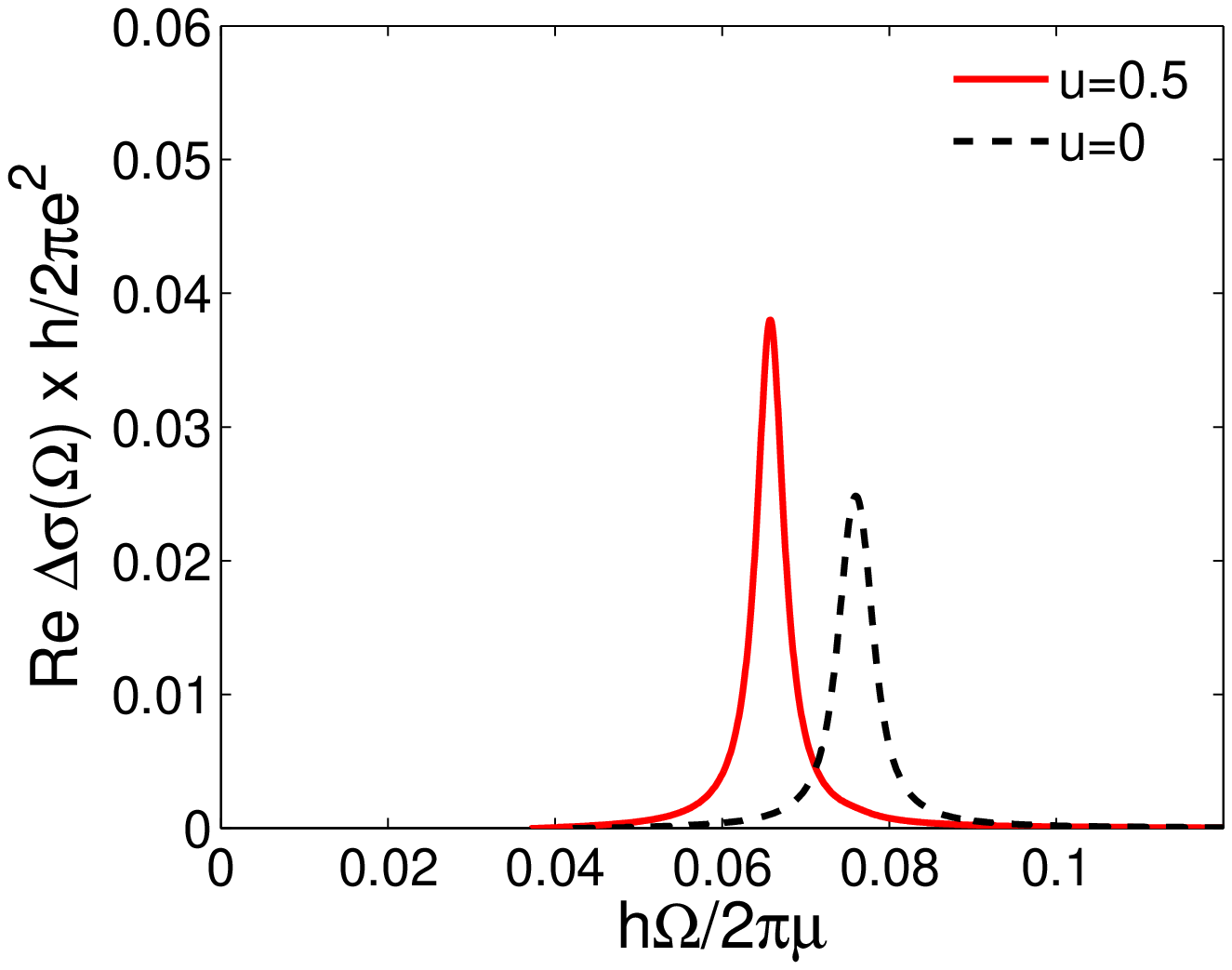}
\end{array}$\caption{
\label{fig:cond_gaas} Top: Real part of the optical conductivity for parameters corresponding to an InGaAs/AlGaAs quantum well. Solid: interacting electrons with $u=0.5$. Dashed: non-interacting electrons.
Bottom:Real part of the optical conductivity with the Drude tail removed. Material parameters: $\alpha=100\; \mathrm{meV}\times\mathrm{\AA}$, $\mu=2\times10^5\;\mathrm{cm}^2/\mathrm{V}\cdot \mathrm{s}$, $n=16\times 10^{11}\; \mathrm{cm}^{-2}$, and\;$m_1=0.042m_e$; for these values of $n$ and $m^*$, $k_F=3.2\times 10^2\;\mu\mathrm{m}^{-1}\;$\AA\; and $\mu=83\;$ meV.}
\end{figure}

\begin{figure}[htp]
\includegraphics[width=0.7\columnwidth]{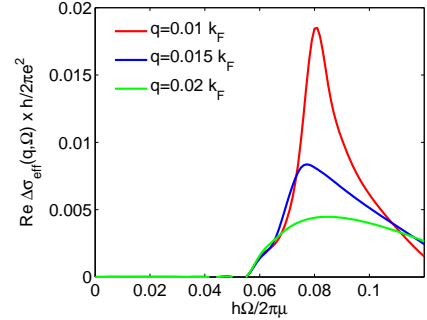}
\caption{
\label{fig:cond_gaas2} Effective conductivity [Eq.~(\ref{eq:eff_cond})] for a range of wave numbers, as indicated in the legend. The effective conductivity for a system without SOC was subtracted for clarity. Material parameters are the same as in Fig. \ref{fig:cond_gaas} except for
$n=10^{12}$cm$^{-2}$.}
\end{figure}

\begin{figure}[htp]
\includegraphics[width=0.8\columnwidth]{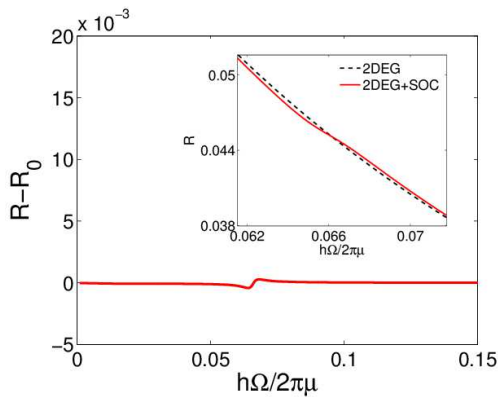}
\caption{
\label{fig:f cond ingaas} Main panel: The reflectance of an interacting 2D electron gas ($ u=0.5$)  minus the same quantity in the absence of SOC. Inset: a zoom of the bare reflectance around the resonance. Material parameters correspond to an InGaAs/AlGaAs quantum well are the same as indicated in the caption to Fig.~\ref{fig:cond_gaas}.}
\end{figure}
Recent advances in inelastic light spectroscopy made it possible to observe spin-plasmon collective modes--as defined in Sec.~\ref{sec:intro}-- in GaAs/AlGaAs\cite{Babu1} and CdMnTe\cite{Babu2} quantum wells with sub-meV resolution. This method also allows to measure dispersion relations directly, without grating, as the scattering wavenumber, $q$, is controlled by the wavenumber of the exciting light (typically, in the near-infrared range) and by the scattering geometry. References \onlinecite{Babu1} and \onlinecite{Babu2} probed the spin-plasmon dispersion at $q$ on the scale of few $\mu$m$^{-1}$, which is exactly the range of $q$ used in bottom panel of Fig.~\ref{fig:cond_gaas}. For reasons explained earlier in this section, the prospects of observing a chiral-spin mode in a GaAs/AlGaAs heterorstructure are not very promising. However, we believe that it can be observed in a high- quality InGaAs/AlGaAs heterostructrure, and inelastic light spectroscopy employed in Refs.~\onlinecite{Babu1} and \onlinecite{Babu2} appears to be well suited for this purpose.

\section{Conclusions}\label{sec:conclusion}

We have studied collective modes in 2D and 3D systems with a linear Rasbha-type SOC of arbitrary strength. For the 3D case, we assumed that the Rashba coupling is in the $xy$ (or $12$, in our notations) plane. Using a perturbative technique which combines an RPA sum for a long-range (Coulomb) and ladder sum for a
short-range (screened)  interaction, we find the collectives modes as poles of susceptibilities. Loosely speaking, the charge susceptibility, obtained by re-summing the RPA series, yields the charge collective modes while the three components of the spin susceptibility, obtained by re-summing the ladder series,
yields three collective modes in the those in the spin channel. More precisely, the charge susceptibility is coupled to the $22$ susceptibility while  the $11$ susceptibility is coupled to the $33$ susceptibility. Intermode couplings are proportional to $q$ and do not affect the masses of the modes but do affect their dispersions.

Our specific results for the 2D case are as follows. The system supports two plasmon modes: one is the usual $\sqrt{q}$ plasmon that splits off the charge continuum at $q=0$. The other one is an intersubband plasmon that splits off the upper edge of the Rashba continuum but remains exponentially close
to it. This closeness is due to logarithmic singularity in the real part of the charge susceptibility at the edges of the Rashba continuum. Thermal smearing and disorder can make it hard to detect this mode experimentally. We have shown that the Drude weight is reduced due to the SOC (this is not unique to Rashba systems but any two or more band system that mixes the bands will show such a reduction). This lost weight is exactly recovered at finite frequencies in agreement with the sum rule.

In the spin sector, we have shown that there are three chiral-spin modes. With the direction of $\bq$ chosen as the $x_1$ axis, these are a longitudinal mode with magnetization along the $x_1$ axis and  two transverse modes with magnetizations along the $x_2$ and $x_3$ axes. We refer to these modes as to $11$, $22$ and $33$ modes, correspondingly, as they occur primarily as poles in corresponding components of the spin susceptibility.  The $11$ and $22$ modes are degenerate at $q=0$ but disperse differently (as was also found in Refs.~\onlinecite{ali_maslov} and \onlinecite{Zhang}) and run quickly into the Rashba continuum. The $33$ mode is more ``robust": disperses all the way down to $\Omega=0$. The $33$ mode does not ``see" the  charge continuum despite its coupling to the $11$ mode. We have shown that the chiral-spin modes, first predicted in Refs.~\onlinecite{shekhter} and \onlinecite {ali_maslov} for the case of weak SOC, exist also for arbitrary SOC, and even if only the lowest spin subband is occupied. In the weak-SOC limit, our results for the frequencies of the modes coincide with those obtained within the FL theory. \cite{shekhter,ali_maslov} However, we have also identified another regime, in which SOC is stronger than the electron-electron interaction. In the this regime, the modes are exponentially close to the boundary of the Rashba continuum at weak electron-electron coupling
but move away from this boundary at stronger interaction. Absorption by the $22$ mode both at $q=0$ and finite $q$ (in the presence of diffraction grating) should by detectable experimentally.

In 3D, the results are qualitatively different. In the charge sector, there is only one (anisotropic) plasmon mode. If $\bq$ if out of the plane, the plasmon dispersion is independent of the SOC. If $\bq$ is in the plane, the plasmon may or may not be damped depending on the ratio of the SOC energy scale to the chemical potential. For parameters characteristic of giant Rashba semiconductors of the BiTeI family, the plasmon is damped (even at $q=0$), which  is a unique feature of 3D system with SOC. This is a consequence of the fact that the Rashba continuum extends all the way to zero frequency, in contrast to the 2D case, when the continuum occupies a finite interval of frequencies. Another consequence of the extended continuum is that the chiral-spin collective modes are Landau-damped. Some features of the 3D system are similar to 2D, e.g., there is a loss of the Drude weight which is recovered at higher frequencies. Although the conductivity spectrum is different in 2D and 3D, the kinks in the spectrum of a 3D system are at the same energies as in 2D.

We have made quantitative estimates regarding the possibility of the 22 mode to be observed in optical measurements in 2D. Our conclusion is that a high-mobility InGaAs/AlGaAs quantum well is an ideal candidate material for this purpose. Recent realization of synthetic SOC in system of cold $^{40}$K and $^6$Li atoms \cite{cold1,cold2} can also serve as an interesting platform to test some of our results.

\acknowledgements

We would like to thank D. G. Cooke, L. W. Engel, K. Ensslin, G. Gervais, J. J. Hamlin, M. Khodas, C. Martin, D. Pesin, F. Perez,  E. I. Rashba, D. B. Tanner, S. Ulloa, W. Wegscheider, and D. Zumb{\"u}hl for stimulating discussions. SM is a Dirac Post-Doctoral Fellow at the National High Magnetic Field Laboratory, which is supported by the National Science Foundation via Cooperative agreement No. DMR-1157490, the State of Florida, and the U.S. Department of Energy. DLM acknowledges support from the National Science Foundation (NSF) from grants NSF DMR-0908029 and NSF DMR-1308972. This work was supported in part by the National Science Foundation under Grant No. PHYS-1066293 and the hospitality of the Aspen Center for Physics; and also by the Kavli Institute for Theoretical Physics via Grant No. NSF PHY11-25915.

\appendix

\section{Explicit forms of $\mathcal{T}_{ij}$}\label{subsec:Tij}
\label{sec:appA}
The matrix elements for the intra/interband transitions $\mathcal{F}_{ij}^{rs}$, defined by Eq.~(\ref{eq:pi}), are
explicitly given by
\bse
\bea
\mathcal{F}_{00}^{rs}&=&1+rs\cos(\theta_\bk-\theta_{\bk+\bq})
%
\\
\mathcal{F}_{01}^{rs}&=&r\sin\theta_\bk+s\sin\theta_{\bk+\bq}
\\
\mathcal{F}_{02}^{rs}&=&-\left(r\cos\theta_\bk+s\cos\theta_{\bk+\bq}\right)
\\
\mathcal{F}_{03}^{rs}&=&irs\sin(\theta_\bk-\theta_{\bk+\bq})
\\
\mathcal{F}_{10}^{rs}&=&\mathcal{F}_{01}
\\
\mathcal{F}_{11}^{rs}&=&1-rs\cos(\theta_\bk+\theta_{\bk+\bq})
\\
\mathcal{F}_{12}^{rs}&=&-rs\sin(\theta_\bk+\theta_{\bk+\bq})
\\
\mathcal{F}_{13}^{rs}&=&-i\left(r\cos\theta_\bk-s\cos\theta_{\bk+\bq}\right)
\\
\mathcal{F}_{20}^{rs}&=&\mathcal{F}_{02}
\\
\mathcal{F}_{21}^{rs}&=&\mathcal{F}_{12}
\\
\mathcal{F}_{22}^{rs}&=&1+rs\cos(\theta_\bk+\theta_{\bk+\bq})
\\
\mathcal{F}_{23}^{rs}&=&-i\left(r\sin\theta_\bk-s\sin\theta_{\bk+\bq}\right)
\\
\mathcal{F}_{30}^{rs}&=&-\mathcal{F}_{03}
\\
\mathcal{F}_{31}^{rs}&=&-\mathcal{F}_{13}
\\
\mathcal{F}_{32}^{rs}&=&-\mathcal{F}_{23}
\\
\mathcal{F}_{33}^{rs}&=&1-rs\cos(\theta_\bk-\theta_{\bk+\bq}),
\eea
\ese
where $r,s=\pm 1$ are the chiralities of the Rashba subbands.
The integrands $\mathcal{T}_{ij}$ of the expressions (\ref{eq:pi}) for polarization bubbles are then given by
\begin{widetext}
\bse
\bea \label{eq:00}
\mathcal{T}_{00}&=&\left(g_+g_++g_-g_-\right)\left\{1+\cos(\theta_\bk-\theta_{\bk+\bq})\right\}
+\left(g_+g_-+g_-g_+\right)\left\{1-\cos(\theta_\bk-\theta_{\bk+\bq})\right\}
\\
\label{eq:0X}
\mathcal{T}_{01}&=&\left(g_+g_+-g_-g_-\right)\left\{\sin\theta_\bk+\sin\theta_{\bk+\bq}\right\}
+\left(g_+g_--g_-g_+\right)\left\{\sin\theta_\bk-\sin\theta_{\bk+\bq}\right\}
\\
\label{eq:0Y}
\mathcal{T}_{02}&=&-\left(g_+g_+-g_-g_-\right)\left\{\cos\theta_\bk+\cos\theta_{\bk+\bq}\right\}
-\left(g_+g_--g_-g_+\right)\left\{\cos\theta_\bk-\cos\theta_{\bk+\bq}\right\}
\\
\label{eq:0Z}
\mathcal{T}_{03}&=&i\left(g_+g_++g_-g_--g_+g_--g_-g_+\right)\sin(\theta_\bk-\theta_{\bk+\bq})
\\
\label{eq:X0}
\mathcal{T}_{10}&=&\mathcal{T}_{01}
\\
\label{eq:XX}
\mathcal{T}_{11}&=&\left(g_+g_++g_-g_-\right)\left\{1-\cos(\theta_\bk+\theta_{\bk+\bq})\right\}
+\left(g_+g_-+g_-g_+\right)\left\{1+\cos(\theta_\bk+\theta_{\bk+\bq})\right\}
\\
\label{eq:XY}
\mathcal{T}_{12}&=&-\left(g_+g_++g_-g_--g_+g_--g_-g_+\right)\sin(\theta_\bk+\theta_{\bk+\bq})
\\
\label{eq:XZ}
\mathcal{T}_{13}&=&-i\left(g_+g_+-g_-g_-\right)\left\{\cos\theta_\bk-\cos\theta_{\bk+\bq}\right\}
-i\left(g_+g_--g_-g_+\right)\left\{\cos\theta_\bk+\cos\theta_{\bk+\bq}\right\}
\\
\label{eq:Y0}
\mathcal{T}_{20}&=&\mathcal{T}_{02}
\\
\label{eq:YX}
\mathcal{T}_{21}&=&\mathcal{T}_{12}
\\
\label{eq:YY}
\mathcal{T}_{22}&=&\left(g_+g_++g_-g_-\right)\left\{1+\cos(\theta_\bk+\theta_{\bk+\bq})\right\}
+\left(g_+g_-+g_-g_+\right)\left\{1-\cos(\theta_\bk+\theta_{\bk+\bq})\right\}
\\
\label{eq:YZ}
\mathcal{T}_{23}&=&-i\left(g_+g_+-g_-g_-\right)\left\{\sin\theta_\bk-\sin\theta_{\bk+\bq}\right\}
-i\left(g_+g_--g_-g_+\right)\left\{\sin\theta_\bk+\sin\theta_{\bk+\bq}\right\}
\\
\label{eq:Z0}
\mathcal{T}_{30}&=&-\mathcal{T}_{03}
\\
\label{eq:ZX}
\mathcal{T}_{31}&=&-\mathcal{T}_{13}
\\
\label{eq:ZY}
\mathcal{T}_{32}&=&-\mathcal{T}_{23}
\\
\label{eq:ZZ}
\mathcal{T}_{33}&=&\left(g_+g_++g_-g_-\right)\left\{1-\cos(\theta_\bk-\theta_{\bk+\bq})\right\}
+\left(g_+g_-+g_-g_+\right)\left\{1+\cos(\theta_\bk-\theta_{\bk+\bq})\right\}
\eea
\ese
Terms that contain $\sin\theta_{\bk}$ and $\sin\theta_{\bk+\bq}$ vanish upon angular integration and, as a result, only
six out of sixteen $\Pi^0_{ij}$ survive.

The charge spin susceptibility was explicitly evaluated in Ref.~\onlinecite{Gritsev} in terms of the elliptic functions.
As our primary goal is obtain numerical results for the dispersions of the various modes, we will evaluate the integrals over the frequency and angle analytically but perform the last integration--over the magnitude over the momentum--numerically.
In what follows, we demonstrate these steps explicitly for an example of $\Pi_{00}^0$.

Carrying the frequency summation in any  $\Pi_{ij}^0$, one arrives at the combination
\beq\frac{n_F(\e^r_\bk)-n_F(\e^s_{\bk+\bq})}{i\Omega + \e^r_{\bk}-\e^s_{\bk+\bq}}.
\eeq
Focusing on the $T=0$ case, we re-group the terms with Fermi functions together and split the resulting expression for $\Pi_{00}^0(q ,\Omega)$ into two terms as
$\Pi_{00}^0(q ,\Omega) = \Pi^{(+)}_{00} + \Pi^{(-)}_{00}$, where
\begin{align}
&\Pi^{+}_{00} = -\int \frac{kdkd\theta_{\bk}}{(2\pi)^2}\Theta(\mu-\e^+_\bk) \sum_{\pm} (\pm) \frac{ \left( A_{\pm} - \frac{kq}{m_1}\cos\theta_{\bk}\right) \pm \alpha |{\bf k} \pm {\bf q}|\cos(\theta_{\bk} - \theta_{\bk \pm \bq}) }{ \left[A_{\pm} - \frac{kq}{m_1}\cos\theta_{\bk}\right]^2 - \alpha^2 |{\bf k} \pm {\bf q}|^2}
\nn\\
& = \mp \int\frac{kdk}{2\pi}\Theta(\mu-\e^{+}_{\bf k})  \frac{1}{z_{2}^{\pm} - z_{1}^{\pm}}\bigg[  \frac{A_{\pm} \pm \alpha k - (1 - \frac{m_1\alpha}{p})z_{1}^{\pm}}{\sqrt{(z_{1}^{\pm})^2 - \left(\frac{kq}{m_1}\right)^2}}\text{sgn}\R(z_{1}^{\pm})
-  \frac{A_{\pm} \pm \alpha k - (1 - \frac{m_1\alpha}{p})z_{2}^{\pm}}{\sqrt{(z_{2}^{\pm})^2 - \left(\frac{kq}{m_1}\right)^2}}\text{sgn}\R(z_{2}^{\pm})\bigg],
\end{align}
where $A_{\pm} = \bar{\Omega} \pm \alpha k \mp \frac{q^2}{2m_1}$, $\bar{\Omega} = \Omega+i\delta$, and
\begin{align}
&z_{1,2}^{+} = A_{+} - m_1\alpha^2 \pm \alpha \sqrt{\bar{k}^2 + 2m_1\bar{\Omega}},
\nn\\
&
z_{1,2}^{-} = A_{-} - m_1\alpha^2 \pm \alpha \sqrt{\bar{k}^2 - 2m_1\bar{\Omega}},
\label{fri1}
\end{align}
with $\bar{k} = k+m\alpha$. In deriving Eq.~(\ref{fri1}), we used an identity $|{\bf k} \pm {\bf q}|\cos(\theta_{\bk} - \theta_{\bk \pm \bq}) = k \pm q\cos\theta_{\bk}$.
The second part of polarization bubble is related to the one calculated above via  $\Pi_{00}^{-}(\alpha) = \Pi_{00}^{+}(-\alpha)$. After straightforward transformations, one obtains for the polarization bubble
\begin{eqnarray}
&
\Pi_{00}^0= \frac{m}{4\pi}\int_{m\alpha}^{p_0} \frac{z_{+}d\bar{k}}{\sqrt{\bar{k}^2 + 2m\bar{\Omega}}  } \bigg[ \text{sgn}\R (z^{+}_{1}) \frac{\sqrt{(2m\alpha z_{+} + \sqrt{2m\bar{\Omega}})^2 - q^2z_{+}^2}}{z_{+}^2\sqrt{2m\bar{\Omega} z_{+}^2 - q^2}}
 -   \text{sgn}\R(z^{+}_{2}) \frac{\sqrt{(\sqrt{2m\bar{\Omega}} z_{+} - 2m\alpha)^2 - q^2}}{\sqrt{2m\bar{\Omega}- q^2z_{+}^2}} \bigg]
\nn\\
& +\frac{m}{4\pi} \int_{m\alpha}^{p_0} \frac{z_{-}d\bar{k}}{\sqrt{\bar{k}^2 - 2m\bar{\Omega}}  } \bigg[  \text{sgn}\R(z^{-}_{1})\frac{\sqrt{(z_{-}\sqrt{2m\bar{\Omega}} -2m\alpha)^2 - q^2}}{\sqrt{2m\bar{\Omega} - q^2z_{-}^2}}
 -  \text{sgn}\R(z^{-}_{2})\frac{\sqrt{(2m\alpha z_{-} - \sqrt{2m\bar{\Omega}})^2 - q^2z_{-}^2}}{z_{-}^2\sqrt{2m\bar{\Omega} z_{-}^2 - q^2}} \bigg]
\nn\\
&
+ (\alpha \rightarrow -\alpha),\label{fri2}
\end{eqnarray}
where $z_{\pm} = \frac{1}{\sqrt{2m\bar{\Omega}}}\left(\bar{k} + \sqrt{\bar{k}^2 \pm 2m\bar{\Omega}}\right)$. It is important to
keep $\delta$ finite  when taking the limit of $\Omega\to 0$ in all components of the polarization bubbles;  otherwise, some contributions can be missed.\cite{Gritsev}
The one-dimensional integral in Eq.~(\ref{fri2}) can be easily calculated numerically.

All other components of the polarization bubble can be obtained in the similar fashion. For example, for $\Pi_{33}^0$ one obtains
$\Pi_{33}^0 = \Pi_{33}^{(+)} + \Pi_{33}^{(-)}$, where
\bea
\Pi^{+}_{33}& = & - \int\frac{kdkd\theta_{\bk}}{(2\pi)^2}\Theta(\mu-\e^{+}_{\bf k}) \sum_{\pm} (\pm) \frac{ A_{\pm} - \frac{kq}{m}\cos\theta_{\bk}  \mp \alpha |{\bf k} \pm {\bf q}|\cos(\theta_{\bk} - \theta_{\bk \pm \bq}) }{ \left[A_{\pm} - \frac{kq}{m}\cos\theta_{\bk}\right]^2 - \alpha^2 |{\bf k} \pm {\bf q}|^2},\nn\\
\Pi_{33}^{-} &=&- \int\frac{kdkd\theta_{\bk}}{(2\pi)^2} \Theta(\mu-\e^{-}_{\bf k}) \sum_{\pm} (\mp) \frac{ B_{\pm} - \frac{kq}{m}\cos\theta_{\bk} \mp \alpha |{\bf k} \mp {\bf q}|\cos(\theta_{\bk} - \phi_{\bk \mp \bq}) }{ \left[B_{\pm} - \frac{kq}{m}\cos\theta_{\bk}\right]^2 - \alpha^2 |{\bf k} \mp {\bf q}|^2}
\eea
where $B_{\pm} = \bar{\Omega} \pm \alpha p \pm \frac{q^2}{2m_1}$, and similarly for other components.
\end{widetext}

\section{
Optical conductivity of a non-interacting electron system with Rashba spin-orbit coupling
}\label{subsec:jj}
The optical conductivity of non-interacting electrons with Rashba SOC was calculated in Ref.~\onlinecite{ } using the Kubo formula
and in Ref.~\onlinecite{eugene_halperin} using the quantum Boltzmann equation. To keep our presentation self-contained, we reproduce the results of these two approaches here and show that resulting equation for the plasmon modes is the same as obtained within our RPA  approach in the main text
[Eq.~(\ref{eq:transcendental2})].

\subsection
{Quantum Boltzmann equation}\label{subsec:j}

The charge-density fluctuation in response to the electric field ${\bf E}$, is obtained by combining the  Poisson's equation ($-\nabla^2\phi=4\pi\rho$), continuity equation ($\dot{\rho}+\boldsymbol{\nabla}\cdot{\bf j}=0$), and Ohm's law (${\bf j}=\sigma {\bf E}$). Plasmon modes are found at the solutions of the equation  $(\Omega+4\pi i \sigma)\rho=0$ in 3D and $(\Omega+2\pi i q \sigma)\rho=0$ in 2D.

In this section, we use the quantum Boltzmann equation to find the optical conductivity $\sigma(\Omega)$. The distribution function
is a $2\times 2$ matrix in the spin basis:
\bea\label{eq:dist_func}
 \hat{f}_0 + \left(
\begin{array}{cc}
f_{11}&f_{12}\\
f_{21}&f_{22}
\end{array}\right) \equiv \hat{f}_0 + \hat{f},\eea
where $\hat{f}_0$ is the equilibrium distribution function given
by $\hat{f}_0=\frac12 + i \int \frac{d\omega}{2\pi}\hat{G}$
$=\frac12(1+\hat{\eta})n_F(\e^+) +
\frac12(1-\hat{\eta})n_F(\e^-)$ with
$\hat{\eta}=\hat\sigma_1\sin\theta-\hat\sigma_2\cos\theta$ with $\theta\equiv \theta_\bp$ and  $\hat f$ is a non-equilibrium part. In the absence of scattering, $\hat f$ satisfies the quantum Boltzmann equation
with a spatially homogeneous ${\bf E}$ in the $x-$direction
\bea\label{eq:BE} &&\frac{\partial \hat{f}}{\partial t} +
\frac{i}{2}\Delta\left[\hat{\eta},\hat{f}\right]+ e
E\frac{\partial \hat{f}_0}{\partial p_1}=0,\eea
where $\Delta\equiv 2\alpha p$.
The analysis is simplified by
switching to the chiral basis via a unitary transformation $\hat M^{\dag}(\dots)\hat M$, where
$\hat M$ is the matrix that diagonalizes the Rashba Hamiltonian (\ref{eq:Hamiltonian}):
\bea\label{eq:M}
\hat M&=&\frac{1}{\sqrt{2}}\left(
\begin{array}{cc}
1&1\\
-ie^{i\theta}&ie^{i\theta}
\end{array}\right).
\eea
Using $\hat M^{\dag}\hat\eta \hat M=\hat\sigma_3$, we obtain for the components of
$\hat{\tilde{f}} =\hat M^{\dag}\hat f \hat M$
\bea\label{eq:f_tilde} \Omega
\tilde{f}_{11}&=&iEe\delta(\e^+_{\bp})\left(\frac{p}{m_1}+
\alpha\right)\cos\theta,\nonumber\\
\Omega \tilde{f}_{22}&=&iEe\delta(\e^-_{\bp})\left(\frac{p}{m_1}-
\alpha\right)\cos\theta,\nonumber\\
(\Omega-\Delta)\tilde{f}_{12}&=&-Ee\frac{n_F(\e^+_{\bp})-n_F(\e^-_{\bp})}{2p}\sin\theta,\nonumber\\
(\Omega+\Delta)\tilde{f}_{21}&=&Ee\frac{n_F(\e^+_{\bp})-n_F(\e^-_{\bp})}{2p}\sin\theta,
\eea
where $\e^{\pm}_\bp$ are given by Eq.~(\ref{eq:Chiral spectrum}).
The current density is found as
\bea\label{eq:j1}
j_1&=&e\int\frac{d^2p}{(2\pi)^2}\text{Tr}\left[\left(\frac{p_1}{m_1}-\alpha\hat\sigma_2\right)\hat{f}\right]\nonumber\\
&=&e\int\frac{d^2q}{(2\pi)^2}\text{Tr}\left[\left(\frac{p_1}{m_1}-\alpha
M^{\dag}\hat\sigma_2 M \right)\hat{\tilde{f}} \right]\eea
Noting that,
\bea\label{eq:chiral sy}
M^{\dag}\hat\sigma_2 M&=&\left(
\begin{array}{cc}
-\cos\theta&i\sin\theta\\
-i\sin\theta&\cos\theta
\end{array}\right),
\eea
we evaluate the trace in Eq.~(\ref{eq:j1}) as
\bea\label{eq:tr}
\text{Tr}[\dots]&=&\left(\frac{q_1}{m_1}+ \alpha\cos\theta\right)\tilde{f}_{11} + \left(\frac{q_1}{m_1}-\alpha\cos\theta\right)\tilde{f}_{22} \nonumber\\
&&+ i\alpha\sin\theta\left(\tilde{f}_{12}-\tilde{f}_{21}\right)
\eea
Using $\hat{\tilde{f}}$ from Eq.~(\ref{eq:f_tilde}), we find
\bea\label{eq:tr}
\text{Tr}[\dots]&=&\frac{iEe}{\Omega}\cos^2\theta\left[\delta(\e^+_\bp)\left(\frac{q}{m_1}+\alpha\right)^2\right.\\
&&\left. + \delta(\e^-_\bp)\left(\frac{q}{m_1}-
\alpha\right)^2\right]\nonumber\\
&&+\frac{iEe}{q}\alpha\sin^2\theta[n_F(\e^+_{\bp})-n_F(\e^-_\bp)]\frac{\Delta}{\Omega-\Delta}.\nn
\eea
Using $\delta(\e^{\pm} _{\bp})= \frac{m_1}{p_0}\delta(p\pm m_1\alpha -p_0)$ and noting that the factor of $n_F(\e^+)-n_F(\e^-)$ restricts integration to the interval between $p_+$ and $p_-$, we
obtain for the conductivity
\bea\label{eq:j}
\sigma(\Omega)&=&i
e^2\left(\frac{p_0^2}{2\pi m_1 \Omega}-\frac{L(\Omega)}{16\pi}\right), \eea
where $L(\Omega)$ is given by Eq.~(\ref{eq:L}).

Plasmon modes correspond to zeros of $\Omega+2\pi i q \sigma(\Omega)$ which, on using Eq.~(\ref{eq:j}), leads to the same transcendental equation (\ref{eq:transcendental2}) as derived using the RPA in the main text.

\subsection{Evaluation of the Kubo formula for the conductivity}\label{subsec:Kubo}
Here, the evaluate the conductivity bubble in Eq.~(\ref{eq:conductivity_bubble}). First, we evaluate the trace
\bea\label{eq:tr2}
&&\text{Tr}[\dots]=\nonumber\\
&&\left[\frac{k_1^2}{m_1^2} +\frac{\alpha^2}{2}(1+\cos2\theta)+\frac{2k_1\alpha}{m_1}\cos\theta\right]\left(g_+g_+ + g_-g_-\right)\nonumber\\
&& + \frac{\alpha^2}{2}(1-\cos2\theta)(g_+g_-+g_-g_+)
\eea
keeping in mind that $g_rg_s\equiv g_r(\bk,\omega)g_s(\bk+\bq,\omega+\Omega)$
and that the conductivity is evaluated at $q=0$.
In this limit,  $\int\frac{d\omega}{2\pi}\left(g_+g_+ + g_-g_-\right)=0$. This leaves us with
\bea\label{eq:pitilde2}
\mathcal{K}(\Omega)&=&e^2\alpha^2\int\frac{kdk}{2\pi}\int\frac{d\theta}{2\pi}\int\frac{d\omega}{2\pi}\left(g_+g_- + g_-g_+\right)\nonumber\\
\eea
The angular integral is trivial as the integrand does not depend on the angle. The remaining integrals give
\beq
\int\frac{kdk}{2\pi}\int\frac{d\omega}{2\pi}\left(g_+g_- + g_-g_+\right)=-\frac{m_1}{2\pi}\left[2+\frac{\Omega}{4m_1\alpha^2}L(\Omega)\right].
\eeq
Substituting this result into $\mathcal{K}$ yields Eq.~(\ref{eq:conductivity_bubble2}) of the main text.

\section{Expansion of $\Pi^0_{00}$ up to fourth order in $q$}\label{subsec:plasmons_rtefact}
In this Appendix, we present details of the calculation leading to Eq.~(\ref{eq:3}) of the main text.
We may write $\Pi_{00}$ as a sum of contributions each pair of the Greens' function : $\Pi_{00}^{++}+\Pi_{00}^{--}+\Pi_{00}^{+-}+\Pi_{00}^{-+}$.
For $q\leq2m_1\alpha$, an exact expression for $\Pi^0_{00}$ reads
\begin{widetext}
\beq\label{eq:exact 1}
\Pi_{00}^{++}=-\frac{m_1}{2\pi^2}\int_0^qdy\int_{-y}^{y}dx \sqrt{\frac{(2p_++x)^2-q^2}{q^2-y^2}}\frac{(2p_0+x)y}{-4m^2\Omega^2 + (2p_0+x)^2y^2}
\eeq
\end{widetext}
We interested in expanding Eq.~(\ref{eq:exact 1}) (and the analogous expressions for other components of $\Pi_{00}^0$) to order $\mathcal{O}(q^4)$.
Notice that phase space of the integral in the $x-y$ plane is  $\mathcal{O}(q^2)$ by itself. This means we need to keep terms up to $\mathcal{O}(q^2)$ in the integrand. We make use of the following expansions:
\bea\label{eq:xpansion1}
\sqrt{(2p_++x)^2-q^2}&\approx & 2p_+ + x-\frac{q^2}{4p_+^2}p_0\nonumber\\
\frac{1}{-4m^2\Omega^2 + (2p_0+x)^2y^2}&\approx & -\frac{1}{4m^2\Omega^2}\left( 1+\frac{y^2p_0^2}{m^2\Omega^2} \right)\nonumber\\
\eea
and arrive at the following result
\bea\label{eq:1}
&&\Pi_{00}^{++}=\frac{m_1}{2\pi^2}\frac{p_0p_+q^2}{m^2\Omega^2}\nonumber\\
&&\left[A-\frac{q^2}{8p_+^2}A + \frac{q^2}{4p_0p_+}B + \frac{q^2p_0^2}{m_1^2\Omega^2} C\right]
\eea
where the numerical coefficients are: $A\equiv\int_0^1dt \int_{-t}^{t} ds \frac {t}{\sqrt{1-t^2}}$, $B\equiv\int_0^1dt \int_{-t}^{t} ds \frac {s^2t}{\sqrt{1-t^2}}$, and $C\equiv\int_0^1dt \int_{-t}^{t} ds \frac {t^3}{\sqrt{1-t^2}}$. Since we are only interested in sorting our the order of terms with various powers of $\Omega$ and $q$, we do not really need to evaluate the coefficients $A,$ $B$, $C$ (although this can performed easily).

The component $\Pi_{00}^{--}$ is obtained by changing $p_+$ to $p_-$ in $\Pi_{00}^{++}$. Thus the sum $\Pi_{00}^{++}+\Pi_{00}^{--}$ may be written as
\beq\label{eq:2}
\Pi_{00}^{++}+\Pi_{00}^{--}=
\frac{m_1}{2\pi}\frac{v^2q^2}{\Omega^2}\left[1+ c_1 \frac{q^2}{p_0^2} + c_2\frac{v^2q^2}{\Omega^2}\right],
\eeq
where $v = \frac{p_0}{m_1}$ and $c_i$'s are numerical coefficients which can be determined but whose particular values are not important.

Similarly, an exact form of  $\Pi_{00}^{+-}+\Pi_{00}^{-+}$ for $q<2m_1\alpha$ is

\begin{widetext}
\beq\label{eq:exact 2}
\Pi_{00}^{+-}+\Pi_{00}^{-+}=-\frac{m_1}{2\pi^2}\int_{-q}^{q}dy\int_{2p_+-y}^{2p_-+y}dx \sqrt{\frac{q^2-y^2}{(x)^2-q^2}}\frac{(2p_0+x)y}{-4m^2\Omega^2 + (2p_0+y)^2x^2}.
\eeq
\end{widetext}
After a few more steps of expansion, the final form can be written as
\bea\label{eq:33}
\Pi_{00}^{+-}+\Pi_{00}^{-+}
&&= \frac{m_1}{2\pi}\left[\left\{\frac{q^2}{m_1\Omega} + d_1 \frac{q^4 (m_1\alpha)^2}{m_1^3\Omega^3}\right\}L(\Omega)+\right.\nonumber\\
&&\left. d_2 \frac{q^4 (m_1\alpha)^2}{p_0^2}\right]
\eea with $d_i$'s being numerical coefficients. Since we need an expression for $\Pi_{00}^0$ at finite $\Omega\approx\Omega_-$, the only small quantity Eqs.~(\ref{eq:2}) and (\ref{eq:33}) is $q$. Adding up all the contributions, we obtain the result presented in Eq.~(\ref{eq:3}).

\section{Splitting of the $11$ and $22$ chiral-spin modes at finite $q$}\label{subsec:split}
In-plane rotational invariance of the Rashba Hamiltonian (\ref{eq:Hamiltonian}) ensures that the chiral-spin modes with in-plane components of magnetization ($11$ and $22$ modes) are degenerate at $q=0$. Once the direction of $\bq$ in the plane is chosen, e.g., as the $x$ axis, the $11$ and $22$ modes become longitudinal and transverse modes, correspondingly, and degeneracy is lifted.
The difference in the dispersions of the $11$ and $22$ modes at finite $q$ occurs naturally within the
FL theory \onlinecite{ali_maslov} and is also evident in the numerical results of
Ref.~\onlinecite{Zhang}. In this Appendix, we provide some details on how the lifting of degeneracy occurs within our approach.

The $q$ dependences of the various components of the polarization tensor to order $q^2$ are presented in Eq.~(\ref{eq:NonZeroPi}). Recall that the dispersions of the collective modes are given by the roots of the equation Det$(\hat\sigma_0+\frac{U}{2}\hat{\Pi}^0)=0$. Making use of the fact that $\hat{\Pi}^0$ is block-diagonal, we obtain the following set of equations
\bea
\left(\frac2U+\Pi_{00}^0\right)\left(\frac2U+\Pi_{22}^0\right)-\left[\Pi_{02}^0\right]^2&=&0,\nonumber\\
\left(\frac2U+\Pi_{11}^0\right)\left(\frac2U+\Pi_{33}^0\right)+\left[\Pi_{13}^0\right]^2&=&0.
\label{eq:zeros2}
\eea
While we can substitute  formulas from
Eqs.~(\ref{eq:NonZeroPi}) into Eq.~(\ref{eq:zeros2}) as they are, it suffices to denote the various components of $\hat\Pi^0$ as
$\Pi_{00}^0 = a_0 q^2$; $\Pi_{11}^0 = -(b+ a_1 q^2)$, $\Pi_{22} ^0= -(b+a_2 q^2)$, $\Pi_{33}^0 = -2b+a_3 q^2$, $\Pi_{02}^0 = c q$, and $\Pi_{13} ^0= d q$.
At $q=0$, we get three solutions  corresponding to $b=2/U$ (two degenerate solutions) and $b=1/U$ (one solution).  The coefficient $b$ is a function of $\Omega$ and thus these there solutions give equations for the frequencies of the three chiral modes at $q=0$. To obtain the dispersions of the modes to $\mathcal{O}(q^2)$, we look for solutions in the form $\Omega^2_j=\Omega^2_j(0)+v_j q^2$ with $j=1,~2,~3$. The exact analysis is quite cumbersome and we refrain from presenting it here as our goal only to see the splitting of the $11$ and $22$ modes. The above {\em Ansatz} for $\Omega_j$ results in $b\rightarrow b_j+ \lambda_j q^2$. Substituting the last equation into Eq.~(\ref{eq:zeros2}) and solving for $\lambda_i$'s, we obtain
\beq\label{eq:lambda}\lambda_1=\frac U2d^2 + a_1; ~\lambda_2=-\frac U2 c^2 + a_2;~ \lambda_3=-\frac U2d^2+\frac{a_3}{2}.\eeq
Since $a_{2,3}$ are coefficients of expansion in $q$ of the bare bubble, it is independent of $U$ and makes $\lambda_{1}$ and $\lambda_2$ to be different. This is indicative of the splitting between the $11$ and $22$ modes. The full effect of this splitting  is presented in Fig.~\ref{fig:velocity}.

\section{Conductivity of an interacting system at finite $q$
}\label{subsec:correction}
The current-current correlation function is found as
\bea\label{eq:cond_corr_U}
{\cal K}^U_{\mathrm{off}}(q,\Omega)&=& e^2\int_K\text{Tr}\left[\hat{v}_1(q)\hat{G}_K \hat\beta(-q)\hat{G}_{K+Q}\right],\eea
where the vertex $\beta(q)$ satisfies the finite-$q$ version of Eq.~(\ref{eq:cond_corr_U})
\bea
\hat\beta(q)=\hat v_1(q) -U\int_P \hat G(P)\hat\beta(q) \hat G(P+Q)
\eea
with $v_1(q)$ defined in Eq.~(\ref{vq}). Expanding $\hat \beta(q)$, as before, over a complete set of Pauli matrices $\hat\beta=N_a\hat\sigma_a$ we find that $N_1,N_3=0$, whereas $N_{0}$ and $N_2$ satisfy a system of integral equations
\bea\label{eq:ms}
N_0(k)&=&\frac{k_1+\frac{q}{2}}{m}-\frac U2\int_P\frac12\mathcal{T}_{00}N_0(p)-\frac U2\Pi_{22}^0 N_2,\nonumber\\
N_2&=&-\alpha-\frac U2\int_P\frac12\mathcal{T}_{02}N_0(p)-\frac U2\Pi_{02}^0N_2,
\eea
where $\mathcal{T}$'s are defined in Appendix \ref{sec:appA}. To solve this system, we
define two yet-to-be determined variables: $Q_{00}\equiv\int_P\frac12 \mathcal{T}_{00}N_0(p)$ and $Q_{02}\equiv\int_P\frac12 \mathcal{T}_{02}N_0(p)$.  The quantities $N_0(k)$ and $N_2$ are determined once $Q_{00}$ and $Q_{02}$ are found. To find the latter, we multiply the equation on $N_0(k)$ by $\mathcal{T}_{00}$ and separately by $\mathcal{T}_{02}$, and integrate over $K$. This leads to
\bea\label{eq:Qs} &&\left(1+\frac U2\Pi_{00}^0\right)Q_{00}=
\int_K\left(\frac{k_1+\frac{q}{2}}{m}\frac12\mathcal{T}_{00}\right)
-\frac U2\Pi_{00}^0\Pi_{02}^0N_2,\nonumber\\
&&Q_{02}=
\int_K\left(\frac{k_1+\frac{q}{2}}{m}\frac12\mathcal{T}_{02}\right)
-\frac U2Q_{00}\Pi_{02}^0-\frac U2\Pi_{02}\Pi_{02}^0N_2.
\nonumber\\
\eea
The integral equation is now reduced to an algebraic one where only $\mathcal {L}_{0p}\equiv \int_K\frac{k_1+\frac{q}{2}}{m}\frac12\mathcal{T}_{0p}$ need to be evaluated. Notice that at the smallest $q$ that $\mathcal{L}_{00}(q)\sim q^3$ because $\mathcal{T}_{00}\propto q^2$ provides and $k_1$ integrates out to null; $\mathcal{L}_{02}(q)\sim q^2$. Solving the Eqs.~(\ref{eq:Qs}) for $Q$'s and them back into Eq.~(\ref{eq:ms}) to find $N$'s, we obtain
\begin{widetext}
\bea\label{eq:ms2}
&&
N_2=-\left(1+\frac U2\Pi_{22}^0-\frac{(\frac U2\Pi_{02}^0)^2}{1+\frac U2\Pi_{00}^0}\right)^{-1}\left(\alpha+\frac U2L_{02}+\frac{\frac U2\Pi_{02}\frac U2 L_{00}}{1+\frac U2\Pi_{00}^0}\right),\nonumber\\
&& N_0(k)=\frac{k_1+\frac q2}{m}-\frac{\frac U2L_{00}}{1+\frac U2\Pi_{00}^0}-\frac{\frac U2\Pi_{02}^0}{1+\frac U2\Pi_{00}^0}N_2.
\eea
\end{widetext}
The pole in the spin susceptibility corresponds to $1+\frac U2 \Pi_{22}^0=0$. Because $N_0$ and $N_2$ have their own poles,
the pole as seen in the conductivity at finite $q$ is shifted with respect to that in the spin susceptibility and is determined from $a+\frac U2 \Pi_{22}^0=0$, where $a=1-{(\frac U2\Pi_{02}^0)^2}({1+\frac U2\Pi_{00}^0})^{-1}$.  For small $q$, $a-1 \propto U^2q^2$ because $\Pi_{02}^0\propto q$.


\begin{thebibliography}{5}
\bibitem{footnote} 2D integration was carried out over a bandwidth $\sim 3 E_F$ to ensure convergence. Finite damping of $1\%$ of the mode mass was added to avoid a numerical instability when integrating around the poles.
\end{thebibliography}
\end{document}